\documentclass[%
 reprint,
 amsmath,amssymb,
prb,
]{revtex4-1}

\usepackage{graphicx}
\usepackage{dcolumn}
\usepackage{bm}

\usepackage{color}

\begin{document}

\title{{A consistent multiphase-field theory
for interface driven multi-domain dynamics}}

\author{Gyula I. T\'oth}
\email{gyula.toth@ift.uib.no}
\affiliation{Department of Physics and Technology, University of Bergen, All\'egaten 55, Bergen, Norway}
\affiliation{Wigner Research Centre for Physics, P.O. Box 49, H-1525 Budapest, Hungary}

\author{Tam\'as Pusztai}
\affiliation{Wigner Research Centre for Physics, P.O. Box 49, H-1525 Budapest, Hungary}

\author{L\'aszl\'o Gr\'an\'asy}
\affiliation{Wigner Research Centre for Physics, P.O. Box 49, H-1525 Budapest, Hungary}
\affiliation{BCAST, Brunel University, Uxbridge, UB8 3PH, Middlesex, United Kingdom}

\date{\today}

\begin{abstract}
We present a new multiphase-field theory for describing pattern formation in multi-domain and/or multi-component systems. The construction of the free energy functional and the dynamic equations is based on criteria that ensure mathematical and physical consistency. We first analyze previous multiphase-field theories, and identify their advantageous and disadvantageous features. On the basis of this analysis, we introduce a new way of constructing the free energy surface, and derive a generalized multiphase description for arbitrary number of phases (or domains). The presented approach retains the variational formalism; reduces (or extends) naturally to lower (or higher) number of fields on the level of both the free energy functional and the dynamic equations; enables the use of arbitrary pairwise equilibrium interfacial properties; penalizes multiple junctions increasingly with the number of phases; ensures non-negative entropy production, and the convergence of the dynamic solutions to the equilibrium solutions; and avoids the appearance of spurious phases on binary interfaces. The new approach is tested for multi-component phase separation and grain coarsening.  

\end{abstract}

\pacs{Valid PACS appear here}
\maketitle


\section{Introduction}

Despite recent advances in atomic scale continuum modeling of crystalline freezing \cite{Elder2002,Elder2007,Wu2010,NanaXPFC2012,Emmerich2012}, and efforts relying on the orientation field models \cite{KobayashiWarren1998, Granasy2002,Plapp2012,Granasy2014,Haataja2014}, the phase-field theoretical methods based on the multiphase-field (MPF) concept remain the method of choice, when addressing complex polycrystalline or multiphase/multi-component problems, such as multi-component phase separation or grain coarsening. A common feature of these models is that the individual physical phases / chemical components / solid grains  are described by separate fields $\mathbf{u}(\mathbf{r},t) = [u_1(\mathbf{r},t),u_2(\mathbf{r},t),\dots,u_N(\mathbf{r},t)]$. A variety of this kind of models is available in the literature ranging from the early formulations by Chen and Yang \cite{ChenYang1994}, Steinbach \cite{Steinbach1996}, Steinbach and Pezzola \cite{SteinbachPezzola1999}, Chen and coworkers \cite{FanChen1997a,FanChen1997b}, via later descendants by Nestler and coworkers \cite{Nest1,Nest2,Nest3,Nest4,Nest5}, Moelans and coworkers \cite{Moelans1,Moelans2,Moelans3,Moelans4}, to more recent developments by Folch and Plapp \cite{FolchPlapp2003,FolchPlapp2005}, Kim {\it et al.} \cite{Kim2006}, Takaki {\it et al.} \cite{Takaki2008}, Steinbach \cite{Steinbach2009}, Ofori-Opoku and Provatas \cite{NanaNick2010}, Cogswell and Carter \cite{Cogswell2011}, Bollada {\it et al.} \cite{Bollada2012}, by Emmerich and coworkers \cite{Kundin2013}, and Kim {\it et al.} \cite{Kim2014}. These models differ in important details that improve the individual models in various respects relative to the others. It is, therefore, desirable to compare them from a theoretical viewpoint, and identify the possible advantages / disadvantages they have relative to each other, to see whether a more general formulation that unifies the advantageous features can indeed be constructed on the basis of the work done so far in this field.\\ 

In attempting to develop a consistent description of interface driven multi-domain dynamics, we need to first identify the criteria the models have to satisfy. A few of such criteria have already been formulated along the development of the MPF models:

(i) The \textit{multiphase-field} descriptions view $u_i(\mathbf{r},t)$ as the local and temporal volume/mass/mole fraction of the component/grain, prescribing thus $\sum_{i=1}^N u_i(\mathbf{r},t)= 1$. (This work concentrates exclusively on these MPF models; the \textit{multi order parameter} theories \cite{ChenYang1994,FanChen1997a,FanChen1997b,Moelans1,
Moelans2,Moelans3,Moelans4,NanaNick2010}, which do not require this criterion, will be addressed elsewhere.) 

(ii) A further natural requirement is that the physical results should be independent of the labeling of the variables. This condition is termed the ''principle of formal indistinguishability'' of the fields.

(iii) The solution of the dynamic equations should tend towards the equilibrium solution obtained from the respective Euler-Lagrange equations (ELEs) based on the free energy functional, where the equilibrium solution minimizes then the free energy of the system. 

(iv) As time evolves the free energy of the total system should decrease monotonically (second law of thermodynamics).

(v) It is an evident requirement that the formulations for different numbers of phases or grains should be consistent with each other; i.e., it should be possible to recover the respective models from each other, when adding or removing a new phase/grain/orientation. It has been suggested recently that the usual variational approach to the MPF model does not satisfy this condition \cite{FolchPlapp2005,Bollada2012}. 

(vi) Another fairly general requirement, formulated by several authors, is that (a) the two-phase planar interfaces should represent a (stable) equilibrium, and should be free of additional phases. This requirement can be extended to the dynamics (b) as follows: if a phase is not present, it should not appear deterministically (in the absence of fluctuations) at any time. This is often called the condition of "no spurious phase generation". The applicability of this requirement makes depends on the problem addressed: In the case of grain coarsening, for instance, an uncontrolled appearance of new grains / orientations at the grain boundaries needs to be avoided. (In other problems, however, metastable structures \cite{tenWolde1995,tenWolde1996,tenWolde1997} or precipitates \cite{Provatas1996} may appear at the solid-liquid interface, requiring models that are able to describe such phenomena \cite{ShenOxtoby1996,GranasyOxtoby2000,Toth2007,Toth2011}.) This criterion for the absence of a third phase has been enforced different ways in different models. For example, Folch and Plapp \cite{FolchPlapp2005} have defined the free energy surface so that in the binary equilibrium the system stays in a two-phase subspace of the respective Gibbs simplex. Bollada {\it et al.} \cite{Bollada2012}, in turn, used the mobility matrix to force the system to avoid the formation of the third phase irrespectively of the free energy surface (and thus the equilibrium states of the system).

Finally, a practical requirement:

(vii) The model should allow the prescription of independent data for the interfacial properties and the kinetic coefficient of the individual phase pairs (including their possible anisotropy).\\

While most of these criteria formulate natural/self-evident requirements, some of them were neglected when developing previous MPF models. 

In the present paper, we formulate an MPF approach that obeys all the criteria defined above. {Herein, we address {\it interface driven} phenomena, in which the free energy density incorporates exclusively the ''interfacial'' contributions, comprising the gradient energy and multi-well terms, whereas driving forces associated with tilting functions\cite{FolchPlapp2005} are not considered. (An extension that includes the latter will be outlined elsewhere.)} The structure of our paper is as follows. In Section II, we present the mathematical formulation of criteria (i) to (vii). In Section III, we first investigate which of these criteria are satisfied and which are not by the existing MPF models. We also point out which features of the individual approaches can be adopted in developing a consistent theory. In Section IV we outline the generalized MPF formulation (henceforth abbreviated as XMPF) that satisfies criteria (i) to (vii). In Section V, we perform illustrative simulations using the XMPF approach to demonstrate the robustness of the theory. Section VI is devoted to a comparison with other models. Finally, in Section VII, we offer a few concluding remarks.

\section{Criteria of physical consistency}

In this section, we present and discuss mathematical formulations of criteria (i) to (vii) identified above in details.

\subsection{Free energy functional formalism}

In the multiphase approach, the following local constraint [criterion (i)] applies for the variables:
\begin{equation}
\label{eq:gencond}
\sum_{i=1}^N u_i(\mathbf{r},t) = 1 \enskip .
\end{equation}
As result of this constraint Eq. (\ref{eq:genfunc}) is not an order parameter model, since normally different order parameters capturing various aspects of symmetry breaking are coupled to each other via physical laws, whereas here Eq. (\ref{eq:genfunc}) prescribes a rather specific relationship: $u_i(\mathbf{r},t)$ represents the local fraction of the $i-$th phase, not identifiable as a quantity, whose magnitude is associated with the extent of symmetry breaking.\\

The general interface contribution of a multiphase free energy functional is usually written as:
\begin{equation}
\label{eq:genfunc}
F[\mathbf{u}] = \int dV \left\{ \frac{\epsilon^2}{2} \sum_{i=1}^N A_{ij}(\nabla u_i \cdot \nabla u_j) + w \, g(\mathbf{u})  \right\} \enskip ,
\end{equation}
where $\mathbf{u}(\mathbf{r},t)$ is the vector of the variables, $g(\mathbf{u})$ is the free energy density landscape, and $\mathbb{A}$ is a coefficient matrix of the general quadratic term for the gradients. For example, choosing $\mathbb{A}=\mathbb{I}$ (where $\mathbb{I}$) is the identity matrix yields a simple sum of the gradient square terms, $\mathbb{A}=\mathbb{I}-\mathbf{e} \otimes \mathbf{e}$ [where $\mathbf{e}=(1,1,\dots,1)$] results in a pure pairwise construction, while $A_{i,i}=\sum_{j \neq i}u_j^2$, $A_{ij \neq i}=-u_i u_j$ corresponds to the anti-symmetrized (Landau-type) gradient term. The (generally) $\mathbf{u}$-dependent coefficients $\epsilon^2$ and $w$ can be related to the pairwise interfacial properties (the interfacial free energy and the interface thickness).\\ 

The equilibrium solution can be found by solving the multiphase Euler-Lagrange equations:
\begin{equation}
\label{eq:genEL}
\frac{\delta F}{\delta u_i} = \lambda(\mathbf{r}) \enskip , \quad i=1\dots N \enskip ,
\end{equation}
where $\delta F/\delta u_i$ is the functional derivative of the free energy functional with respect to $u_i(\mathbf{r})$, and $\lambda(\mathbf{r})$ is a Lagrange multiplier emerging from the local constraint described by Eq. (\ref{eq:gencond}). Eliminating the Lagrange multiplier results in
\begin{equation}
\label{eq:genELv}
\frac{\delta F}{\delta u_i} = \frac{\delta F}{\delta u_j} \quad \text{for} \quad i,j=1\dots N \enskip ,
\end{equation}
thus offering the most general description of equilibrium.\\

For non-conserved variables, the dynamic equations are written in the following general form:
\begin{equation}
\label{eq:gendin}
- \frac{\partial u_i}{\partial t} = \sum_{j=1}^N L_{ij} \frac{\delta F}{\delta u_j} \enskip ,
\end{equation}
where the mobility matrix can be determined from the following conditions:
\begin{enumerate}
\item The time derivatives sum up to zero [it follows from criterion (i)]: 
\begin{equation*}
\quad \sum_{i=1}^N \frac{\partial u_i}{\partial t} = 0 \enskip .
\end{equation*}
\item The variables are not labeled, i.e., none of them is distinguished on the basis of its index [criterion (ii)].
\item The solutions of the Euler-Lagrange equations must be stationary solutions of the dynamic equations [a requirement that follows from criterion (iii)]: 
\begin{equation*}
\left. \frac{\partial \mathbf{u}}{\partial t}\right|_{\mathbf{u}^*(\mathbf{r})}=0 \enskip ,
\end{equation*} 
where $\mathbf{u}^*(\mathbf{r})$ stands for a solution of Eq. (\ref{eq:genELv}).
\end{enumerate}
Applying condition 1 for Eq. (\ref{eq:gendin}) results in the general form
\begin{equation}
\label{eq:gendin2}
- \frac{\partial u_i}{\partial t} = \sum_{j=1}^N \kappa_{ij} \left( \frac{\delta F}{\delta u_i} - \frac{\delta F}{\delta u_j} \right) \enskip ,
\end{equation}
where $\kappa_{ij}>0$ still can be arbitrary. Furthermore, using condition 2 yields
\begin{equation}
\label{eq:gendin3}
\sum_{i=1}^N \kappa_{ij} = 0
\end{equation}
for $j=1 \dots N$. Note that Eq. (\ref{eq:gendin2}) and (\ref{eq:gendin3}) resulted in a mobility matrix, whose elements sum up to zero in each row and column. Finally, condition 3 results in a symmetry condition, i.e. 
\begin{equation}
\kappa_{ij}=\kappa_{ji}
\end{equation} 
(for the derivation, see Appendix A). Since $\kappa_{ii}=-\sum_{j\neq i}\kappa_{ij}$, $N(N-1)/2$ mobilities can be chosen arbitrarily.\\

Next we have to test the mobility matrix against the time dependence of the total free energy. The main reason of applying a linear approximation for the dynamic equations is to establish a free energy minimizing behavior [criterion (iv)]. Using Eq. (\ref{eq:gendin}), the condition for the time derivative of the total free energy reads as
\begin{equation*}
\begin{split}
\frac{dF}{dt} &= \int dV \left\{ \sum_{i=1}^N \frac{\delta F}{\delta u_i} \frac{\partial u_i}{\partial t} \right\} =\\
&= - \int dV \left\{ (\delta F_\mathbf{u}) \mathbb{L} (\delta F_\mathbf{u})^T \right\} \leq 0 \enskip ,
\end{split}
\end{equation*}
where $\mathbb{L}$ is the mobility matrix, and we use the short notation $\delta F_\mathbf{u}=\left( \frac{\delta F}{\delta u_1},\frac{\delta F}{\delta u_2},\dots,\frac{\delta F}{\delta u_N} \right)$. Since the free energy must decrease in any volume, we can write
\begin{equation}
\label{eq:Lcond}
 (\delta F_\mathbf{u}) \mathbb{L} (\delta F_\mathbf{u})^T \geq 0 \enskip ,
\end{equation}
indicating that the mobility matrix must be \textit{positive semidefinite}. A special, frequently made choice for the mobility matrix is the \textit{Lagrangian} mobility $\kappa_{ij}\equiv 1/N$. Although this matrix is used quite widely, the dynamic equations are necessarily $N-dependent$ [i.e., criterion (v) is not satisfied], and it does not solve the problem of spurious phase appearance in non-equilibrium processes in general, as will be discussed later.    

\subsection{Spurious phases}

Herein we address pattern coarsening phenomena, for which the requirement of 'no spurious phase appearance' needs to be satisfied [criterion (vi)]. In general, this criterion means that any $p$--phase equilibrium solution (i.e., when exactly $p$ fields are present) must be stable against the appearance of a new phase. Accordingly, assuming that $q=N-p$ of the $N$ phases (namely, $i_1,i_2,\dots,i_q$) are missing in an equilibrium solution $\mathbf{u}^*(\mathbf{r})$, the condition can be re-formulated as:
\begin{equation*}
\delta F = F[\mathbf{u}^*(\mathbf{r})+\delta \mathbf{u}(\mathbf{r})]-F[\mathbf{u}^*(\mathbf{r})] \geq 0
\end{equation*}
for \textit{any} small perturbation for which $\sum_{k=1}^N\delta u_k=0$ and at least one of $\delta u_{i_1}(\mathbf{r}),\delta u_{i_2}(\mathbf{r}),\dots,\delta u_{i_q}(\mathbf{r})$ is not equal to zero. In other words, leaving the $p=N-q$ dimensional subspace (together with keeping the local constraint, naturally) always has to result in higher energy. This condition is satisfied for the equilibrium solutions representing minima of the free energy functional: Since
\begin{equation*}
F[\mathbf{u}^*(\mathbf{r})+\delta \mathbf{u}(\mathbf{r})] = F[\mathbf{u}^*]+\frac{\delta F}{\delta \mathbf{u}}\cdot \delta \mathbf{u}^T + \frac{\delta \mathbf{u} \cdot \mathbb{D} \cdot \delta \mathbf{u}^T}{2} + \dots \enskip ,
\end{equation*}
the second term on the right hand side vanishes for a solution of the Euler-Lagrange equation: $(\delta F/\delta \mathbf{u})\cdot\delta\mathbf{u}=\lambda(\mathbf{r})\sum_{i=1}^N\delta u_i=0$. Therefore, if $\mathbb{D}$ is positive definite, the equilibrium solution is a minimum. Consequently, if the binary planar interfaces represent minima of the multiphase functional, they are stable against the appearance of additional phases, which is a crucial requirement from the viewpoint of criterion (vi).\\

To fulfill criterion (vi), we also have to ensure the proper dynamic behavior of the system. The condition of 'no spurious phase appearance' can be generalized for the non-equilibrium regime as follows: If a phase is not present in the system, it must not appear deterministically, i.e. 
\begin{equation}
\label{eq:spuricondd}
\left.\frac{\partial u_i}{\partial t}\right|_{u_i(\mathbf{r},t)=0}=0 \enskip .
\end{equation} 
Unfortunately, in an \textit{arbitrary}, $p$--phase non-equilibrium state $\mathbf{u}(\mathbf{r})$, the condition $\delta F=F[\mathbf{u}(\mathbf{r})+\delta\mathbf{u}(\mathbf{r})]-F[\mathbf{u}(\mathbf{r})]>0$  cannot be satisfied for all possible perturbations. Therefore, Eq. (\ref{eq:spuricondd}) cannot be guaranteed on the level of the free energy functional \textit{in general}. Note, however, that the mobility matrix defines a 'conditional functional derivative', which allows the system to leave the $p$--dimensional subspace only in particular directions [or, in other words, not any $\mathbf{u}(\mathbf{r})+\delta\mathbf{u}(\mathbf{r})$ state is available from $\mathbf{u}(\mathbf{r})$ in the dynamics]. For special mobility matrices, like the Lagrangian matrix, one may find such a free energy functional that satisfies Eq. (\ref{eq:spuricondd}) \cite{FolchPlapp2005}. Nevertheless, the free energy functional should not depend on the form of the mobility matrix in general. This problem is resolved in a recent work \cite{Bollada2012}, in which the authors choose a mobility matrix having vanishing rows (and columns) for the fields not being present. {Although this concept trivially results in Eq. (\ref{eq:spuricondd}), the application of such a matrix together with a particular free energy functional can be 'dangerous' in the sense that it may generate stationary solution from a non-equilibrium state, a possibility that has to be checked for all solutions obtained.}

\section{Analysis of previous MPF descriptions}
 
In this section, we analyze the most frequently used multiphase-field theories from the viewpoint of equilibrium solutions. We check whether the trivial extension of the planar interface emerging from the binary reduction of the free energy functional is a solution of the multiphase problem too. Summarizing, we require that the equilibrium solution for the pure binary planar interface in the multiphase-field problem ($N \le 3$) coincides with the equilibrium solution for the binary planar interface of the binary ($N$ = 2) problem [criterion (iv)]. 
The methodology for testing this feature consists of the following steps: 
\begin{enumerate}

\item Take the free energy functional in the two-phase limit;

\item Solve the respective Euler-Lagrange equation of the two-phase problem for planar interface geometry, resulting in $u(x)$;

\item Make a natural multiphase extension of $u(x)$ via adding zero additional fields as needed for the multiphase case, i.e. $u_i(x):=u(x)$, $u_{j\neq i}(x):=1-u(x)$, and $u_{k \neq i,j}(x)=0$, where $1\leq i,j,k \leq N$;

\item Plug the extended solution into the Euler-Lagrange equations of the multiphase problem described by Eq. (\ref{eq:genEL}), and check whether it satisfies them.     
\end{enumerate}

The test can be simplified in case of free energy functionals constructed exclusively from pairwise contributions, i.e.
\begin{equation*}
F = \int dV \left\{ \sum_{i<j} \hat{f}(u_i,u_j) \right\} \enskip ,
\end{equation*}
where $\sum_{i<j}=\sum_{i=1}^{N-1} \sum_{j=i+1}^N$. $\hat{f}(u,v)$ is called \textit{generator}. The functional derivatives read as:
\begin{equation}
\label{eq:pairfunc}
\frac{\delta F}{\delta u_i} = \sum_{j \neq i} \delta\hat{f}(u_i,u_j) \enskip , 
\end{equation}
where
\begin{equation*}
\delta\hat{f}(u_i,u_j) = \frac{\partial \hat{f}}{\partial u_i} - \nabla \frac{\partial \hat{f}}{\nabla\partial u_i} \enskip .
\end{equation*}
\textit{Assuming that $\delta\hat{f}(u_i,0)=0$}, and plugging in the extended planar interface solution into the Euler-Lagrange equations of the multiphase problem yield
\begin{eqnarray*}
\frac{\delta F}{\delta u_i} &=& \delta\hat{f}[u(x),1-u(x)] = \lambda(x) \\
\frac{\delta F}{\delta u_j} &=& \delta\hat{f}[1-u(x),u(x)] = \lambda(x) \\
\frac{\delta F}{\delta u_k} &=& \delta\hat{f}[0,u(x)]+\delta\hat{f}[0,1-u(x)] = \lambda(x) \enskip ,
\end{eqnarray*}
where $u_i(x)=u(x)$, $u_{j\neq i}(x)=1-u(x)$, and $u_{k\neq i,j}(x)=0$, and $u(x)$ represents the planar binary solution. From the equations above follows that
\begin{equation}
\label{eq:simplified}
\delta\hat{f}(u,1-u) = \delta\hat{f}(1-u,u) = \delta\hat{f}(0,u)+\delta\hat{f}(0,1-u) 
\end{equation}
must apply. If Eq. (\ref{eq:simplified}) does not apply, the binary planar interface is not an equilibrium solution of the multiphase problem.

\subsection{Steinbach et al.}

In 1996, Steinbach {\it et al.} \cite{Steinbach1996} proposed the following free energy functional for multiphase systems, which serves as the basis for the worldwide used phase-field software MICRESS \cite{MICRESS}:
\begin{equation}
\label{eq:sfunc}
F = \int dV \left\{ f_{\rm intf}(\mathbf{u},\nabla\mathbf{u}) + f_{\rm df}(\mathbf{u})  \right\} \enskip ,
\end{equation}
where $\sum_{i=1}^N u_i(\mathbf{r},t)=1$, whereas the term $f_{\rm df}(\mathbf{u})$ is responsible for the thermodynamic driving force (i.e. free energy difference between the bulk phases) and $f_{\rm intf}(\mathbf{u},\nabla\mathbf{u})$ denotes the interface energy consisting of a gradient [$f_{\rm gr}(\mathbf{u},\nabla\mathbf{u})$] and a multi-well [$f_{\rm mw}(\mathbf{u})$] contribution:
\begin{equation}
\label{eq:sintf}
f_{\rm intf}(\mathbf{u},\nabla\mathbf{u})=f_{\rm gr}(\mathbf{u},\nabla\mathbf{u})+f_{\rm mw}(\mathbf{u}) \enskip ,
\end{equation}
where the terms are given in the following specific forms:
\begin{eqnarray}
\nonumber f_{\rm gr}(\mathbf{u},\nabla\mathbf{u}) &=& \sum_{i<j} \frac{\epsilon_{ij}^2}{2}(u_i\nabla u_j-u_j\nabla u_i)^2 \\
\label{eq:smulti} f_{\rm mw}(\mathbf{u}) &=& \sum_{i<j} \frac{w_{ij}}{2}(u_i u_j)^2 \enskip .
\end{eqnarray}
The functional naturally reduces to the standard binary form 
\begin{equation*}
F = \int dV \left\{ \frac{\epsilon^2_{ij}}{2}(\nabla u)^2 + \frac{w_{ij}}{2} [u(1-u)]^2 \right\} \enskip .
\end{equation*}
The 1D Euler-Lagrange equation reads as: $\delta F/\delta u=w_{ij}u(1-u)(1-2u)-\epsilon_{ij}^2 \partial_x^2 u=0$, thus resulting in the usual
\begin{equation*}
u(x) = \frac{1+\tanh[x/(2\,\delta_{ij})]}{2}
\end{equation*}
planar interface solution, where $\delta_{ij}^2=\epsilon_{ij}^2/w_{ij}$. As a first step, we investigate whether the extension of this solution minimizes the multiphase problem. Since $f_{intf}(\mathbf{u},\nabla\mathbf{u})$ is a pure pairwise construction, it is enough to take the generator, which reads as:
\begin{equation*}
\hat{f}(u_i,u_j) = \frac{\epsilon_{ij}^2}{2}(u_i\nabla u_j-u_j\nabla u_i)^2 + \frac{w_{ij}}{2}u_i^2 u_j^2 \enskip , 
\end{equation*}
yielding
\begin{equation}
\label{eq:selgen}
\begin{split}
\delta\hat{f}(u_i,u_j) &= w_{ij} u_i u_j^2 + \epsilon^2_{ij} \left[ 2(u_i\partial_x u_j-u_j\partial_x u_i)\partial_x u_j+ \right. \\
& + \left. (u_i\partial_x^2 u_j - u_j\partial_x^2 u_i)u_j \right] \enskip .
\end{split}
\end{equation}
Note that $\delta\hat{f}(u,0)=0$. Substituting $u_i(x)=u(x)$, $u_{j\neq i}(x)=1-u(x)$ and $u_{k \neq i,j}(x)=0$ into Eq. (\ref{eq:selgen}), yields 
\begin{eqnarray}
\label{eq:sbfd1}&&\frac{\delta F}{\delta u_i} = \delta\hat{f}(u,1-u) = \frac{w_{ij}}{4} \text{sech}^4[x/(2\,\delta_{ij})] \\
\label{eq:sbfd2}&&\frac{\delta F}{\delta u_j} = \delta\hat{f}(1-u,u) = \frac{w_{ij}}{4} \text{sech}^4[x/(2\,\delta_{ij})] \\
\label{eq:sbfd3}&&\frac{\delta F}{\delta u_k} = \delta\hat{f}(0,u)+\delta\hat{f}(0,1-u) = 0 \enskip .
\end{eqnarray}
These equations clearly show that \textit{the planar binary interfaces are not equilibrium solutions of the multiphase problem}.\\

The dynamic equations read as:
\begin{equation*}
-\frac{\partial u_i}{\partial t}=\sum_{i=1}^N \kappa_{ij} \left( \frac{\delta F}{\delta u_i}-\frac{\delta F}{\delta u_j} \right) \enskip ,
\end{equation*}
which are variational, therefore, the planar binary interfaces are not a stationary solutions of these. To avoid the problem, the authors used a "binary approximation" of $\frac{\delta F}{\delta u_i}-\frac{\delta F}{\delta u_j}$, in which all terms of $k \neq i,j$ indices are neglected \cite{Steinbach1996}:
\begin{equation*}
\frac{\delta F}{\delta u_i}-\frac{\delta F}{\delta u_j} \approx w_{ij}u_i u_j (u_j-u_i)+\epsilon_{ij}^2(u_i\nabla^2 u_j - u_j \nabla^2 u_i)
\end{equation*}
 Although the planar binary interfaces represent stationary solutions of the resulting \textit{non-variational} dynamics, Eq. (\ref{eq:Lcond}) does not apply, therefore, \textit{the dynamics is not energy minimizing in principle}, as it will be demonstrated later.

\subsection{Steinbach-Pezzolla}

In the Steinbach-Pezzolla formalism (published first in 1999 \cite{SteinbachPezzola1999}, and adopted in various works \cite{Takaki2008,Steinbach2009,Cogswell2011} including the OpenPhase software \cite{OpenPhase}), the interface contribution to the free energy is given by the following simplified form of Eq. (\ref{eq:sintf}):
\begin{equation}
\label{eq:spmodel}
f_{\rm intf}^{\rm SP} = \sum_{i<j} \frac{4 \sigma_{ij}}{\eta} \left( |u_i| \, |u_j|  - \frac{\eta^2}{\pi^2} \nabla u_i \cdot \nabla u_j \right) \enskip ,
\end{equation}
where the gradient term $-\nabla u_i \cdot \nabla u_j$ is the linear approximation of $(u_i\nabla u_j-u_j\nabla u_i)^2$, $\sigma_{ij}$ the interfacial free energy of the equilibrium $(i,j)$ planar interface, while the local term $|u_i|\, |u_j|$ is responsible for the finite interface width given by $\eta$. Reducing the model to the $N=2$ case yields
\begin{equation}
\label{eq:sp2red}
f_{\rm intf}^{\rm SP} = \frac{4 \sigma_{ij}}{\eta} \left[ \frac{\eta^2}{\pi^2} (\nabla u)^2 + | u |\, | 1-u | \right] \enskip ,
\end{equation}
where we used $u(x):=u_i(x)$, $u_j(x):=1-u(x)$, and $u_{k \neq i,j}(x)=0$. The 1D Euler-Lagrange equation reads as
\begin{equation*}
 {\rm sign}(u)\, |1-u| - {\rm sign}(1-u)\, |u| = \left( \frac{2 \eta^2}{\pi^2} \right) \partial_x^2 u(x) \enskip ,
\end{equation*}
from which
\begin{equation*}
u(x) = \frac{1+\sin[(\pi/\eta) \cdot x]}{2}
\end{equation*}
emerges for the planar binary interface. The generator function reads as
\begin{equation*}
\hat{f}(u_i,u_j) = \frac{4 \sigma_{ij}}{\eta} \left( |u_i|\,|u_j| - \frac{\eta^2}{\pi^2} \nabla u_i \cdot \nabla u_j \right) \enskip ,
\end{equation*}
therefore,
\begin{equation*}
\delta\hat{f}(u_i,u_j) = \frac{4\sigma_{ij}}{\eta} \left( \text{sign}(u_i)\,|u_j| + \frac{\eta^2}{\pi^2} \nabla^2 u_j \right) \enskip ,
\end{equation*}
yielding
\begin{eqnarray*}
&& \delta\hat{f}(u,1-u) = \frac{2}{\eta}\sigma_{ij} \\
&& \delta\hat{f}(1-u,u) = \frac{2}{\eta}\sigma_{ij} \\
&& \delta\hat{f}(0,u)+\delta\hat{f}(0,1-u) = \frac{2}{\eta}\sin\left(\frac{\pi}{\eta}x\right)(\sigma_{kj}-\sigma_{ki}) \quad\quad
\end{eqnarray*}
for the extension $u_i(x)=u(x)$, $u_{j=\neq i}(x)=1-u(x)$ and $u_{k \neq i,j}(x)=0$, showing that the planar binary interfaces do not minimize the free energy functional for non-zero interfacial free energies.\\

The dynamic equations of Steinbach and Pezzolla \cite{SteinbachPezzola1999} also read as:
\begin{equation*}
-\frac{\partial u_i}{\partial t} = \sum_{j \neq i} \kappa_{ij} \left[ \frac{\delta F}{\delta u_i} -  \frac{\delta F}{\delta u_j} \right] \enskip ,
\end{equation*}
with $\kappa_{ij}=\kappa_{ji}>0$ constants, therefore, the mobility matrix satisfies conditions 1--3 of Section II.C. Considering that the planar binary interfaces represent no equilibrium solution, they are not stationary. The problem is resolved again by replacing the gradient term $-\nabla u_i \cdot \nabla u_j$ by $(u_i\nabla u_j-u_j\nabla u_i)^2$, and using the 'binary approximation' \cite{Steinbach2009}:
\begin{equation}
\label{eq:twophaseapprox}
\begin{split}
\frac{\delta F}{\delta u_i}-\frac{\delta F}{\delta u_j} \approx \enskip & \text{sign}(u_i)|u_j|-\text{sign}(u_j)|u_i| \\
& - (u_j\Delta u_i - u_i\Delta u_j) \enskip .
\end{split}
\end{equation} 
Although the resulting non-variational dynamics stabilizes the extension of the planar interface solution $u(x)=\{1+\sin[(\eta/\pi)x]\}$, unfortunately it does not minimize the free energy functional. A remarkable improvement of the Steinbach-Pezzola model has been put forward by Kim {\it et al.} \cite{Kim2006}. Introducing step functions that are $S_i = 1$ for $u_i > 0$ and $S_i = 0$ otherwise, the mobility matrix has been assumed to have the form shown below 
\begin{equation*}
-\frac{\partial u_i}{\partial t} = \sum_{j \neq i} \kappa_{ij} S_i S_j \left[ \frac{\delta F}{\delta u_i} -  \frac{\delta F}{\delta u_j} \right] \enskip .
\end{equation*}
This change leads to an important step ahead: it retains the variational formalism, while stabilizing the flat interface. Nevertheless, it is not yet a solution of the Euler-Lagrange equation of the multiphase problem, therefore, this mobility matrix is 'dangerous' in the sense that it generates stationary solution from a non-equilibrium state. This approach has recently been applied for an asymmetric case taking the grain boundary energies from a database \cite{Kim2014}.\\

Finally, we mention that the derivations presented above can be trivially repeated for using $u_i\,u_j$ instead of $|u_i|\,|u_j|$ in the free energy functional described by Eq. (\ref{eq:spmodel}), resulting in the same qualitative results, i.e. the planar binary interfaces do not minimize the multiphase functional. In addition, the absence of the absolute value function terminates the bulk $u_i=1$ equilibrium solution too.

\subsection{Nestler-Wheeler}

Another descendant of the original Steinbach {\it et al.} model \cite{Steinbach1996} is the general Nestler-Wheeler type formalism \cite{Nest1, Nest2, Nest3, Nest4, Nest5}:
\begin{equation}
\label{eq:nwfunc}
f_{\rm intf}^{\rm NW} = \sum_{i<j} \left[ \frac{\epsilon^2_{ij}}{2} (u_i \nabla u_j-u_j\nabla u_i)^2 + \frac{w_{ij}}{2} (|u_i|\cdot |u_j|)^p  \right] \enskip ,
\end{equation}
where $p=1$ or $2$. For $p=2$ \cite{Nest1, Nest2, Nest3, Nest4, Nest5}, Eq. (\ref{eq:nwfunc}) recovers Eqs. (\ref{eq:sfunc})-(\ref{eq:smulti}) (Steinbach \textit{et al.}), whereas in case of $p=1$ \cite{Nest2,Nest4,Nest5}, it reduces to Eq. (\ref{eq:sp2red}) for $N=2$ with the solution
\begin{equation*}
u(x) = \frac{1+\sin(x/\delta_{ij})}{2} \enskip .
\end{equation*}
For $p=1$, the derivative of the generator function reads as
\begin{equation*}
\begin{split}
\delta\hat{f}(u_i,u_j) =& (w_{ij}/2) \, {\rm sign}(u_i) |u_j| + \\
 &+ \epsilon^2_{ij} \left[ 2(u_i\partial_x u_j-u_j\partial_x u_i)\partial_x u_j+ \right. \\
& + \left. (u_i\partial_x^2 u_j - u_j\partial_x^2 u_i)u_j \right] \enskip ,
\end{split}
\end{equation*} 
which, in the case of $u_i(x):=u(x)$, $u_{j\neq i}(x):=1-u(x)$, and $u_{k \neq i,j}(x):=0$, yields
\begin{eqnarray*}
&&\delta\hat{f}(u,1-u) = \frac{3w_{ij}}{4} \cos^2(x/\delta_{ij}) \\
&&\delta\hat{f}(1-u,u) = \frac{3w_{ij}}{4} \cos^2(x/\delta_{ij}) \\
&& \delta\hat{f}(0,u) + \delta\hat{f}(0,1-u) = 0 \enskip ,
\end{eqnarray*}
showing that the binary planar interfaces do not minimize the free energy functional again. This, together with the fact that the mobility matrix was chosen to be Lagrangian, means that we have here the same problem as in the case of the Steinbach-Pezzolla formalism, which can be resolved by using Eq. (\ref{eq:twophaseapprox}), i.e., by adopting non-variational dynamics.\\

In a recent variant of the $p=1$ Nestler-Wheeler model by Ankit {\it et al} \cite{Ankit2013} the multi-obstacle free energy landscape contains a triplet term of the form:
\begin{equation}
\label{eq:triplet}
f_3 = \sum_{i<j<k} \gamma_{ijk} |u_i|\,|u_i|\,|u_k| \enskip ,
\end{equation}
where the triple sum runs for all different $(i,j,k)$ triplets, and the authors use $\gamma_{ijk}$ to control the appearance of the third phase at the binary interfaces. We note, however, that Eq. (\ref{eq:triplet}) has no effect on the existence of the planar interface solution, since the derivative $\partial f_3/\partial u_i=\text{sign}(u_i)\sum_{j<k}|u_i|\,|u_j|$ vanishes for binary planar interfaces. In other words, $f_3$ is not suitable for generating equilibrium planar interface solutions. Nevertheless, choosing $\gamma_{ijk}\to \infty$ results in two-phase interfaces free of additional fields, but for any finite $\gamma_{ijk}$, additional fields are always present at the interfaces mathematically.\\

We note that, in models relying on the multi-obstacle potential, $f_{\rm intf} \to \infty$ is often prescribed out of the physical regime to prevent the evolution of the fields into the "unphysical" states $u_i<0$ and $u_i>1$. This is another way to stabilize the (otherwise non-equilibrium) two-phase interfaces. We recall furthermore that there is physical interpretation for $u_i<0$ and $u_i>1$. Comparison of the phase-field models to the Ginzburg-Landau model and/or to amplitude equations emerging from classical density functional theories \cite{TothProvatas} implies that $u_i<0$ can be simply associated with a negative amplitude of the first reciprocal lattice vector set in the crystal, which is a real perturbation of the liquid state for cubic crystal structures. Similarly, $u_i>1$ is nothing more than an amplitude larger than the equilibrium crystal amplitude.\\

\subsection{Folch-Plapp}

The term "Lagrange multiplier formalism" originates from Folch and Plapp \cite{FolchPlapp2005}. In their multiphase description, the free energy functional is based on several theoretical considerations including binary equilibrium solutions and the condition of no spurious phase generation. For three phases the interface contribution of the free energy functional reads as:
\begin{equation}
\label{eq:FPfunc}
f_{\rm intf}^{FP} =  \frac{\epsilon^2}{2} \sum_{i=1}^3 (\nabla u_i)^2 + \frac{w}{2} f_{TW}(\mathbf{u}) \enskip ,
\end{equation}
where the "triple-well" free energy density is $f_{TW}(\mathbf{u})=\sum_{i=1}^3 g(u_i)$, .where $g(u_i)=[u_i(1-u_i)]^2$. First, we analyze the binary planar interfaces. For $N=2$ the free energy functional reduces to
\begin{equation*}
f_{\rm intf}^{FP} =  \epsilon^2 (\nabla u)^2 + w g(u) \enskip .
\end{equation*}
generating the usual 1D Euler-Lagrange equation
\begin{equation}
\label{eq:FPRED}
\frac{\delta F}{\delta u}=w g'(u) - \epsilon^2 \partial_x^2 u=0 \enskip .
\end{equation}
The equilibrium planar interface solution is then $u(x)=\{1+\tanh[x/(2\delta)]\}/2$ with $\delta^2=\epsilon^2/w$. Since  Eq. (\ref{eq:FPfunc}) is not a pairwise construction, the generator function technique does not apply here. The general Euler-Lagrange equations read as:
\begin{equation}
\label{eq:FPEL}
\frac{\delta F}{\delta u_i} = \frac{1}{2} \left[ w \, g'(u_i)-\epsilon^2 \, \partial_x^2 u_i \right]  =\lambda(\mathbf{r}) \enskip .
\end{equation}
Comparing Eqs. (\ref{eq:FPRED}) and  (\ref{eq:FPEL}) results in 2 important properties of the model:
\begin{equation*}
\frac{\delta F}{\delta u_i} \propto \left. \frac{\delta F}{\delta u}\right|_{u_i} \quad \text{and} \quad \left. \frac{\delta F}{\delta u_i}\right|_{u_i=0}=0 \enskip ,
\end{equation*}
indicating that \textit{the equilibrium planar binary interfaces are solutions of the multiphase problem with $\lambda(x)=0$}. The proposed dynamics uses the {\it Lagrangian mobility matrix}, therefore, the Folch-Plapp model is the first model, which passes the binary interface criterion.\\

Next, we discuss the appearance of spurious phases. The general condition reads as $u_k(\mathbf{r},t) = 0$, i.e. if phase $k$ is not present at $t=0$, it must not appear at any time. In the Folch-Plapp model, the time evolution of a phase, which is apparently not present reads as
\begin{equation*}
\left. \frac{\partial u_k}{\partial t} \right| _{u_k=0} \propto \left( \frac{\delta F}{\delta u_i} + \frac{\delta F}{\delta u_j} \right) \enskip ,
\end{equation*}
where $k \neq i,j$ and $i\neq j$. Using Eq. (\ref{eq:FPEL}) it is obvious that $\delta F/\delta u_i+\delta F/\delta u_j \equiv 0$ for $u_i(\mathbf{r},t)+u_j(\mathbf{r},t)=1$, therefore, no spurious phases appear.\\

The authors have also worked out an asymmetric version of the model, in which different binary interfacial free energies can be used, which also satisfies the basic criteria of physical consistency together with no spurious phase generation. Despite its advantageous features compared to former approaches, there are a few weaknesses of the model: (a) to avoid the appearance of spurious phases, a free energy functional is used whose form depends on the particular choice of the mobility matrix, which is clearly not physical, (b) no $N>3$ generalization of the model is available. Furthermore, (c) as the authors suggested, the necessary minimum exponent in the multi-well term might be proportional to the number of phases, making the model practically useless when a large number (e.g., thousands) of differently oriented dendrites or grains have to be simulated.\\

\textit{At this stage, it is clear that the Folch-Plapp model is a large step towards constructing a consistent multiphase description, since it satisfies almost all the criteria of physical consistency. Unfortunately, the price is high: it is not immediately clear how to generalize this model to $N > 3$, together with avoiding the appearance of spurious phases via introducing special terms in the free energy functional, which follow from the form of the dynamic equations.}

\subsection{Bollada-Jimack-Mullis}

In the previous sections it has been demonstrated that the condition of no spurious phase generation works at two levels. First, if a solution of the Euler-Lagrange equation represents minimum of the free energy functional, it is stable against small perturbations (assuming variational dynamics with a positive semi-definite mobility matrix). In addition, to avoid the appearance of spurious phase outside of equilibrium may necessitate the adjustment of the free energy functional. In a recent work \cite{Bollada2012}, however, Bollada {\it et al.} avoided the problem by introducing a field-dependent mobility matrix
\begin{equation}
\label{eq:BJMmob}
L_{ii} = \sum_{j \neq i}^N h(u_i,u_j) \quad \text{and} \quad L_{ij} = - h(u_i,u_j) \enskip \text{for} \enskip i \neq j \enskip ,
\end{equation} 
where
\begin{equation*}
h(u_i,u_j) = \left(\frac{u_i}{1-u_i}\right)\left(\frac{u_j}{1-u_j}\right) \enskip .
\end{equation*} 
Apparently, Eq. (\ref{eq:BJMmob}) satisfies the conditions of Section II.C; i.e., the dynamics ensures non-negative entropy production inside the simplex (i.e., when all $u_i \in [0,1]$, all $h(u_i,u_j) \geq 0$). In addition, the spurious phase generation is excluded, since for $u_k(\mathbf{r},t)=0$ the $k^{th}$ row (and column) of the mobility matrix vanishes, yielding $(\partial u_k/\partial t)|_{u_k(\mathbf{r},t)=0}=0$. Note that this is achieved without revising the free energy functional. Moreover, the description is $N$-independent, since the mobility matrix consistently reduces to the $N-1$ case. Despite the significant improvement, one has to be careful with the Bollada-Jimack-Mullis mobility matrices, since they can be dangerous with respect to the free energy functional in the sense that the mobility matrix may generate stationary solution of the dynamics from a non-equilibrium state (as it happens in case of a multi-obstacle potential). In addition, outside of the simplex (i.e., for $u_i<0$), the mobility matrix is indefinite, thus violating criterion (iv).\\

{The results of the above analysis of the MPF theories are summarized in Table I. The following has been found:} 

{(A) The criteria for the sum of the fields and for labeling are satisfied by all the models investigated.} 

{(B) The lack of equilibrium planar two-phase interfaces (in the $N$-phase problem) can be resolved by employing: (a) non-variational dynamics (the models by Steinbach {\it et al.}\cite{Steinbach1996}, Steinbach and Pezzola\cite{SteinbachPezzola1999}, Nestler and Wheeler\cite{Nest1, Nest2, Nest3, Nest4, Nest5}), (b) a degenerate mobility matrix (the models by Kim {\it et al.}\cite{Kim2006}, and by Bollada {\it et al.}\cite{Bollada2012}), or by (c) penalizing the triplet term (the model by Ankit {\it et al.}\cite{Ankit2013}). We identified the following problems associated with methods (a)-(c): the solution does not converge to the equilibrium solution; furthermore, in case (c) the third phase is unavoidably present (even if in a small amount).} 

{(C) When the equilibrium conditions are satisfied (as in the model by Folch and Plapp\cite{FolchPlapp2005}), we obtain an $N$-dependent approach without the possibility of prescribing freely the pairwise interfacial data.} 

{Considering these, we conclude that none of the MPF models investigated here satisfy all the consistency criteria specified. We stress furthermore that the introduction of additional thermodynamical driving force via an appropriate tilting function (as needed for describing polycrystalline solidification) would influence neither the validity of these criteria, nor the outcome of this analysis.}

\begin{table}[t]
\caption{Properties of different multiphase-field models from the viewpoint of criteria defined in this work.}
\begin{center}
\begin{tabular}{|c|c|c|c|c|c|c|c|c|}
\hline
model $\backslash$ criterion & i & ii & iii & iv & v & vi(a) & vi(b) & vii \\
\hline
\hline
Steinbach {\it et al.} & x & x & & & & & x & x \\
\hline
Steinbach \& Pezzola & x & x & & & & & x & x\\
\hline
Nestler \& Wheeler & x & x & & & & & x & x\\
\hline
Kim {\it et al.} & x & x & & x & x & & x & x\\
\hline
Bollada {\it et al.} & x & x & & x & x & & x & x\\
\hline
Ankit {\it et al.} & x & x & x & x & & & & x\\
\hline
Folch \& Plapp & x & x & x & x & & x & x & \\
\hline
\end{tabular}
\end{center}
\end{table} 
 
\section{Consistent multiphase formalism}

\subsection{General framework}

Herein, we derive a multiphase description that satisfies criteria (i) to (vii). It is useful to start with the condition of formal reducibility [criterion (iv)]. First, setting $u_N(\mathbf{r},t)=0$ in the $N$-phase free energy functional $F^{(N)}$ should result in the $N-1$ phase functional, $F^{N-1}$. The same should apply for the dynamic equations
\begin{equation*}
-\frac{\partial u_i}{\partial t} = \sum_{i \neq j} \kappa_{ij}\left( \frac{\delta F}{\delta u_i}-\frac{\delta F}{\delta u_J}\right)
\end{equation*}
as well; i.e., $-\dot{\mathbf{u}}^{(N)}=\mathbb{L}^{(N)} \cdot \delta F_{\mathbf{u}}^{(N)}|_{u_{N}(\mathbf{r},t)=0}$ should reduce to $-\dot{\mathbf{u}}^{(N-1)}=\mathbb{L}^{(N-1)} \cdot \delta F_{\mathbf{u}}^{(N-1)}$, and $\dot{u}_N = 0$. Note that the latter satisfies the condition of no spurious phase generation [criterion (ii)]), since $\dot{u}_i|{u_i=0} = 0$. Here $\mathbb{L}^{(N)}$ is a general, symmetric, positive semidefinite $N$--phase mobility matrix, i.e. $L_{ij}=-\kappa_{ij}$ for $i \neq j$, while $L_{ii}=\sum_{j\neq i} \kappa_{ij}$, where $i,j=1\dots N$, and $\kappa_{ij}=\kappa_{ji}>0$. Since the (modified) Bollada-Jimack-Mullis matrix defined by Eq. (\ref{eq:BJMmob}) satisfies the condition of formal reducibility, we choose this mobility matrix, namely,
\begin{equation*}
\kappa_{ij} := \kappa_{ij}^0 \left| \frac{u_i}{1-u_i}\right| \, \left| \frac{u_j}{1-u_j}\right| \enskip ,
\end{equation*} 
where constant positive prefactors $\kappa_{ij}^0$ accounting for the mobility of the planar $i,j$ interface are also incorporated. Furthermore, we prescribe the condition of reduction also for the functional derivatives; i.e., the first $N-1$ components of $\delta F_{\mathbf{u}}^{N}$ should reduce to $\delta F_\mathbf{u}^{(N-1)}$ for $u_N(\mathbf{r},t)=0$. Since the $N^{th}$ row (and column) of the reduced Bollada-Jimack-Mullis matrix is $0$, it always results in $\dot{u}_N=0$, therefore, $(\delta F^{N}/\delta u_N)|_{u_N(\mathbf{r},t)=0}$ can be arbitrary. Since the $n$--phase Bollada-Jimack-Mullis matrix is positive semidefinite on an $n$--phase state (i.e. when none of the $n$ components vanishes) with multiplicity 1 for the eigenvalue 0 (it can be proven numerically), the $n$--phase stationary solutions coincide with the $n$ phase equilibrium states. Moreover, since the dynamic equations reduce naturally to the $(N-1)$--phase case, the stationary solutions of the reduced dynamics include the natural $N$ phase extensions of the $(N-1)$--phase equilibrium solutions (where none of the $N-1$ phases is missing). Therefore, the natural $N$--phase extensions of all $(N-1)$--phase equilibrium solutions emerging from $F^{(N-1)}$ should represent extrema of $F^{N}$. If this is true, the Bollada-Jimack-Mullis matrix is not dangerous with respect to the free energy functional, since all stationary states of the dynamics represent equilibrium. Since the condition must apply for arbitrary $N$, the general condition for the free energy functional reads as follows:\\

\textit{The $(p+q)$--phase trivial extensions of all $p$--phase equilibrium solutions (where all the $p$ phases are non-vanishing) emerging from the $p$--phase free energy functional $F^{(p)}$ must represent extrema of the $(p+q)$--phase free energy functional $F^{(p+q)}$ too, for any $p>0,q>0$.}\\

For practical reasons, we introduce the following condition: for a field, which is not present, the functional derivative vanishes, i.e.
\begin{equation*}
\left.\frac{\delta F}{\delta u_i}\right|_{u_i(\mathbf{r},t)=0}=0 \enskip .
\end{equation*}
If the Lagrange multiplier also vanishes [$\lambda(\mathbf{r})=0$] for \textit{all} $p-$phase equilibrium solutions of $F^{(p)}$ (where $p>1$ arbitrary), while the free energy functional (and the functional derivative) reduces naturally, all trivial extensions of all $p$-phase equilibrium solutions remain equilibrium solutions of the $N$-phase free energy functional. Consequently, here the Bollada-Jimack-Mullis matrix does not stabilize non-equilibrium solutions, while preventing the appearance of spurious phases. {Note, that this is obviously not true for the models of Steinbach \textit{et al.}, Steinbach and Pezzola, Nestler and Wheeler, Ankit {\it et al.}, and for the model potential used in the work of Bollada, Jimack, and Mullis. Although there the free energy functionals and the functional derivatives reduce naturally, and $(\delta F/\delta u_i)|_{u_i=0}=0$ also applies, the planar two-phase interface solution generates different Lagrange multipliers for the vanishing and non-vanishing fields in the complete ($N$-phase) Euler-Lagrange problem, indicating that the natural extensions of the planar two-phase interfaces do not represent equilibrium. Yet, the application of the Bollada-Jimack-Mullis mobility matrix transforms them into stationary solutions, showing that in these cases the application of the Bollada-Jimack-Mullis matrix is ''dangerous''.} 

\subsection{Free energy functional}

\subsubsection{Symmetric system}

The main question is, how one should construct an interface term that satisfies the conditions given above. First, we consider the symmetric case, where all interface thicknesses and interfacial free energies are equal. Following Chen and co-workers \cite{ChenYang1994,FanChen1997a,FanChen1997b} and Moelans and co-workers \cite{Moelans1,Moelans2,Moelans3,Moelans4}, the interface term of the free energy functional is constructed as follows
\begin{equation}
\label{eq:tothmodel}
f_{\rm intf} = \frac{\epsilon^2}{2} \sum_{i=1}^N (\nabla u_i)^2 + w\, g(\mathbf{u}) \enskip ,
\end{equation}
where we use the following new Ansatz for the multiphase barrier function:
\begin{equation}
\label{eq:tothpoly}
g(\mathbf{u}) := \frac{1}{12} + \sum_{i=1}^N \left( \frac{u_i^4}{4} - \frac{u_i^3}{3} \right) + \frac{1}{2} \sum_{i<j} u_i^2 u_j^2 \enskip .
\end{equation}
The functional derivative reads as
\begin{equation}
\label{eq:tothfder}
\frac{\delta F}{\delta u_i} =w \, [u_i(\mathbf{u}^2-u_i)] - \epsilon^2 \nabla^2 u_i \enskip ,
\end{equation}
which vanishes for $u_i=0$. The binary planar interface solution is $u(x)=\{1+\tanh[x/(2 \delta)]\}/2$ (where $\delta^2=\epsilon^2/w$), for which the multiphase Euler-Lagrange equations reduce to
\begin{eqnarray*}
\frac{\delta F}{\delta u_i} &=& -\frac{\delta F}{\delta u_j} = w \, u(1-u)(1-2u) - \epsilon^2 \partial_x^2 u =0 \\
\frac{\delta F}{\delta u_k} &=& 0 \enskip .
\end{eqnarray*}
Here we used the trivial extension $u_i:=u(x)$, $u_{j\neq j}:=1-u(x)$ and $u_{k\neq i,j}=0$. Since the free energy functional and the functional derivatives reduce naturally, and $\delta F/\delta u_i$ vanishes for $u_i=0$, together with the fact that any trivial extension of the planar interface solution represents equilibrium, it is reasonable to assume that the Bollada-Jimack-Mullis matrix is not dangerous considering at least the planar interfaces, i.e., it does not stabilize a non-equilibrium planar interface, since all planar interfaces represent equilibrium. Naturally, the same investigation should be repeated for all $n$--phase equilibrium solutions of $F^{(n)}$ for any positive $n$, however, this kind of study is out of the scope of the present paper.
\\

It is important to mention, that Eq. (\ref{eq:tothpoly}) shows a very practical feature
\begin{equation}
\label{eq:tothtendency}
g(\{1/N,1/N,\dots,1/N\}) = \frac{1}{12} \left( 1-\frac{1}{N^2} \right) \enskip,
\end{equation}
i.e., the higher-order junctions are energetically increasingly less favorable. Note that this is not true for Eq. (\ref{eq:smulti}). The tendency of increasing free energy is also ensured by the multi-well term defined in Eq. (\ref{eq:spmodel}), however there, as we have shown previously, the binary planar interfaces do not minimize the free energy functional in the general $N$-phase case. It is worth noting that Eq. (\ref{eq:tothtendency}) contradicts Folch and Plapp \cite{FolchPlapp2005}, who expect that the polynomial degree of the $g(\mathbf{u})$ function that penalizes the high-order multiple junctions would increase with $N$. Eq. (\ref{eq:tothtendency}) shows that the double-obstacle function is not the only one that realizes this tendency, furthermore, here [see Eq. (\ref{eq:tothpoly})] the planar binary interface represents an equilibrium solution of the multiphase problem.

\subsubsection{Asymmetric extension}

Following Moelans \cite{Moelans2}, the asymmetric extension of Eq. (\ref{eq:tothmodel}) can be obtained by employing the Kazaryan-polynomials \cite{Kazaryan2000}
\begin{eqnarray}
\label{eq:totheps2}
\epsilon^2(\mathbf{u}) &:=& \frac{\sum_{i,j} \epsilon^2_{ij} u_i^2 u_j^2}{\sum_{i,j} u_i^2 u_j^2}\\
w(\mathbf{u}) &:=& \frac{\sum_{i,j} w_{ij} u_i^2 u_j^2}{\sum_{i,j} u_i^2 u_j^2} \enskip .
\end{eqnarray}
The free energy density then reads as
\begin{equation}
\label{eq:tothasymm}
f_{\rm intf} = \frac{\epsilon^2(\mathbf{u})}{2} \sum_{i=1}^N (\nabla u_i)^2 + w(\mathbf{u}) \, g(\mathbf{u}) \enskip ,
\end{equation}
where $g(\mathbf{u})$ is defined by Eq. (\ref{eq:tothpoly}). Since the Kazaryan polynomial is "quasi-constant" (the nominator and the denominator contain the same terms with different coefficients), it is reasonable to assume that this modification does not change the structure of extrema of $g(\mathbf{u})$. Although it will be demonstrated for asymmetric $N=4$ and $N=5$ systems, a strict mathematical derivation for arbitrary number of phases is out of the scope of the present paper.  

The binary reduction of Eq. (\ref{eq:tothasymm}) reads as
\begin{equation*}
f_{\rm intf} = \epsilon_{ij}^2 (\nabla u)^2 + w_{ij} [u(1-u)]^2 \enskip ,
\end{equation*}
generating the equilibrium planar binary interface with $\delta^2 = \epsilon^2_{ij}/w_{ij}$. The general functional derivative is
\begin{equation}
\label{eq:tothgender}
\begin{split}
\frac{\delta F}{\delta u_i} & =  w \,\frac{\partial g}{\partial u_i} + \frac{\partial w}{\partial u_i} \, g(\mathbf{u}) - \epsilon^2 \, \nabla^2 u_i \\
& +  \sum_{j=1}^N \left[ \frac{1}{2} \frac{\partial \epsilon^2}{\partial u_i} \nabla u_j - \frac{\partial \epsilon^2}{\partial u_j} \nabla u_i \right] \cdot \nabla u_j \enskip ,
\end{split}
\end{equation}
where
\begin{eqnarray*}
\frac{\partial \epsilon^2}{\partial u_j} &=& (2 u_j) \frac{\sum_{l \neq j} (\epsilon_{lj}^2-\epsilon^2)u_l^2}{\sum_{k,l} u_k^2 u_l^2}\\
\frac{\partial w}{\partial u_j} &=& (2 u_j) \frac{\sum_{l \neq j} (w_{lj}-w)u_l^2}{\sum_{k,l} u_k^2 u_l^2} \enskip .
\end{eqnarray*}
Note that $\left. \frac{\partial \chi}{\partial u_k} \right|_{u_k=0}=\left. \frac{\partial \chi}{\partial u_{i,j}} \right|_{u_i+u_j=1}=0$, where $\chi=\epsilon^2,w$, therefore, $(\delta F/\delta u_k)_{u_k=0}=0$, and the second line of Eq. (\ref{eq:tothgender}) also vanishes for $u_i=u$, $u_j=1-u$, $u_{k \neq i,j}=0$, therefore, \textit{the planar binary interfaces are equilibrium solutions of the multiphase problem for arbitrary pairwise $\epsilon_{ij}^2$ and $w_{ij}$} fitted to the interfacial free energy $\gamma_{ij}$ and interface thickness $\delta_{ij}$ as follows:
\begin{equation*}
\epsilon_{ij}^2 = 3 \, (\delta_{ij} \cdot \gamma_{ij}) \quad \text{and} \quad w_{ij} = 3 \, (\gamma_{ij}/\delta_{ij}) \enskip .
\end{equation*}

\subsubsection{Introducing anisotropy}

In various practically important cases, including dendritic solidification and grain coarsening, the interfacial free energy between two phases displays anisotropy, which can be formulated mathematically as:
\begin{equation*}
\epsilon_{ij} \to \epsilon_{ij} \left[ 1 + a_{ij} \cdot \eta_{ij}(\mathbf{n}_{ij}) \right] \enskip ,
\end{equation*}
where $a_{ij}$ is the amplitude (strength) of the anisotropy,
\begin{equation*}
\mathbf{n}_{ij} = \frac{\nabla u_i - \nabla u_j}{| \nabla u_i - \nabla u_j |}
\end{equation*}
is a unit vector characterizing the $(i,j)$ binary interface, while $\eta_{ij}(\mathbf{n}_{ij})$ reflects the crystal symmetry. This extension modifies Eq. (\ref{eq:totheps2}) and the functional derivative as follows
\begin{equation*}
\begin{split}
\frac{\delta F}{\delta u_i} &  =  g(\mathbf{u}) \frac{\partial w}{\partial u_i}  + w(\mathbf{u}) \frac{\partial g}{\partial u_i} + \frac{\partial \epsilon^2}{\partial u_i} \left[ \frac{1}{2} \sum_{j=1}^N (\nabla u_j)^2 \right] -\\
& - \nabla \cdot \left\{ \frac{\partial \epsilon^2}{\partial \nabla u_i} \left[ \frac{1}{2} \sum_{j=1}^N (\nabla u_j)^2 \right] + \epsilon^2 \cdot \nabla u_i \right\} \enskip .
\end{split}
\end{equation*}
Here the extra term reads as
\begin{equation*}
\frac{\partial \epsilon^2}{\partial \nabla u_i} = \frac{\sum_{k,l} \frac{\partial \epsilon_{kl}^2}{\partial \nabla u_i} u_k^2 u_l^2}{\sum_{k,l} u_k^2 u_l^2} \enskip ,
\end{equation*}
where $\frac{\partial \epsilon_{kl}^2}{\partial \nabla u_i} \propto \delta_{ki}+\delta_{li}$, therefore, $\frac{\partial \epsilon^2}{\partial \nabla u_i} \propto u_i^2$, which means $(\delta F/\delta u_k)|_{u_k=0}=0$. In addition, for $u_i+u_j=1$ $\frac{\partial \epsilon^2}{\partial \nabla u_{i,j}} = \frac{\partial \epsilon_{ij}^2}{\partial \nabla u_{i,j}}$, therefore, $\frac{\delta F}{\delta u_{i,j}} = \frac{1}{2} \frac{\delta F}{\delta u}$, where $\delta F/\delta u$ is the functional derivative in the reduced model, therefore, the equilibrium binary interfaces emerging from $\delta F/\delta u=0$ are stationary solutions of the multiphase problem.\\

\section{Testing the XMPF model}

In this section, we review whether the proposed model satisfies indeed criteria (i) to (vii). Several of these criteria are satisfied owing to the specific formulation of our model [these are (i), (iii) -- (vi), and (vii)] as summarized in sub-section A. Fulfillment of practical criterion (ii), however, needs further investigation, which is undertaken in sub-section B. {Finally, illustrative simulations are presented for grain coarsening in sub-section C.}

\subsection{Consistency criteria satisfied}

It can be shown that the proposed model satisfies the following criteria:\\ 

(i) $\sum_{i=1}^N u_i(\mathbf{r},t)= 1$ (see Section IV.A);\\

(ii) Since the mobility matrix is symmetric, the physical results are invariant to formally exchanging pairs of field indices, $i \leftrightarrow j$, i.e. the variables are not labeled (see Section IV.A);\\

(iii) Under appropriate boundary conditions, any trivial multiphase extension of the equilibrium binary solution represents equilibrium solution of the multiphase problem, which is then a stationary solution of the dynamic equations towards which the time dependent solution evolves (see Section IV.D);\\

(iv) Since the mobility matrix is positive semidefinite, non-negative entropy production is ensured for both conserved and non-conserved dynamics (see Sections IV.A and VI.B);\\ 

(v) Reduction/extension of the $N$-field theory to $N-1$ or $N+1$ fields is trivial on the level of both the free energy functional (and the functional derivative) and the mobility matrix (see Sections IV.A and IV.D);\\ 

(vi) No additional phases appear at the equilibrium binary interfaces (see Sections  IV.B, IV.C, IV.D and VI.A), and the dynamic spurious phase generation is also excluded (see Sections IV.A and V.B);\\

(vii) Freedom for choosing independent interfacial ($\epsilon_{ij}$ and $w_{ij}$) and kinetic properties ($\kappa_{ij}$) for the individual binary boundaries, including their anisotropy (see the formulation in Sections IV.C and IV.D).
\\

Since the dynamic generation of spurious phases is excluded by the modified Bollada-Jimack-Mullis mobility matrix, the trivial extensions of the equilibrium planar solutions represent equilibrium, and the stationary solutions of the dynamic equations coincide with the equilibrium solutions, hence the mobility matrix is regarded as 'not dangerous', when considering planar interfaces. Nevertheless, it still remains unclear, whether the same applies for {\it all} equilibrium solutions, such as the trivial extensions of equilibrium trijunctions, etc.. Therefore, next we investigate the time evolution of multi-domain systems in this respect.

\begin{figure}[t] 
  (a)\includegraphics[width=0.44\linewidth]{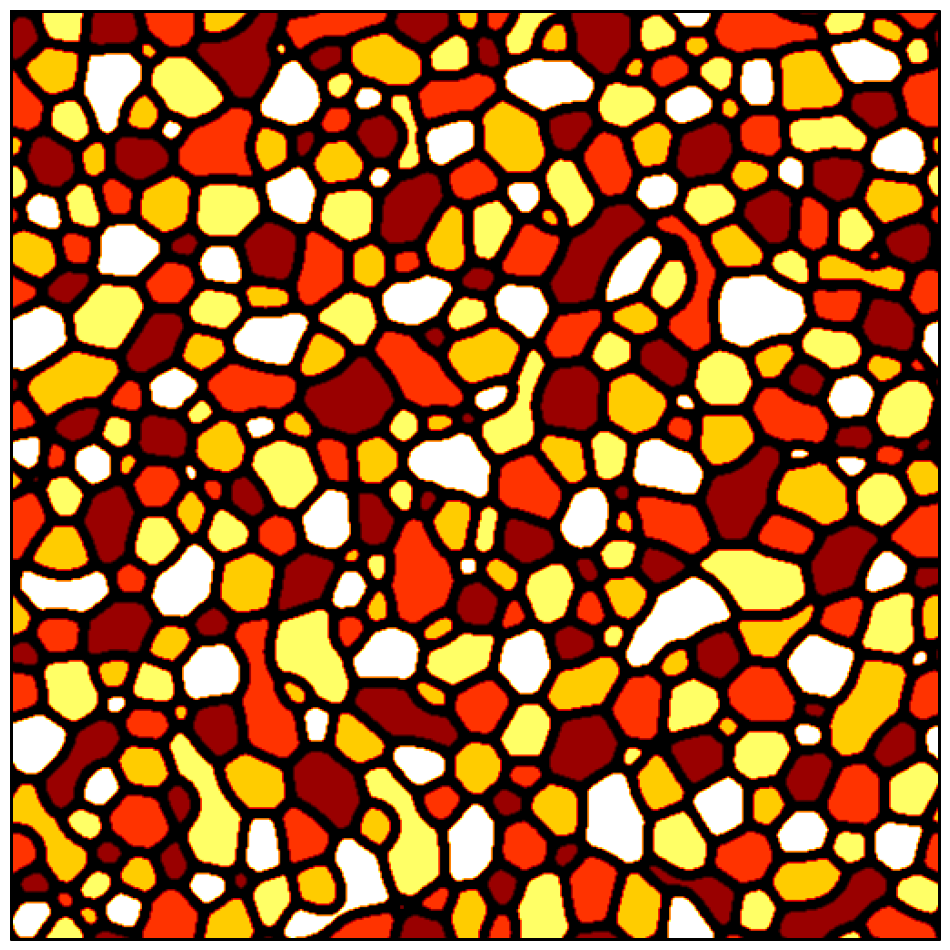} (b)\includegraphics[width=0.44\linewidth]{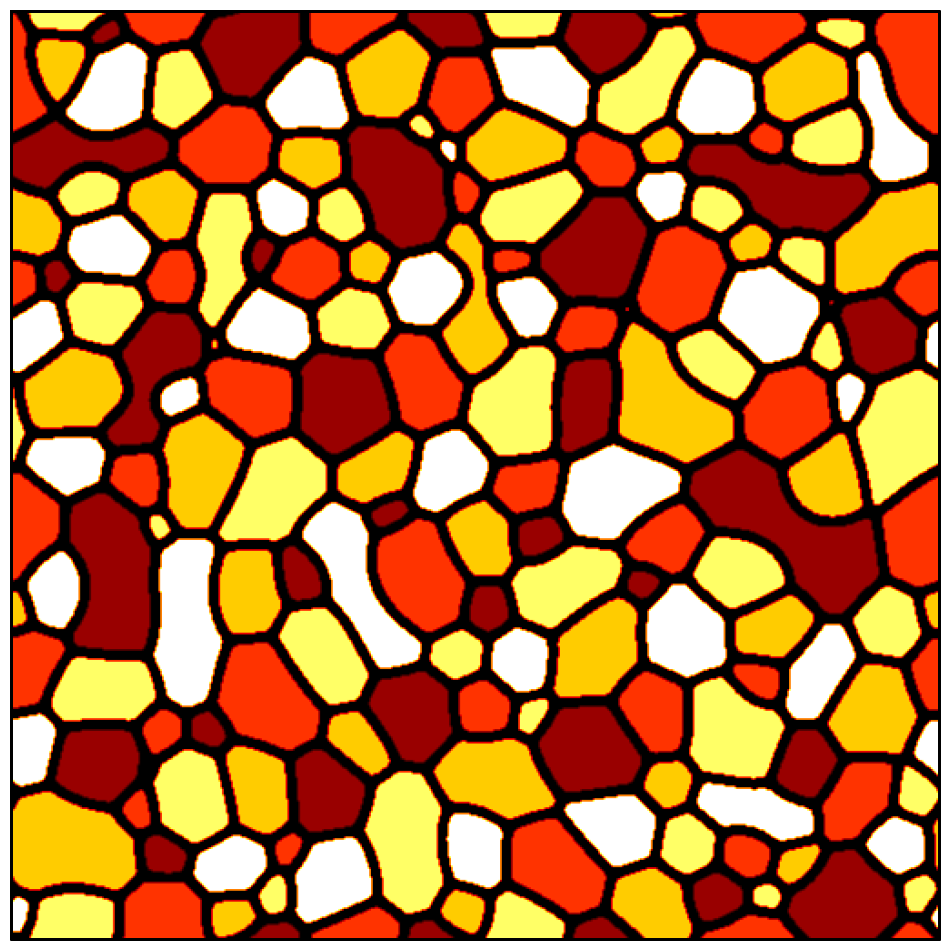}\\
  (c)\includegraphics[width=0.44\linewidth]{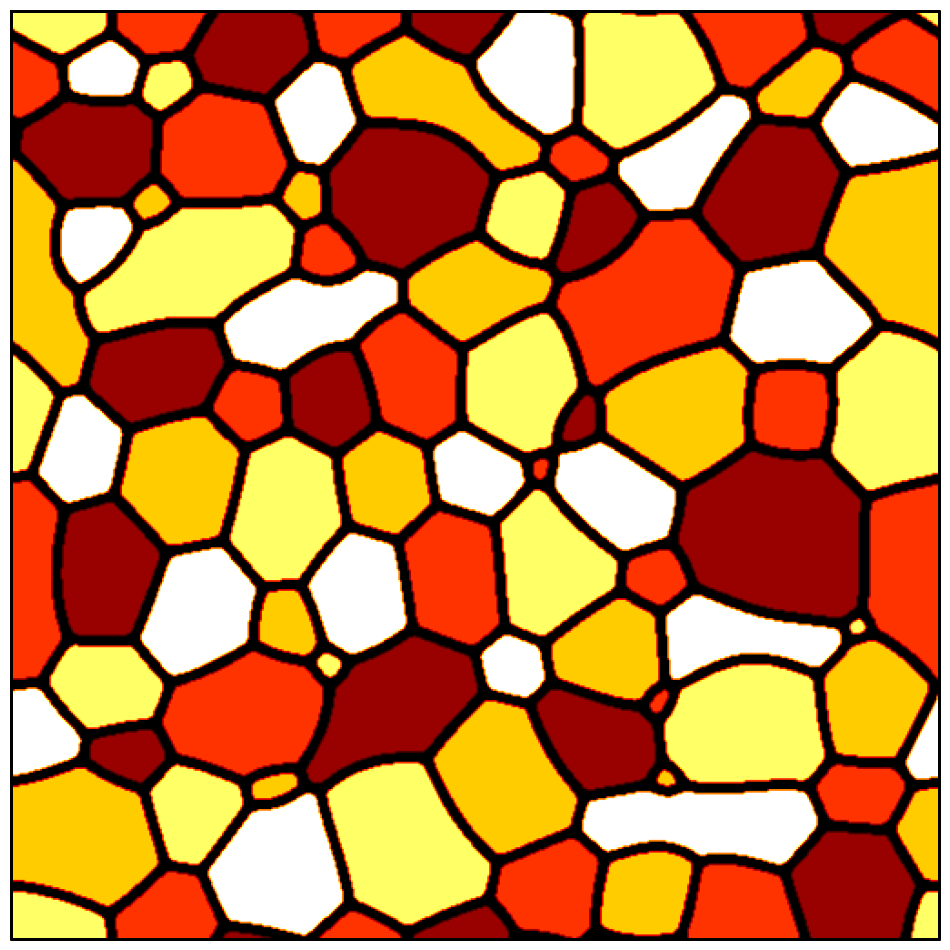} (d)\includegraphics[width=0.44\linewidth]{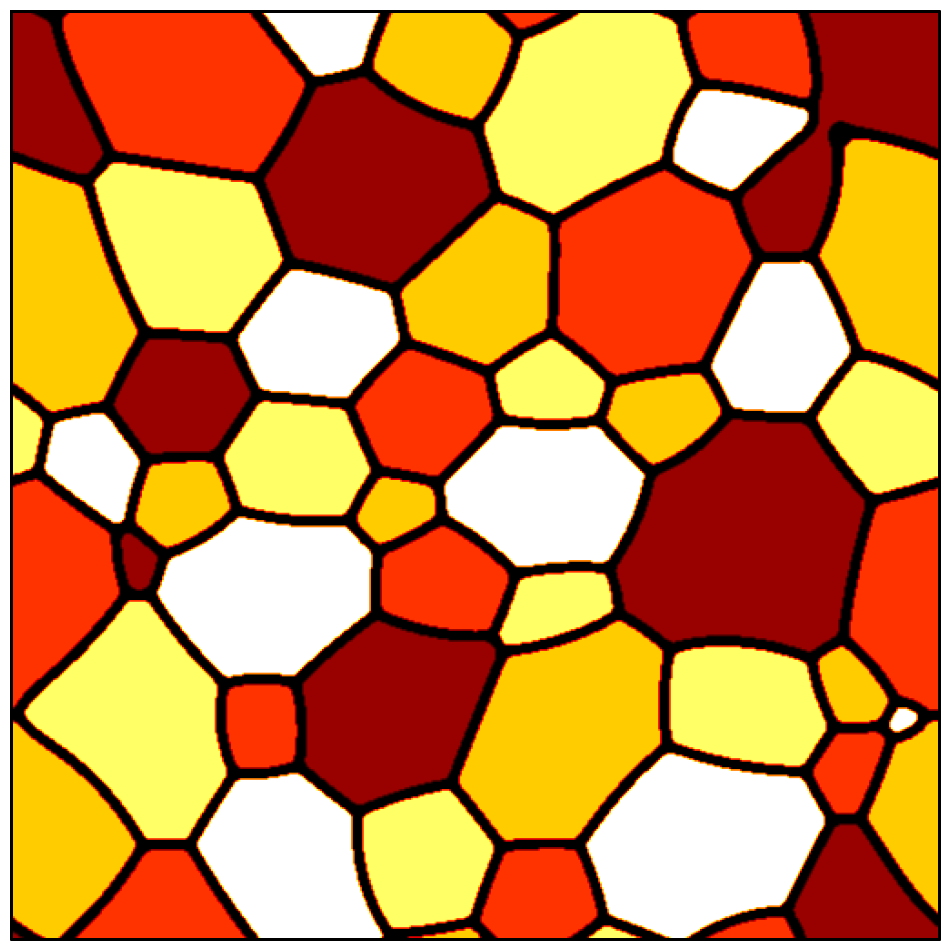}\\
  (e)\includegraphics[width=0.44\linewidth]{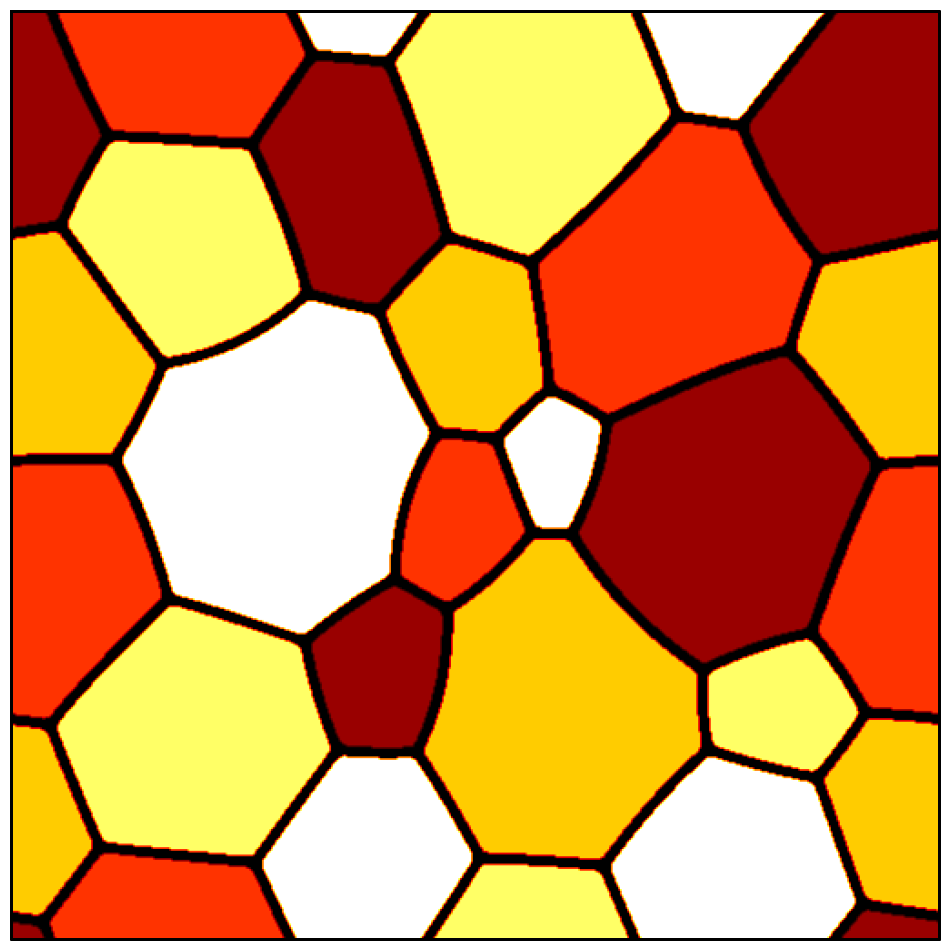} (f)\includegraphics[width=0.44\linewidth]{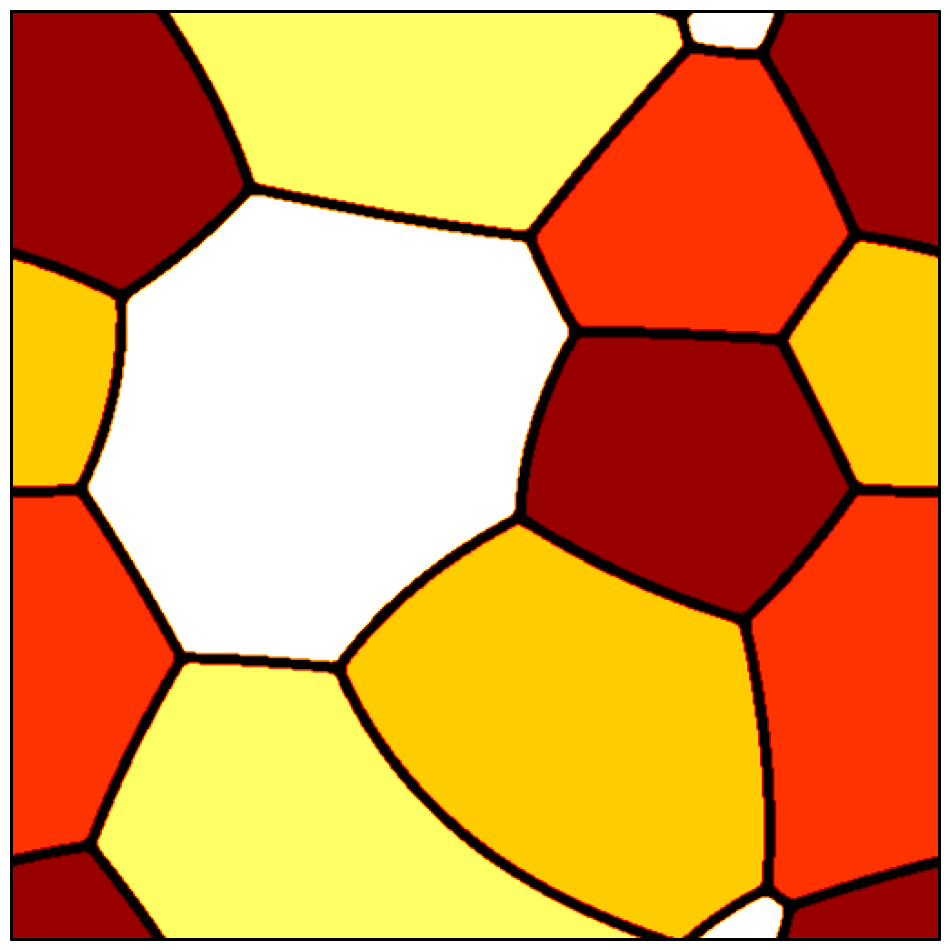}
 \caption{\label{fig:timeevol} (Color online) Time evolution in an $N = 5$ symmetric Cahn-Hilliard problem with starting conditions $u_1 = u_2 = u_3 = u_4 = u_5 = 0.2$: Different colors stand for different phases (where $u_i>0.99$). Snapshots taken at $t = (1, 3, 10, 30, 100, 300) \times 25,000$ dimensionless time are displayed. (Time increases from left to right, and from top to bottom.)}
\end{figure}

\subsection{Multiphase separation}

For practical reasons, the time evolution of multi-domain systems is investigated in two dimensions using conserved dynamics ($N-component$ Cahn-Hilliard systems), for which the dimensionless equations of motion read as
\begin{equation*}
\frac{\partial u_i}{\partial t} = \nabla \cdot \left[ \sum_{j\neq i} \kappa_{ij} \nabla \left( \frac{\delta F}{\delta u_i} - \frac{\delta F}{\delta u_j} \right) \right] \enskip .
\end{equation*}
Our choice is motivated by the fact that simulations with conserved dynamics are much less forgiving to possible numerical errors than those with non-conserved dynamics.\\

First, we consider a symmetric system, and demonstrate that our free energy functional prescribes the hierarchy $F_{bulk} < F_{interface} < F_{trijunction} < F_{quadruple} < \dots$ for the equilibrium solutions, therefore, a 2-dimensional, multi-domain system displays a bulk--interface(--trijunction) topology. Accordingly, the additional phases near any bulk / interface / trijunction vanish with time, since these states are energetically not preferred in the vicinity of the equilibrium solutions, to which the system converges. \textit{This happens independently of the particular choice of the mobility matrix};  the only requirement is that the mobility matrix has to be symmetric and positive semidefinite. Since $\sum_{i=1}^N (\delta F/\delta u_i) \equiv 0$ applies for the symmetric model, the Lagrangian mobility matrix $\kappa_{ij}=1$ yields:
\begin{equation*}
\frac{\partial u_i}{\partial t} = \nabla^2 \frac{\delta F}{\delta u_i} \enskip ,
\end{equation*} 
a fairly simple system of dynamic equations. Since $(\delta F/\delta u_i)_{u_i=0}=0$, this dynamics satisfies the condition of "no spurious phase generation". Indeed, we demonstrate that if a phase is not present in the beginning of the calculation, it never appears, even if the other phases are not in equilibrium.\\

Next, we perform simulations in an $N =  4$ asymmetric system with two types of mobility matrices (the Lagrangian and the Bollada-Jimack-Mullis type) to demonstrate spurious phase generation. We show that the spurious phase appears with the Lagrangian mobility, whereas in the case of the Bollada-Jimack-Mullis mobility matrix the zero-amplitude initially prescribed for one of the field is retained throughout the simulation, even if the other fields are not in equilibrium. Finally, a similar behavior is demonstrated for anisotropic systems.\\

If not stated otherwise, the dynamic equations were solved numerically, using a pseudo-spectral semi-implicit method based on operator splitting (see Appendix C), on a rectangular grid of size $512 \times 512$, applying periodic boundary conditions at the perimeters. The dimensionless time and spatial steps were $\Delta t = 1$, and $\Delta x = 1$. The computations were performed in double precision on GTX Titan GPU cards. As starting condition, we prescribe a spatially homogeneous state containing equal quantities of all the components, supplemented by a small amplitude of flux noise. For this composition, equilibrium requires the coexistence of $N$ pure phases. Accordingly, the transition process requires the formation and coarsening of these phases. We present illustrations for symmetric ($N = 5$), asymmetric ($N = 4$), and anisotropic ($N = 4$) cases.  

\begin{figure}[t]
\begin{center}
(a)\includegraphics[width=0.44\linewidth]{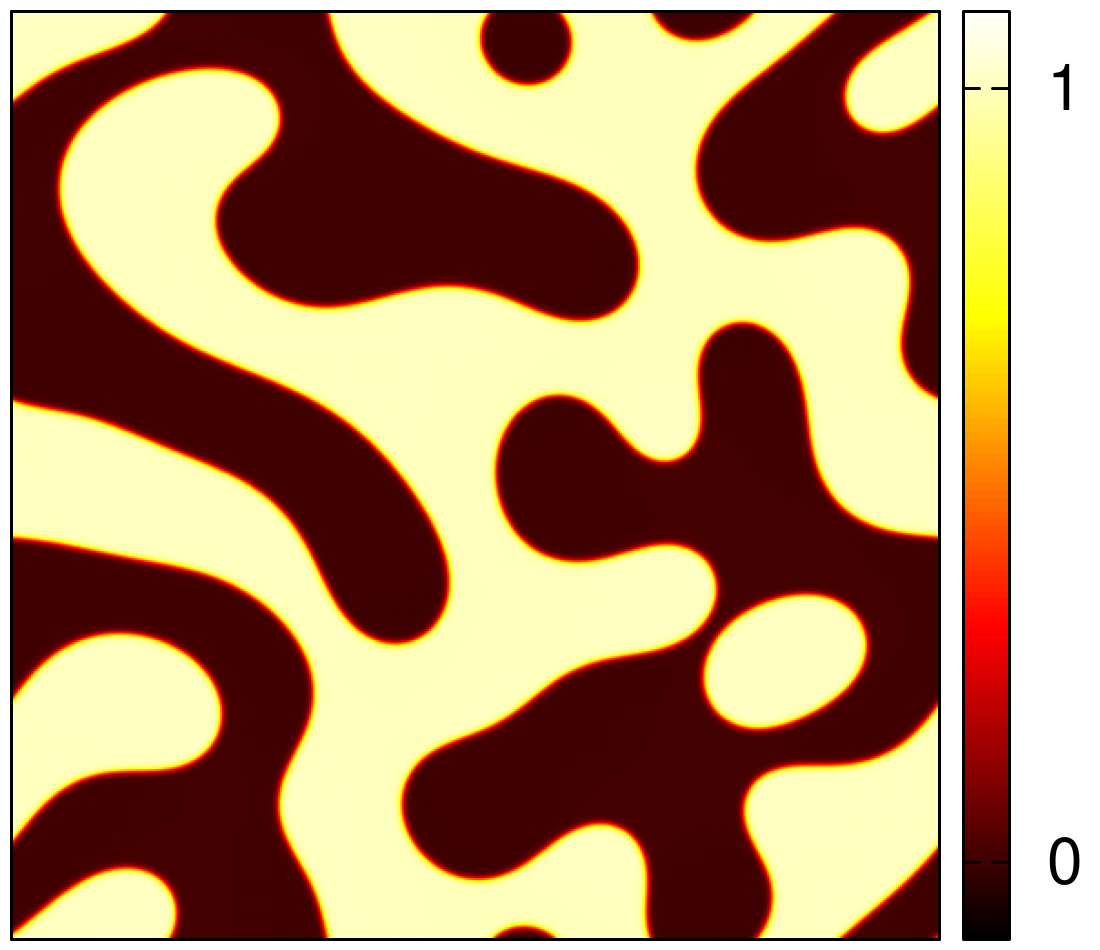} (b)\includegraphics[width=0.44\linewidth]{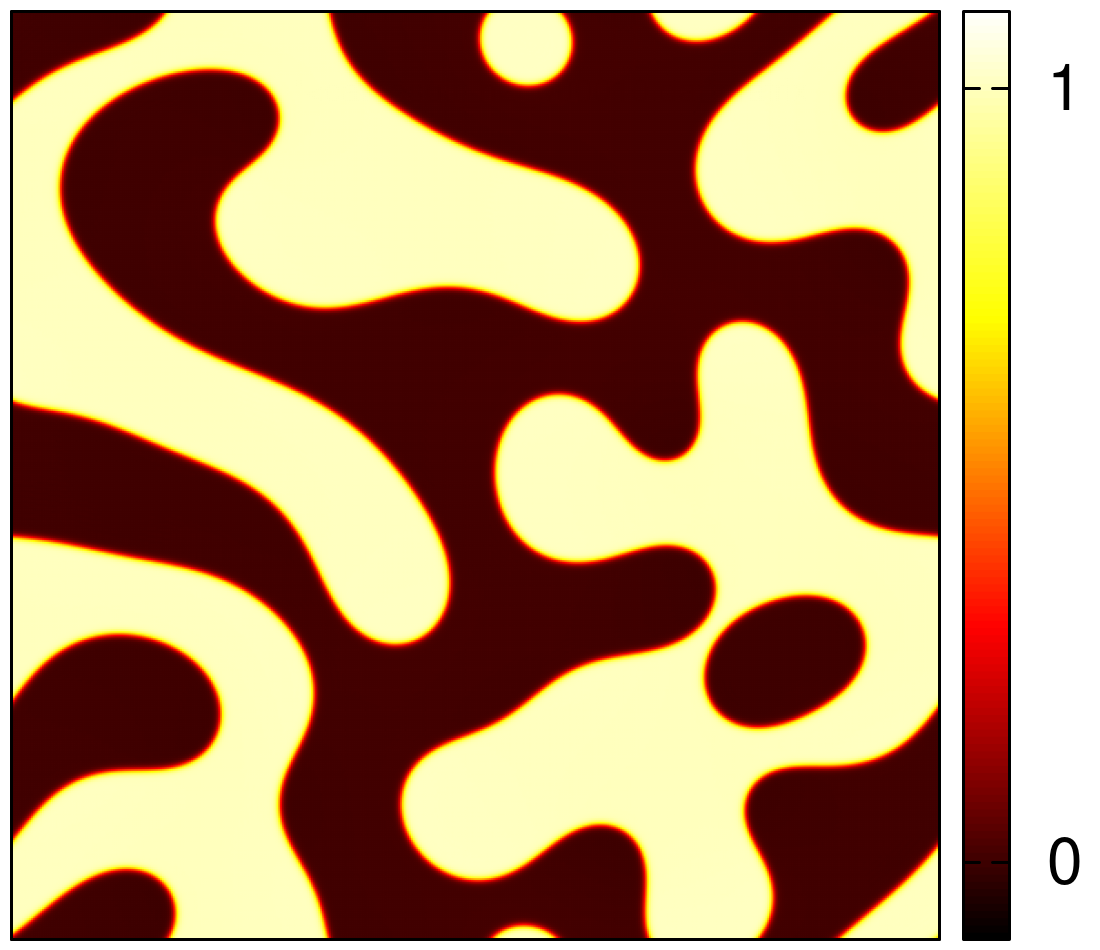}\\
(c)\includegraphics[width=0.44\linewidth]{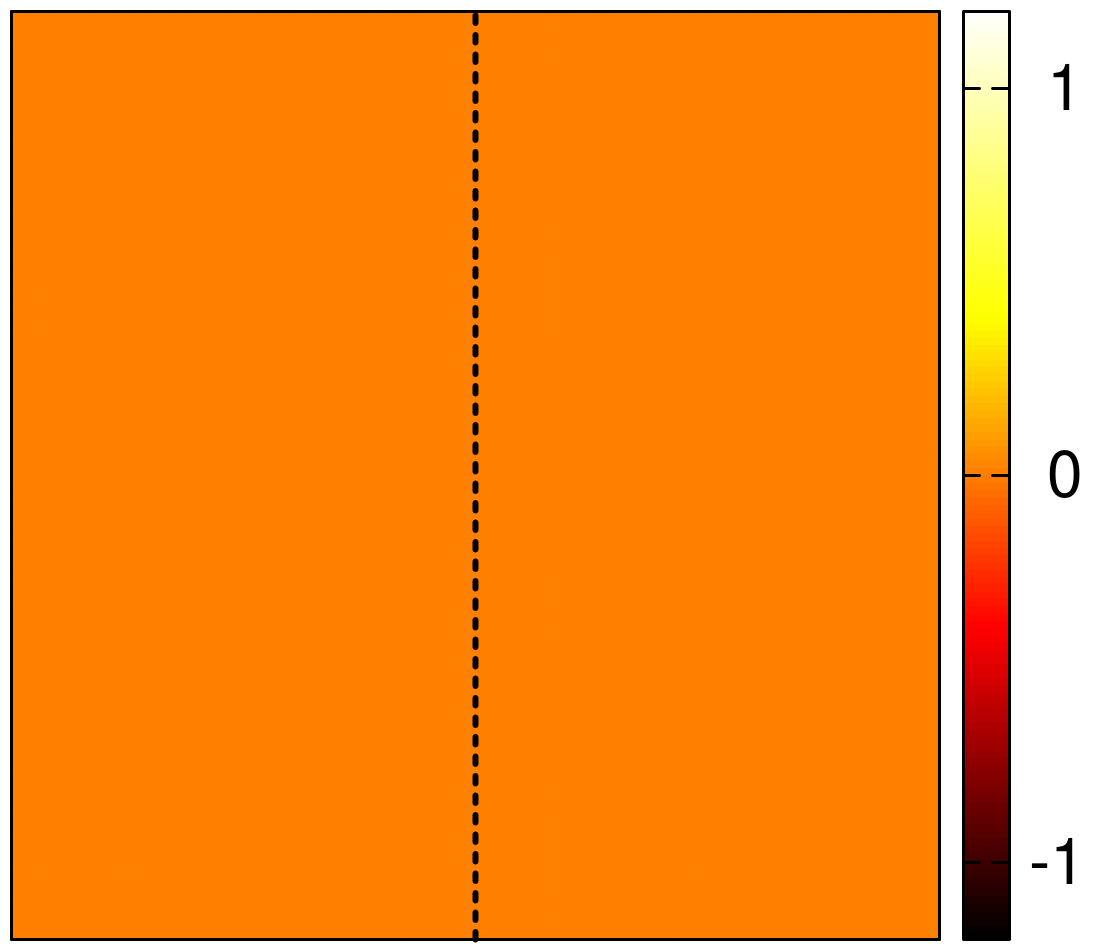} (d)\includegraphics[width=0.44\linewidth]{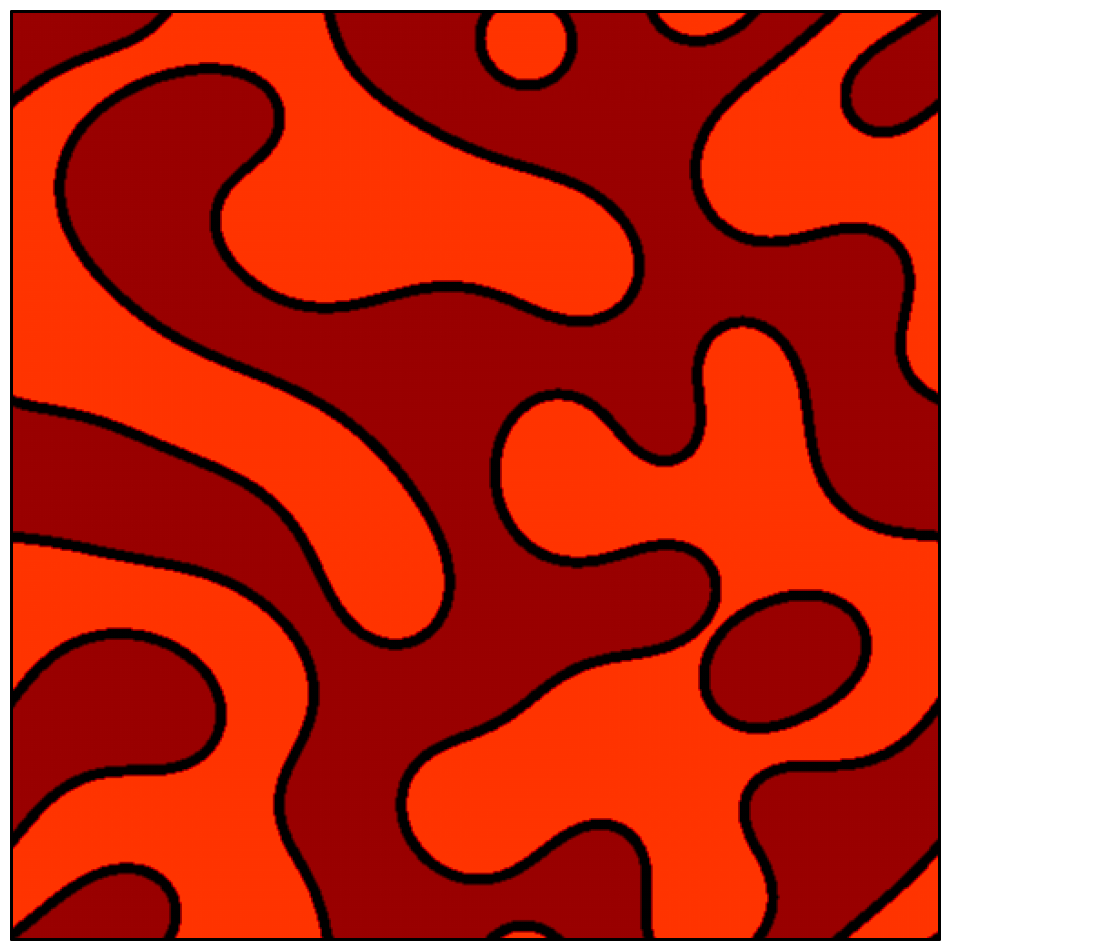} \caption{\label{fig:2nonzero} (Color online)  $N = 5$ symmetric Cahn-Hilliard problem with starting conditions $u_1 = u_2 = 0.5$ and $u_3 = u_4 = u_5 = 0$. (a), (b): Snapshots of the phase fields $u_1$ and $u_2$ at $t = 2.5\time10^5$ dimensionless time; (c) in the left half panel $|u_3|+|u_4|+|u_5|$ is shown, whereas the right half panel shows $\sum_{i=1}^{5}u_i-1$. Note the absence of spurious phases [criterion (ii)] and that the sum of the phase fields is 1 [criterion (i)]. (d) Multiphase-field map at the same instant.}
  \end{center}
\end{figure}
       
\subsubsection{Symmetric case}

\begin{figure}[t]
  \begin{center}
  (a)\includegraphics[width=0.44\linewidth]{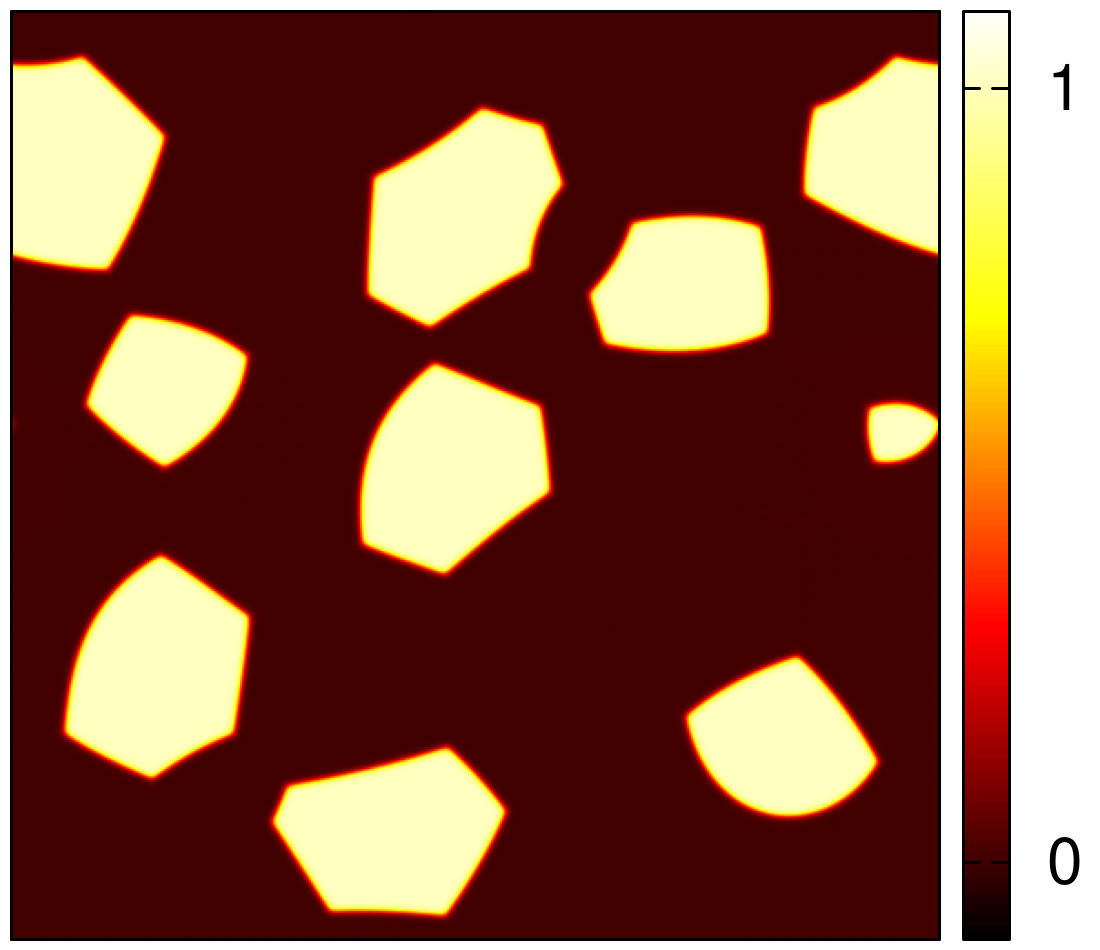} (b)\includegraphics[width=0.44\linewidth]{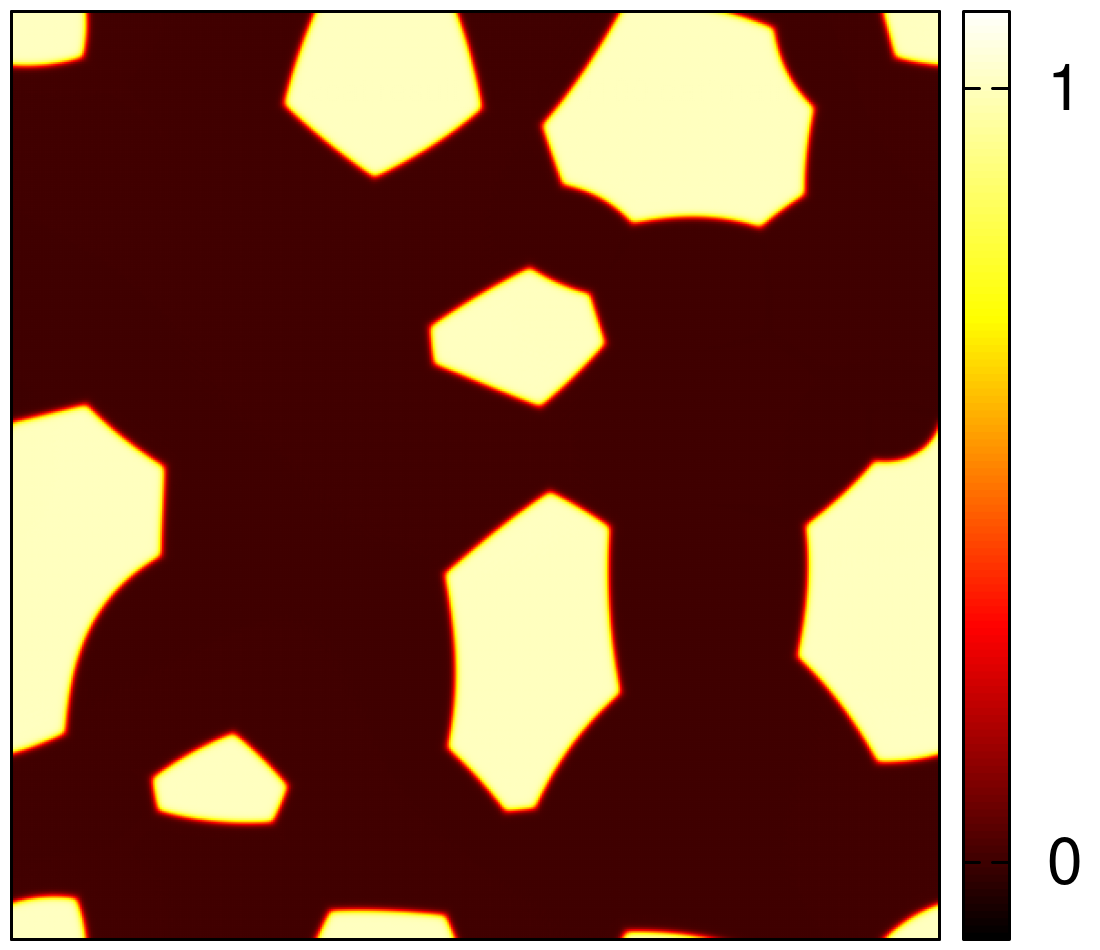}\\
  (c)\includegraphics[width=0.44\linewidth]{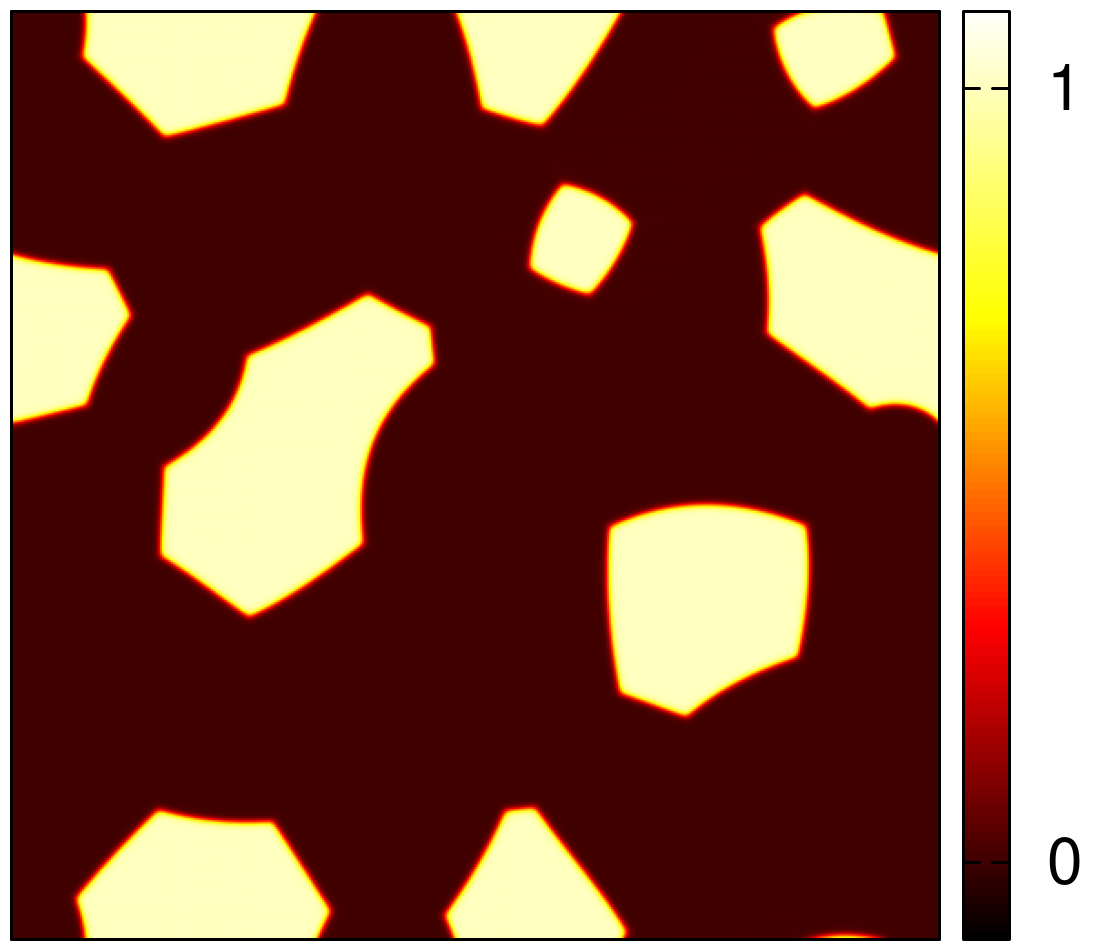} (d)\includegraphics[width=0.44\linewidth]{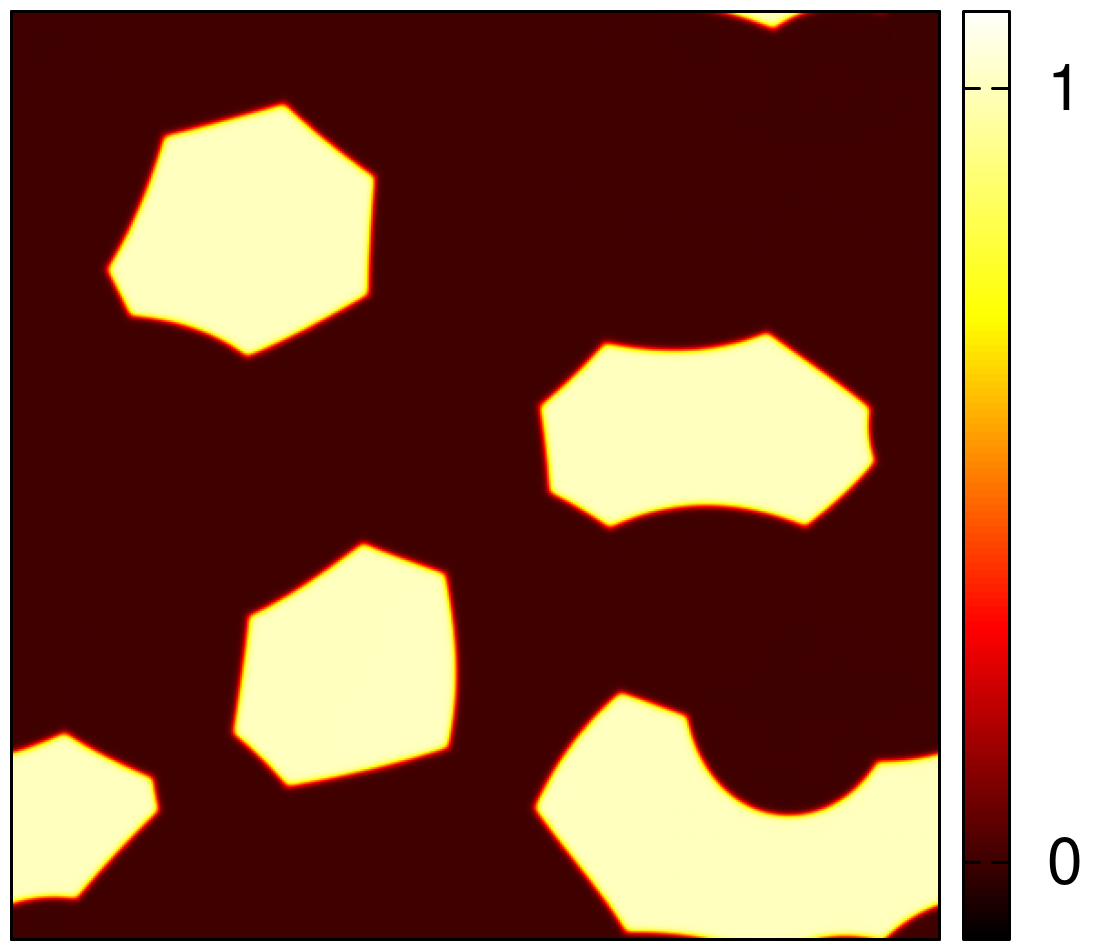}\\
  (e)\includegraphics[width=0.44\linewidth]{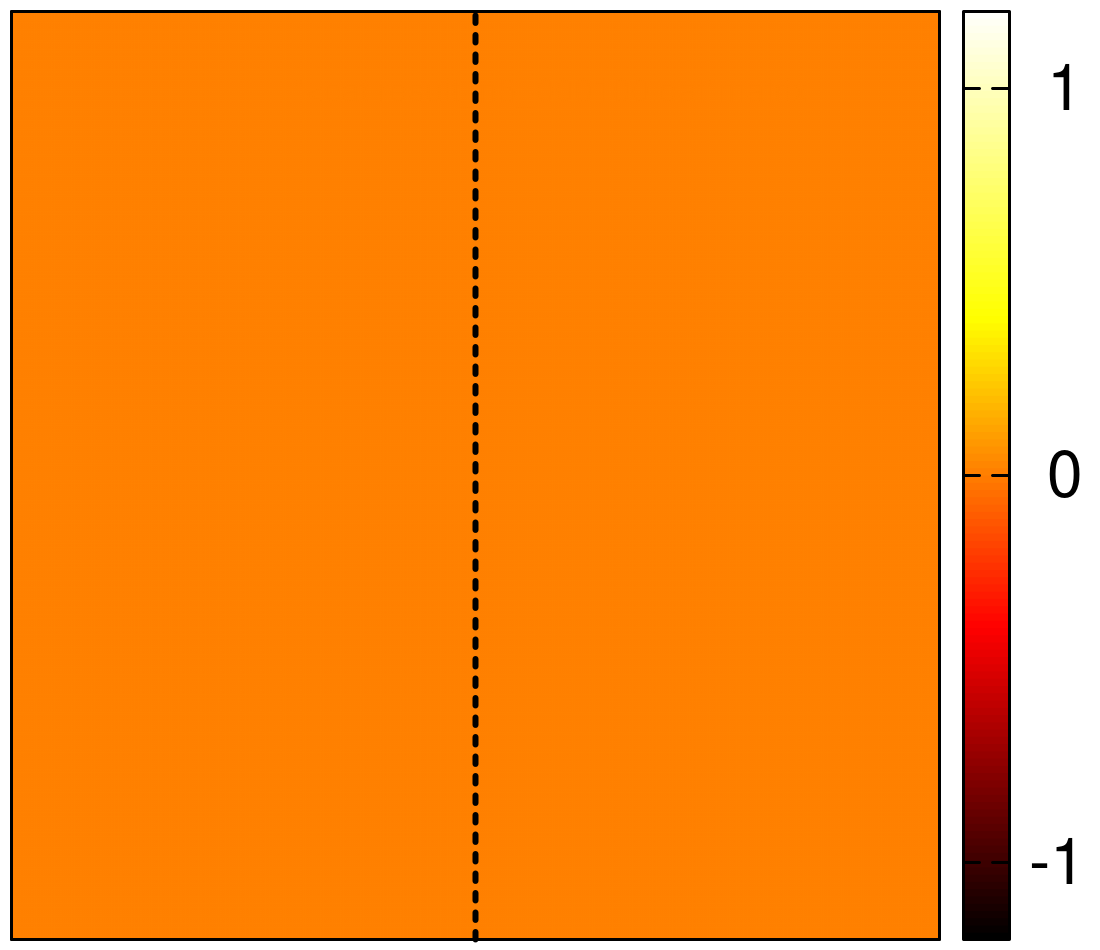} (f)\includegraphics[width=0.44\linewidth]{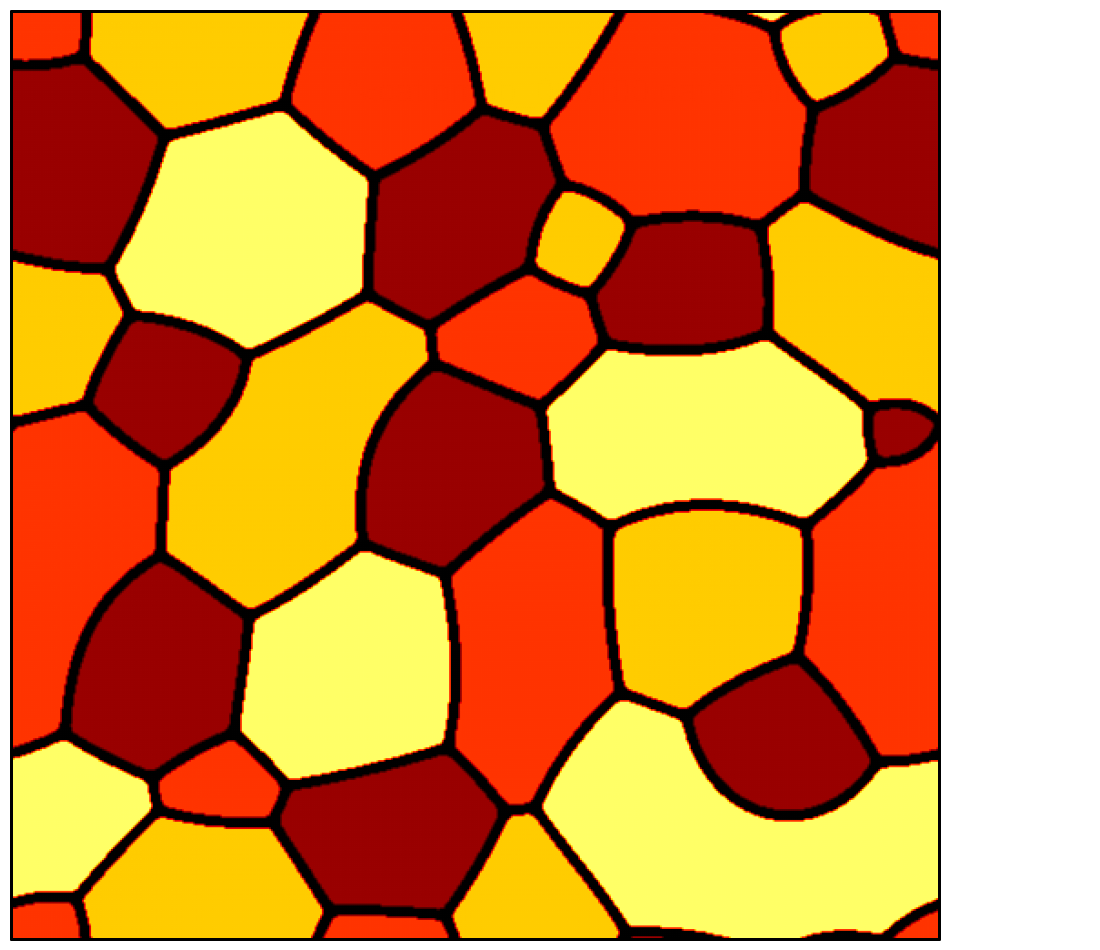} 	
  \caption{\label{fig:4nonzero} (Color online)  $N = 5$ symmetric Cahn-Hilliard problem with starting conditions $u_1 = u_2 = u_3 = u_4 = 0.25$ and $u_5 = 0$. (a) -- (b): Snapshots of the phase fields $u_1, u_2, u_3 $ and $u_4$ at $t = 10^6$ dimensionless time; (e) in the left half panel $u_5$ is shown, whereas the right half panel shows $\sum_{i=1}^{5}u_i-1$. Note the absence of spurious phases [criterion (ii)] and that the sum of the phase fields is 1 [criterion (i)]. (f) Multiphase-field map at the same instant.}
  \end{center}
\end{figure}

Snapshots illustrating the time evolution of an $N = 5$ symmetric Cahn-Hilliard problem (where $\epsilon_{ij}^2 = 1$ and $w_{ij} = 1$ for all phase pairs, $i,j = 1$ to $N$) are displayed in Fig. 1. The simulation started from a spatially homogeneous initial condition $\mathbf{u}(\mathbf{r},0) = \lbrace 0.2, 0.2, 0.2,$ $ 0.2, 0.2 \rbrace$, which was perturbed by a small amplitude of initial noise to induce phase separation. Note the coarsening of the various types of grains with time. To test whether spurious phase generation takes place, we have performed two $N = 5$ simulations with initial conditions: (1) $\mathbf{u}(\mathbf{r},0) =  \lbrace 0.5, 0.5, 0, 0, 0 \rbrace$ and for (2) $\mathbf{u}(\mathbf{r},0) =  \lbrace 0.25, 0.25, 0.25, 0.25, 0 \rbrace$. The individual phase-field maps are shown for $t = 250,000$ dimensionless time in Figs. 2 and 3. Apparently, $\sum u_j = 1$ is satisfied with a high accuracy [criterion (i), see Figs. 2(c) and 3(f)]. Furthermore, the zero amplitude fields retained accurately their zero amplitude status throughout the simulations, i.e., no third phase generation took place at the phase boundaries, indicating that, in agreement with the expectations, criterion (ii) is also fully satisfied. While in the effectively two-component case (Fig. 2), we observe the usual binary phase separation pattern, the patterns appearing in Figs. 1 and 3 are significantly different: multi-grain networks appear that are dominated exclusively by trijunctions and binary boundaries; higher-order junctions have not been observed at all. 

\begin{figure}[ht!]
  \begin{center}
  (a)\includegraphics[width=0.44\linewidth]{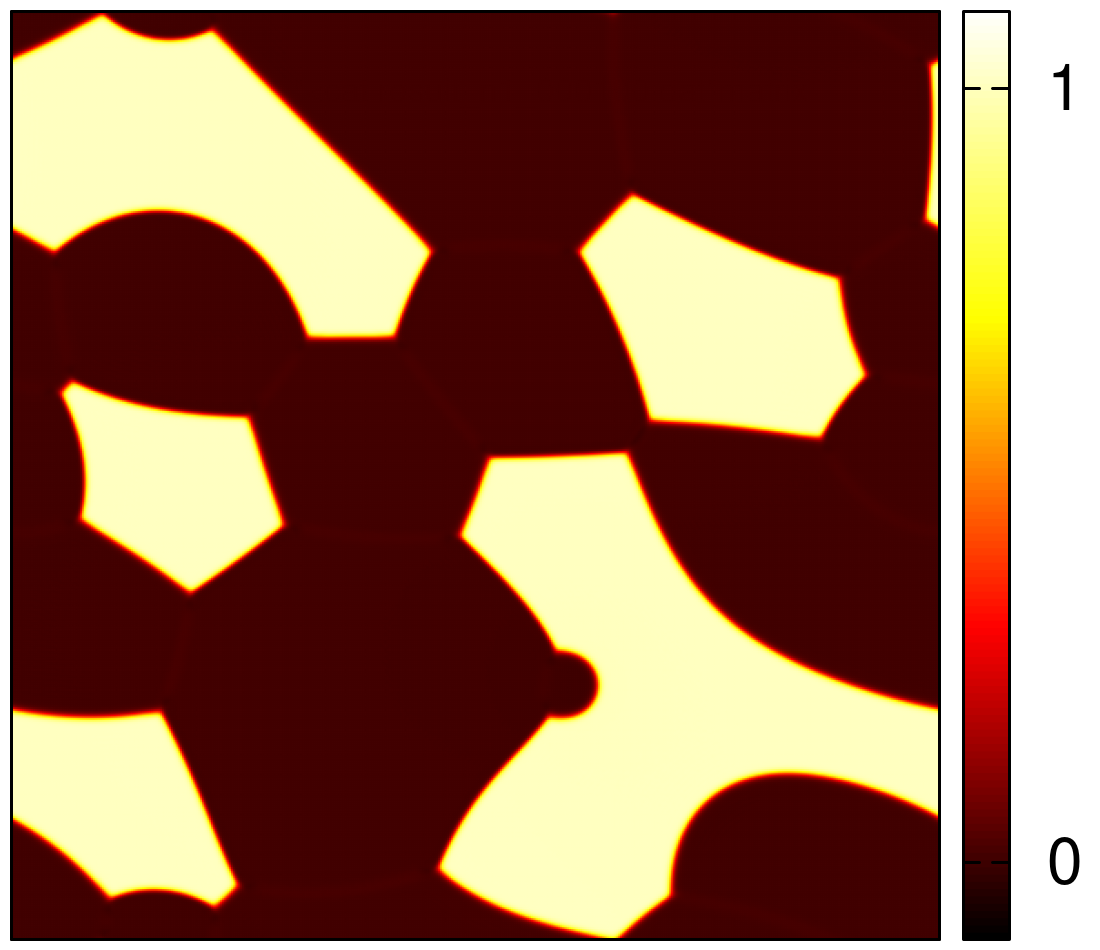} (b)\includegraphics[width=0.44\linewidth]{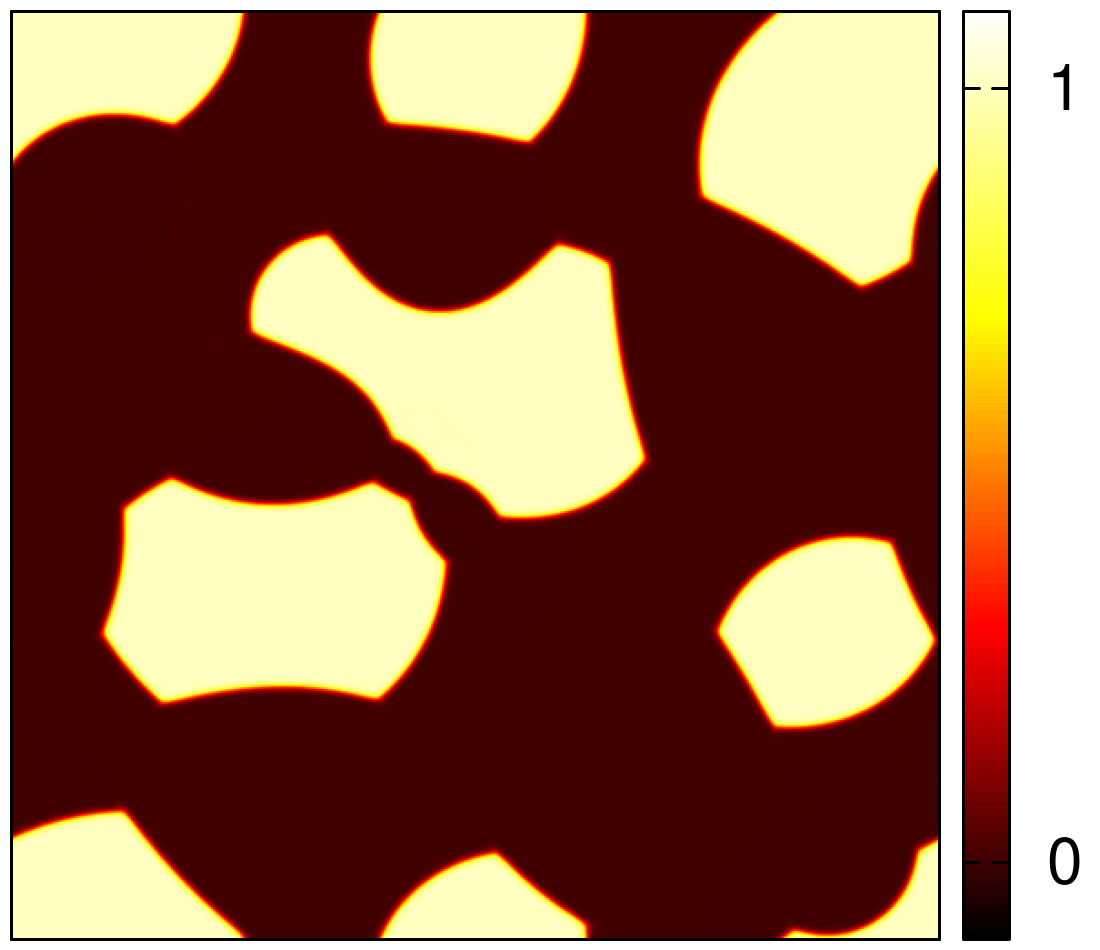}\\
  (c)\includegraphics[width=0.44\linewidth]{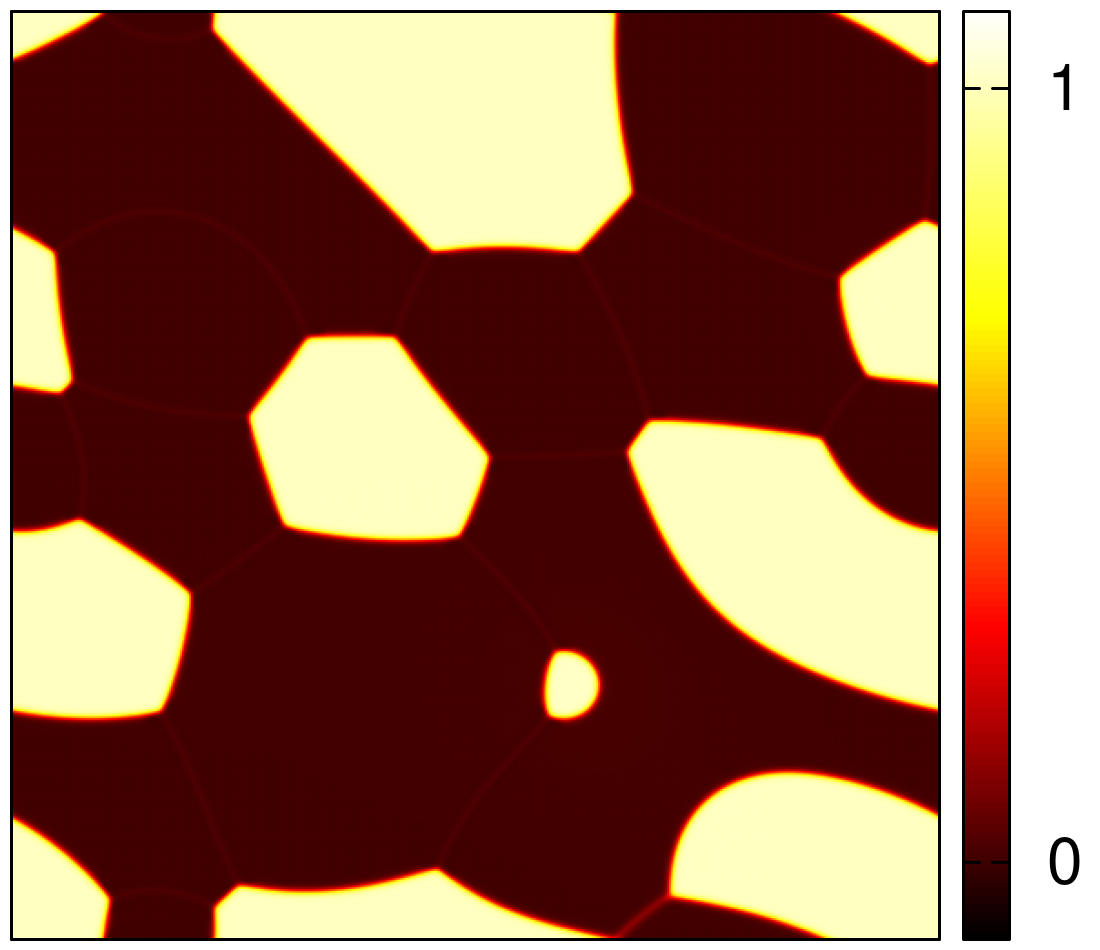} (d)\includegraphics[width=0.44\linewidth]{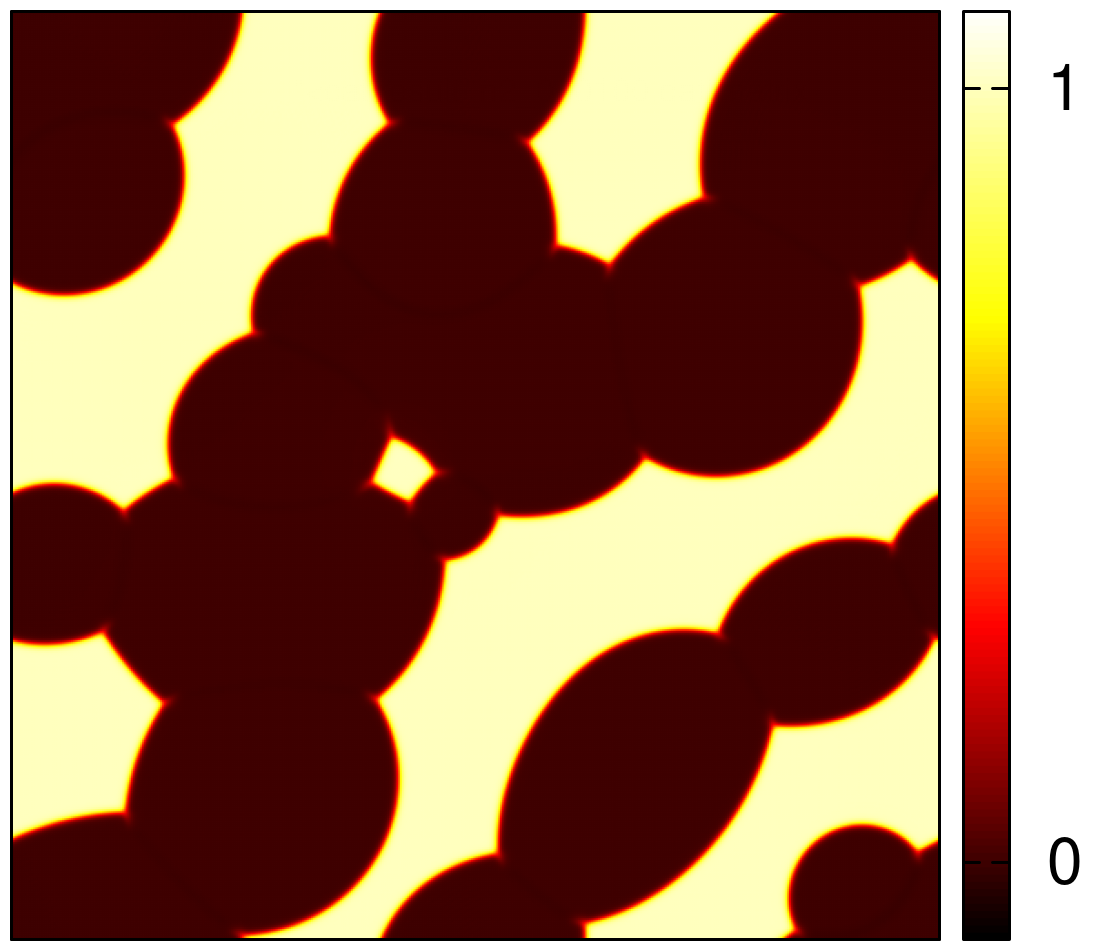}\\
  (e)\includegraphics[width=0.44\linewidth]{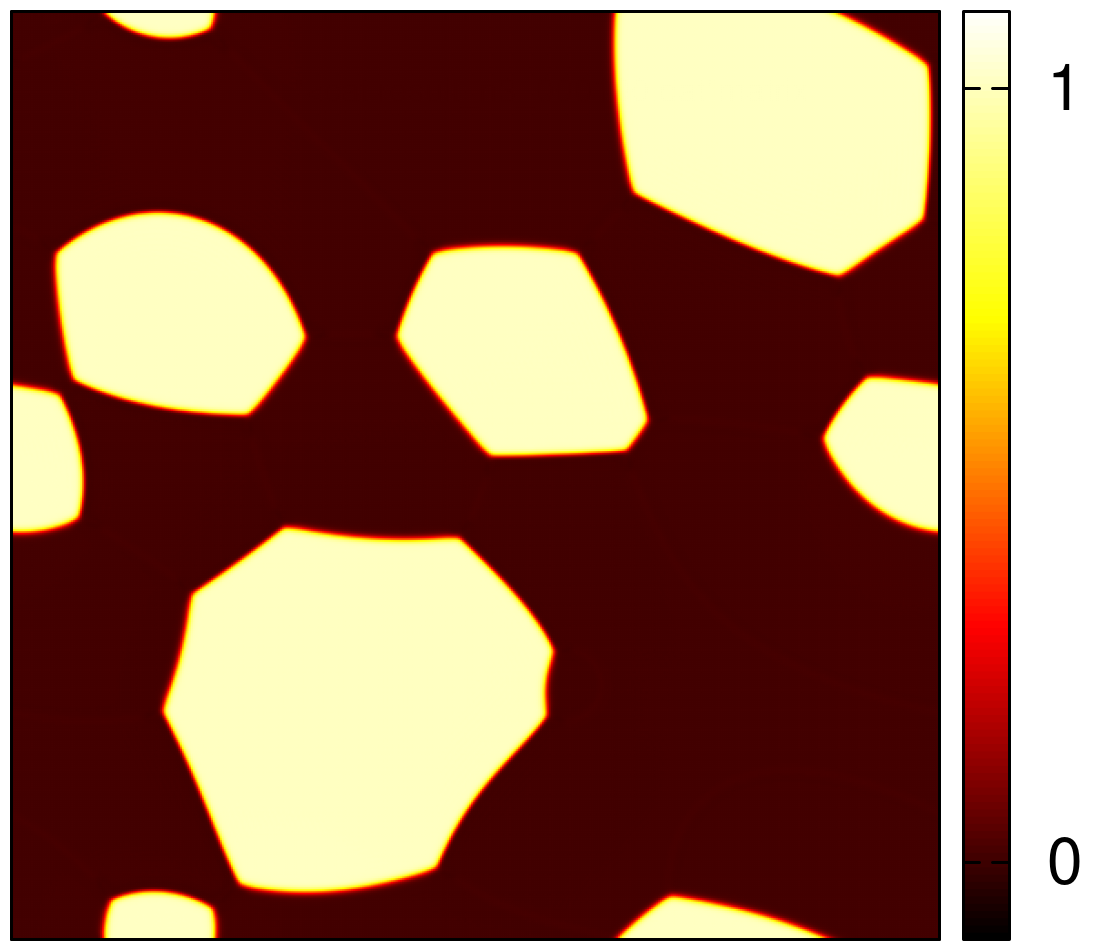} (f)\includegraphics[width=0.44\linewidth]{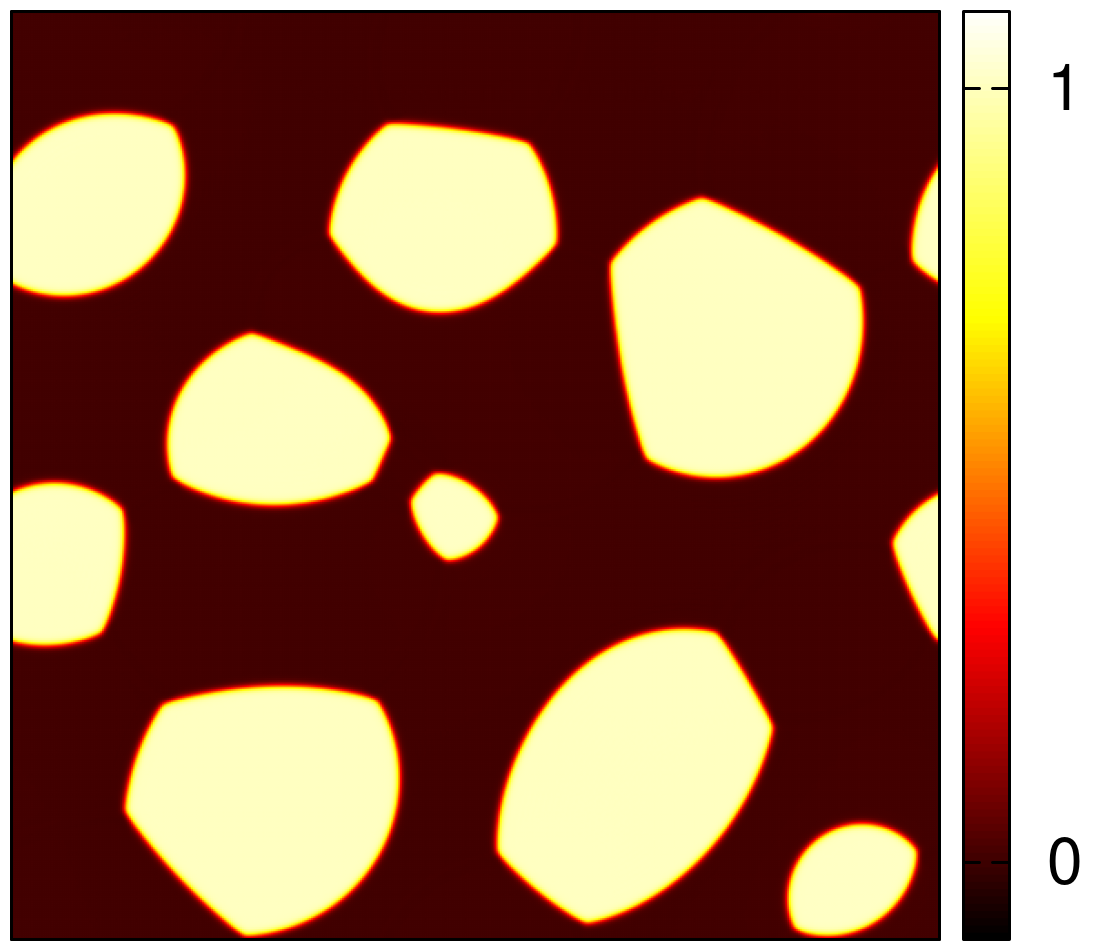}\\ 	
  (g)\includegraphics[width=0.44\linewidth]{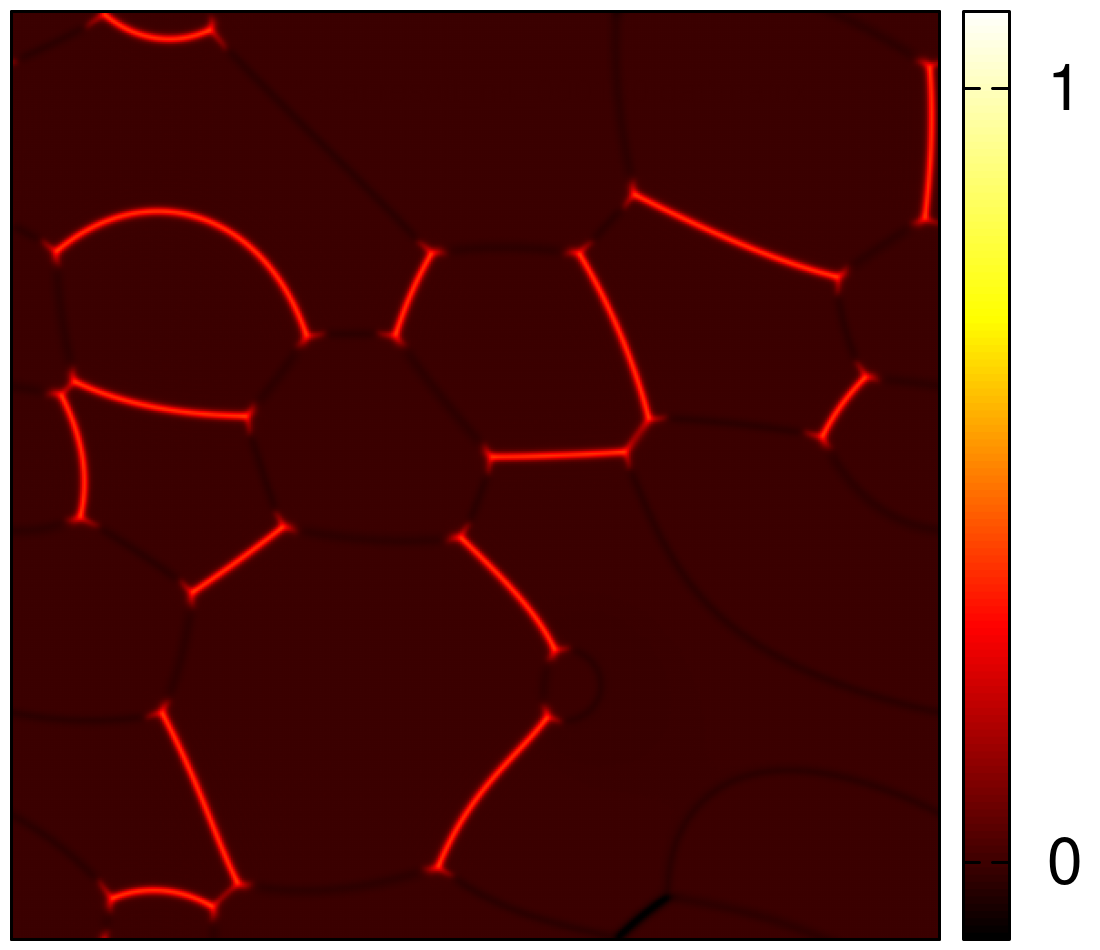} (h)\includegraphics[width=0.44\linewidth]{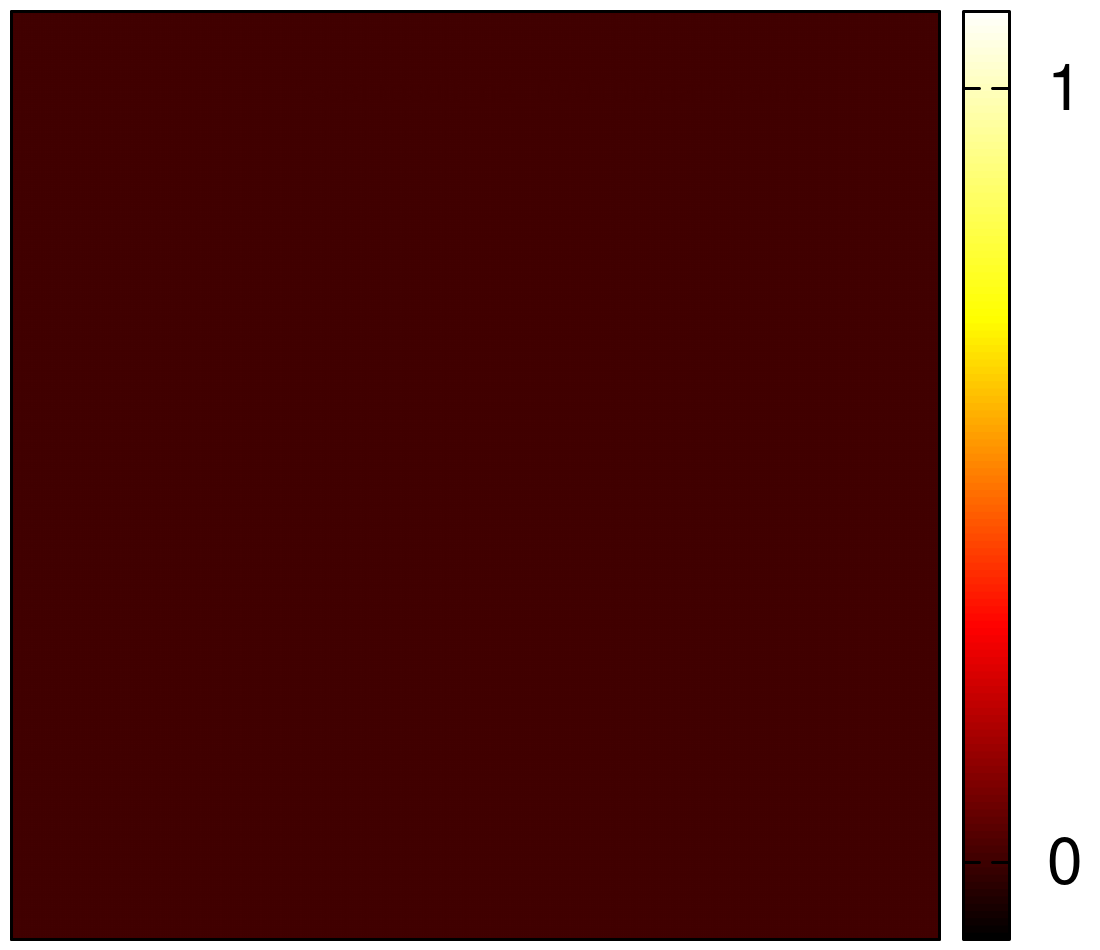}\\
  (i)\includegraphics[width=0.44\linewidth]{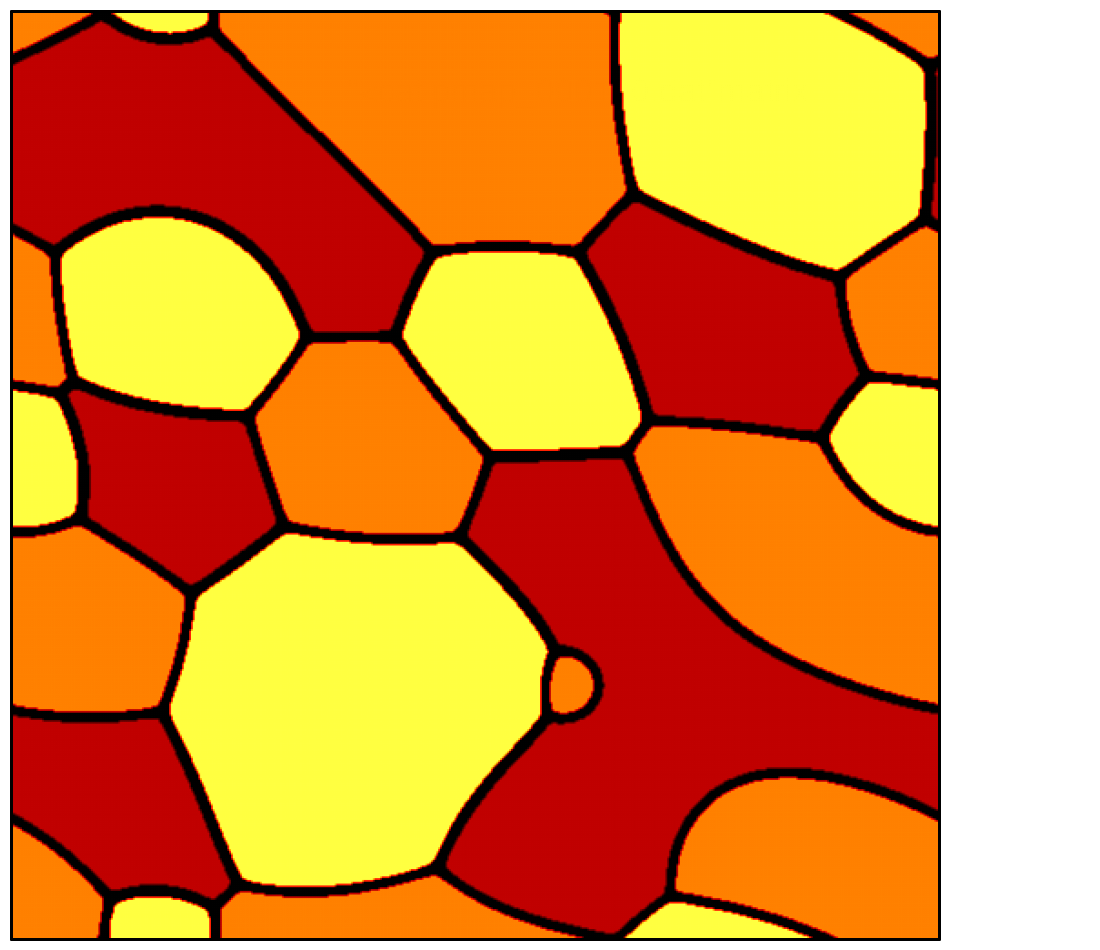} (j)\includegraphics[width=0.44\linewidth]{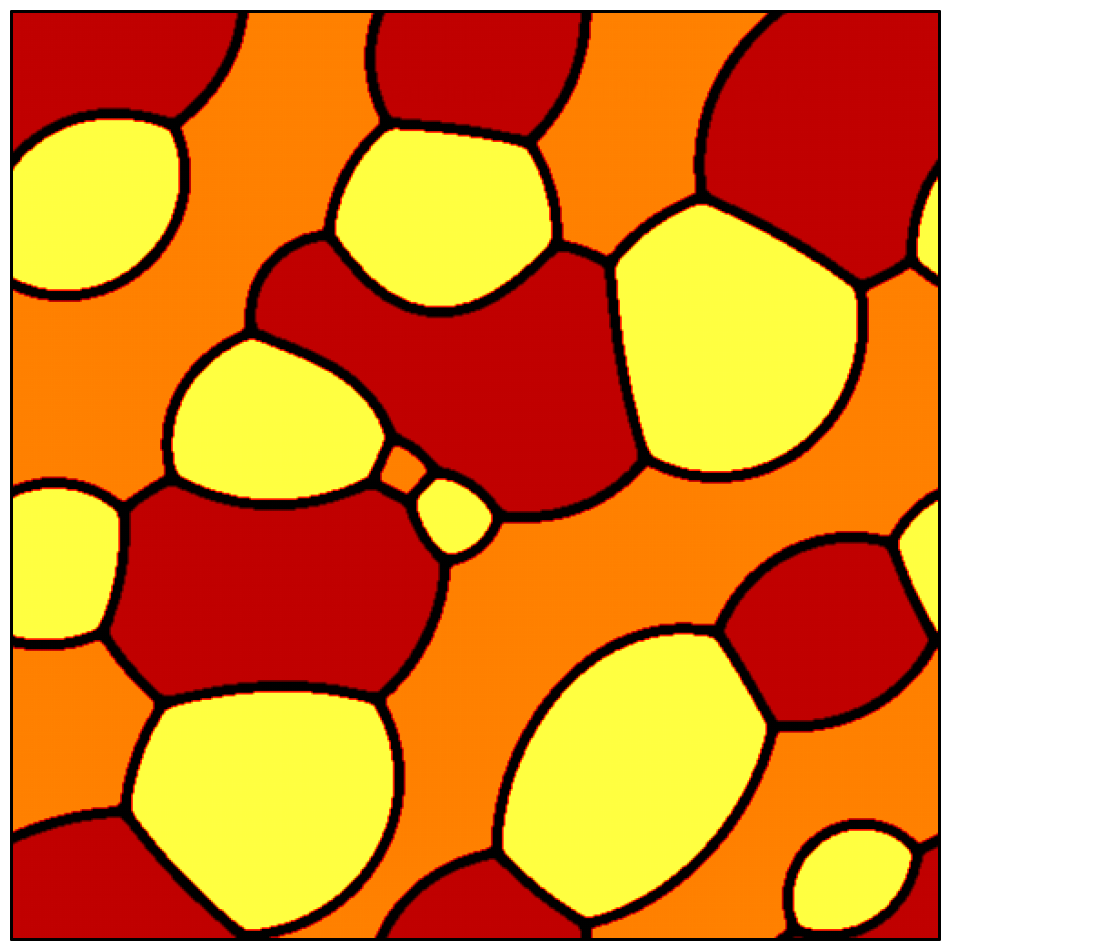}
  \caption{\label{fig:timeevol} (Color online)  $N = 4$ asymmetric Cahn-Hilliard problem with initial condition $\mathbf{u} =  \lbrace 1/3, 1/3, 1/3, 0 \rbrace$. The results on the left were obtained with a Lagrangian mobility matrix, whereas those on the right were obtained using the Bollada-Jimack-Mullis type mobility matrix. The upper four rows [panels (a) -- (h)] show the maps for $u_1, u_2, u_3$ and $u_4$, respectively, whereas in the lowermost row [panels (i) and (j)] phase distribution maps are displayed. Results corresponding to $t = 10^6$ dimensionless time are presented. Note that using the Lagrangian mobility matrix spurious phase generation in the vicinity of the phase boundaries could not be avoided for $u_4$ [see panel (g)], whereas the Bollada-Jimack-Mullis mobility matrix suppresses the formation of spurious phases entirely [see panel (h)].}
  \end{center}
\end{figure}

\subsubsection{Asymmetric case}

Here, the parameters $\epsilon_{ij}^2$ and $w_{ij} $ are different for the individual binary interfaces (for the matrices used in the present work see Ref. \cite{matrices}). The simulations were performed for an $N = 4$ Cahn-Hilliard model. $\mathbf{u}(\mathbf{r},0) =$ $ \lbrace 1/3, 1/3, 1/3, 0\rbrace$ has been chosen as the initial condition, with a small amplitude of initial noise to induce phase separation. First a Lagrangian mobility matrix has been used. The phase-field maps corresponding to $t = 10^6$ dimensionless time are displayed in the left column of Fig. 4. Apparently, the solution is less satisfactory than the results for the symmetric case: substantial deviation from $u_4(\mathbf{r},t) = 0$ is observed [Fig. 4(g)]. However, this spurious phase generation disappears entirely, if the Bollada-Jimack-Mullis mobility matrix is used [see Fig. 4(h)], as expected. Again, the multiphase domain structure is dominated by trijunctions and binary boundaries at all times. It appears though that the structure obtained with the Bollada-Jimack-Mullis type mobility matrix contains chains of alternating $u_1$ and $u_4$ "bubbles" [see Fig. 4(j)], a feature that can be associated with the asymmetry of the kinetic coefficients ($\kappa_{ij}$) applied in this simulation.

\begin{figure}[ht!]
  \begin{center}
 (a)\includegraphics[width=0.9\linewidth]{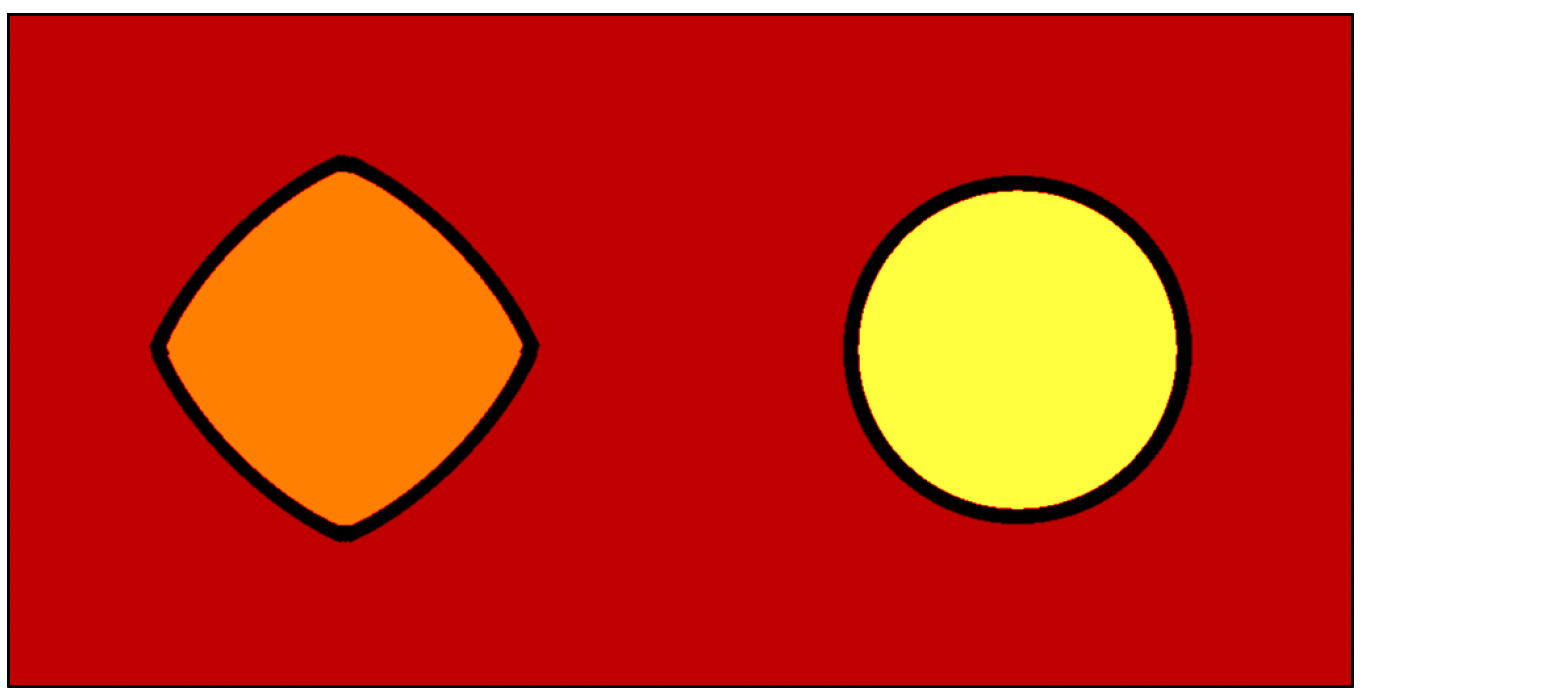}
 (b)\includegraphics[width=0.9\linewidth]{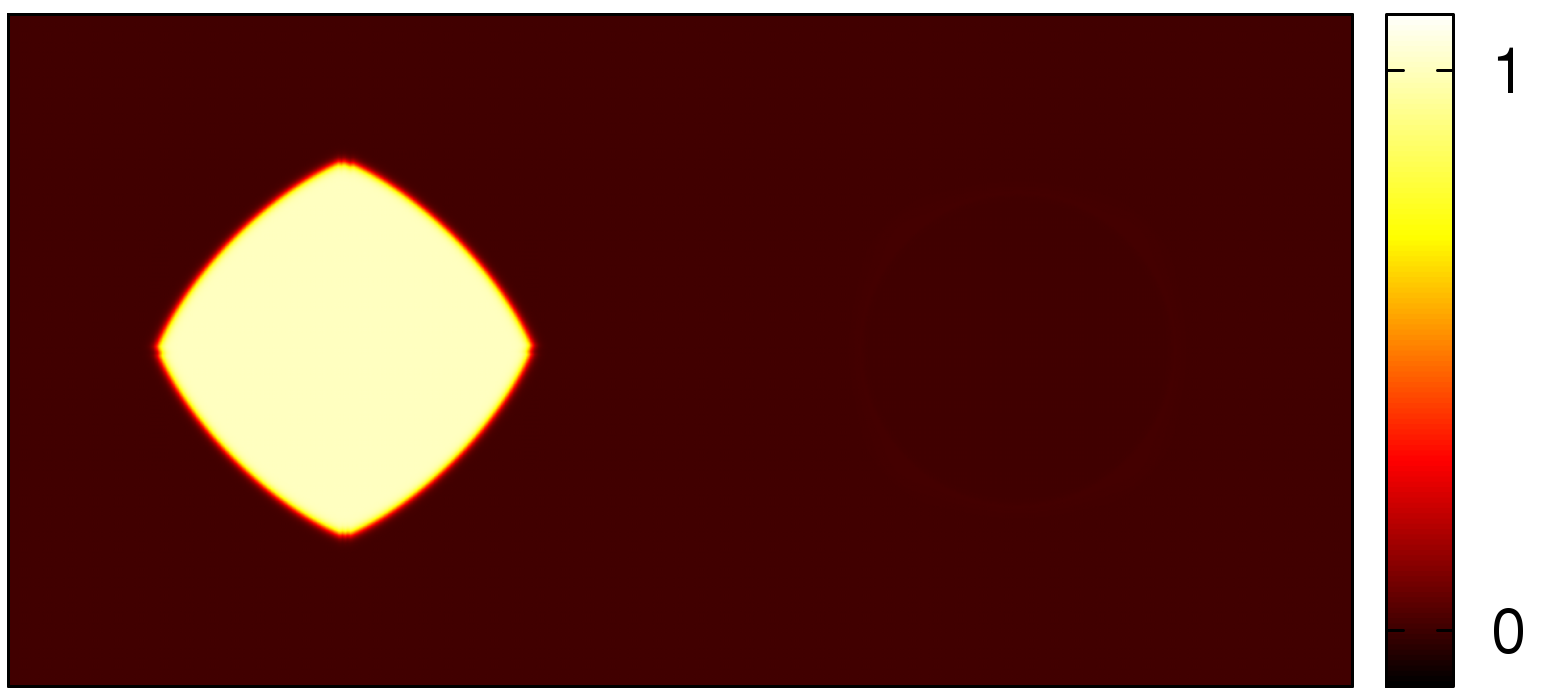}  
 (c)\includegraphics[width=0.9\linewidth]{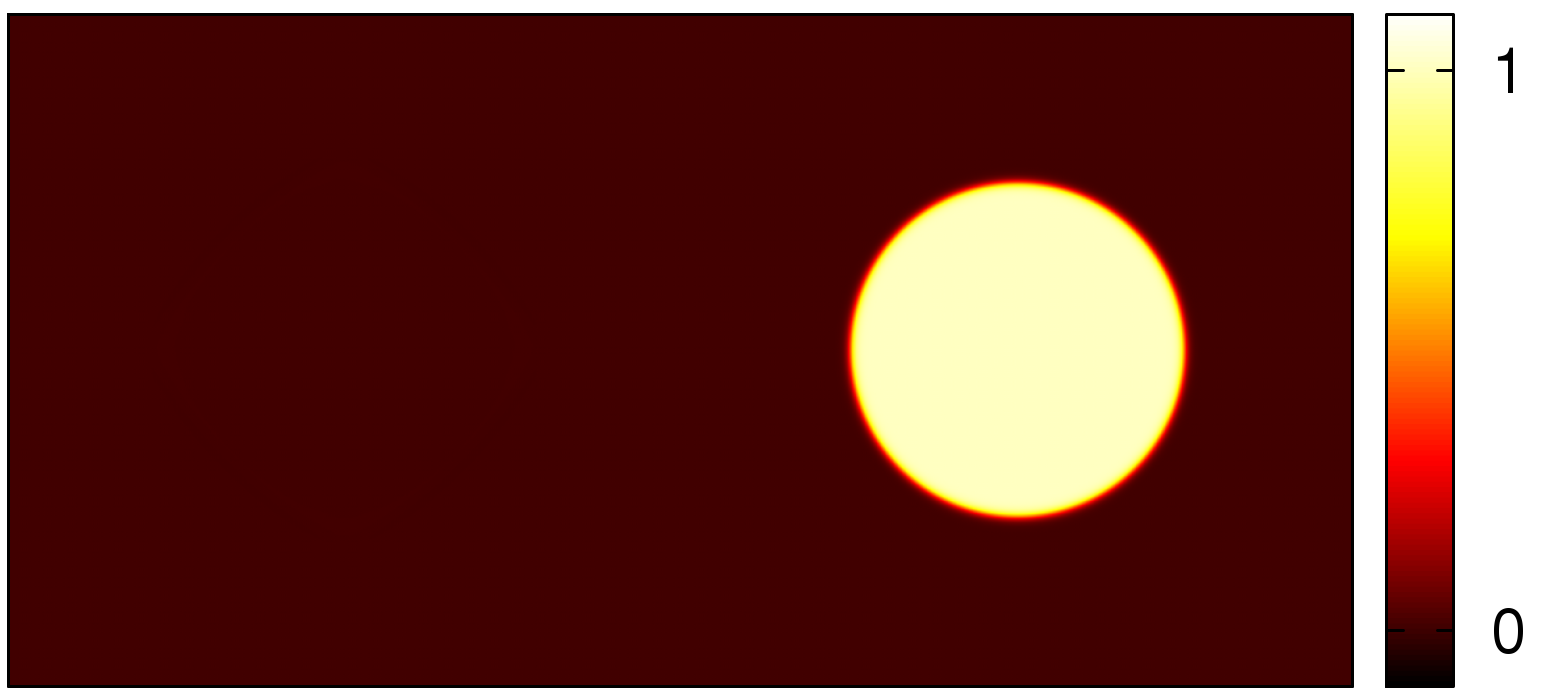} 	
 (d)\includegraphics[width=0.9\linewidth]{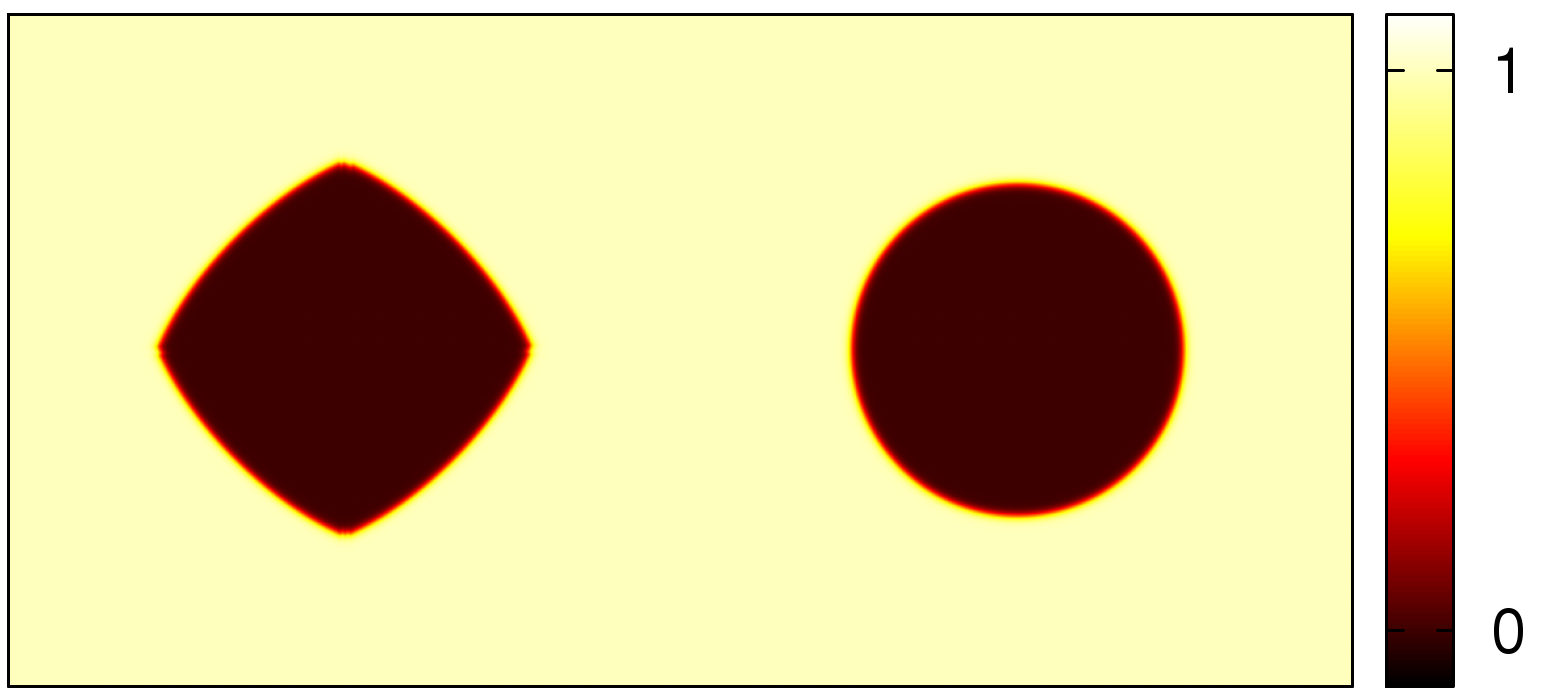}  
 (e)\includegraphics[width=0.9\linewidth]{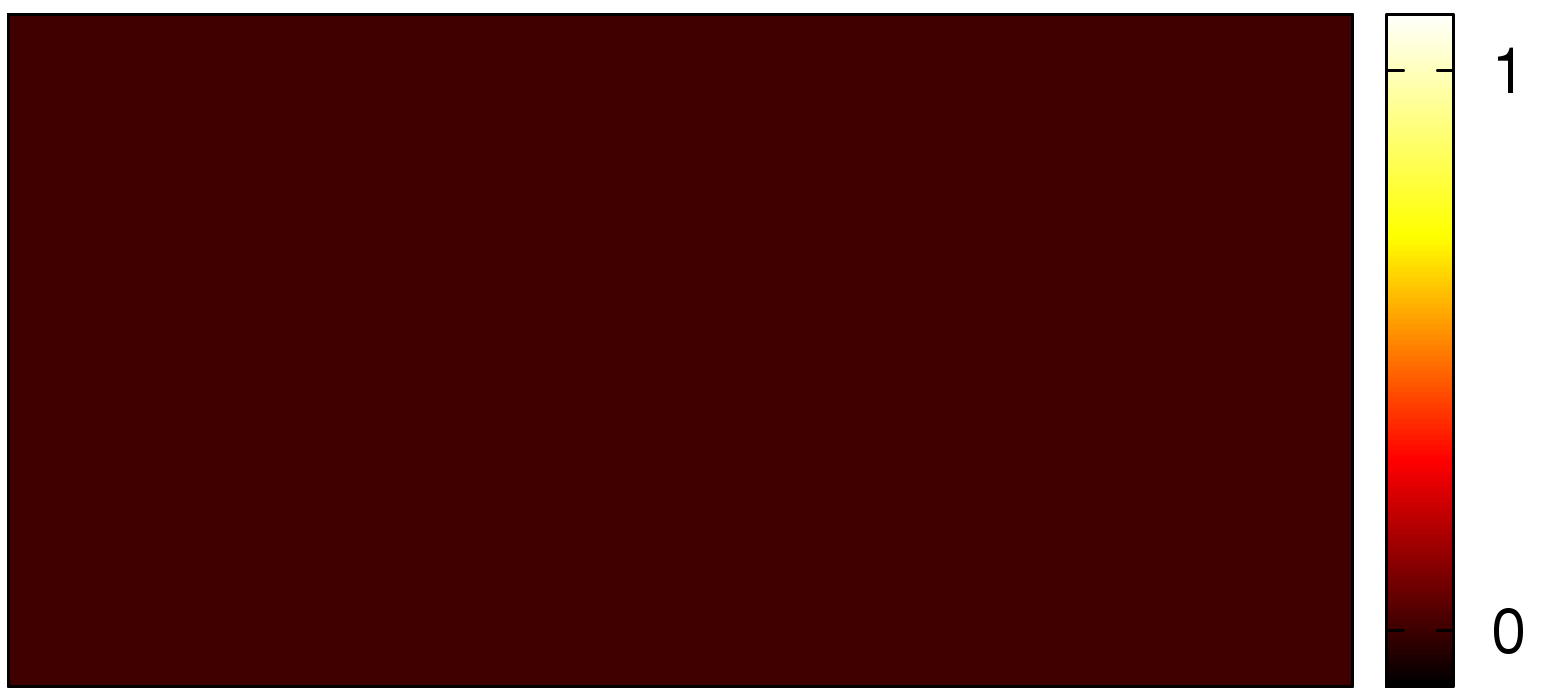} 	
 \caption{\label{fig:timeevol} (Color online)  $N = 4$ asymmetric Cahn-Hilliard problem with anisotropy: (a) Time evolution towards the equilibrium shapes of $u_1$  and $u_2$ embedded into $u_3$ at $t=3.125\times10^6$ dimensionless time. (b)-(e): Snapshots of the phase fields $u_1, u_2, u_3,$ and $u_4$, respectively, taken at $t = 2\times10^5$ dimensionless time. Note the lack of third phase generation at the phase boundaries, and that $u_4$ remains absent even at the phase-boundaries.}
  \end{center}
\end{figure}

\subsubsection{Anisotropic case}

Next, we investigated an asymmetric $N = 4$ Cahn-Hilliard theory \cite{matrices}, however, with anisotropic interface free energies and a Bollada-Jimack-Mullis type mobility matrix. The dynamic equations were solved on a rectangular grid of size $1024 \times 512$. Time and spatial steps of $\Delta t = 0.25$ and $\Delta x = 0.5$ were used. The starting conditions were as follows: Two circles filled with $u_1 = 1$ (left) and $u_2 = 1$ (right) were placed besides each other, while a zero value was assigned to these fields outside the circles. In contrast, $u_3 = 1$ was prescribed in the background, and $u_3 = 0$ inside the circles, whereas $u_4 = 0$ was assigned to the whole simulation domain (i.e., the fourth field was missing everywhere). All interfaces were assumed isotropic, except for the 1--3 interface, for which an anisotropy of $a_{13} = 0.1$ was prescribed that is larger than the critical anisotropy $a_c = 1/(2^k - 1) = 1/15$ for fourfold ($k =4$) symmetry \cite{Kobayashi2001,Eggleston2001}. With elapsing time, the circle on the left evolved into a square-like object of curved sides, and four pointed corners (see Fig. 5), displaying missing orientations (following from $a_{13} > a_c$), as expected on the basis of the prescribed anisotropy function. Apparently, as found for the central finite differencing scheme \cite{Eggleston2001}, the spectral discretization regularized the high anisotropy problem: the predicted numerical shape is very close to the analytical solution corresponding to this anisotropy. Remarkably, no spurious phase appearance was observed at the phase-boundaries, and $u_4 = 0$ has been satisfied throughout the simulation. We have obtained similar results using finite difference discretization.


\begin{figure}
\includegraphics[width=1.0\linewidth]{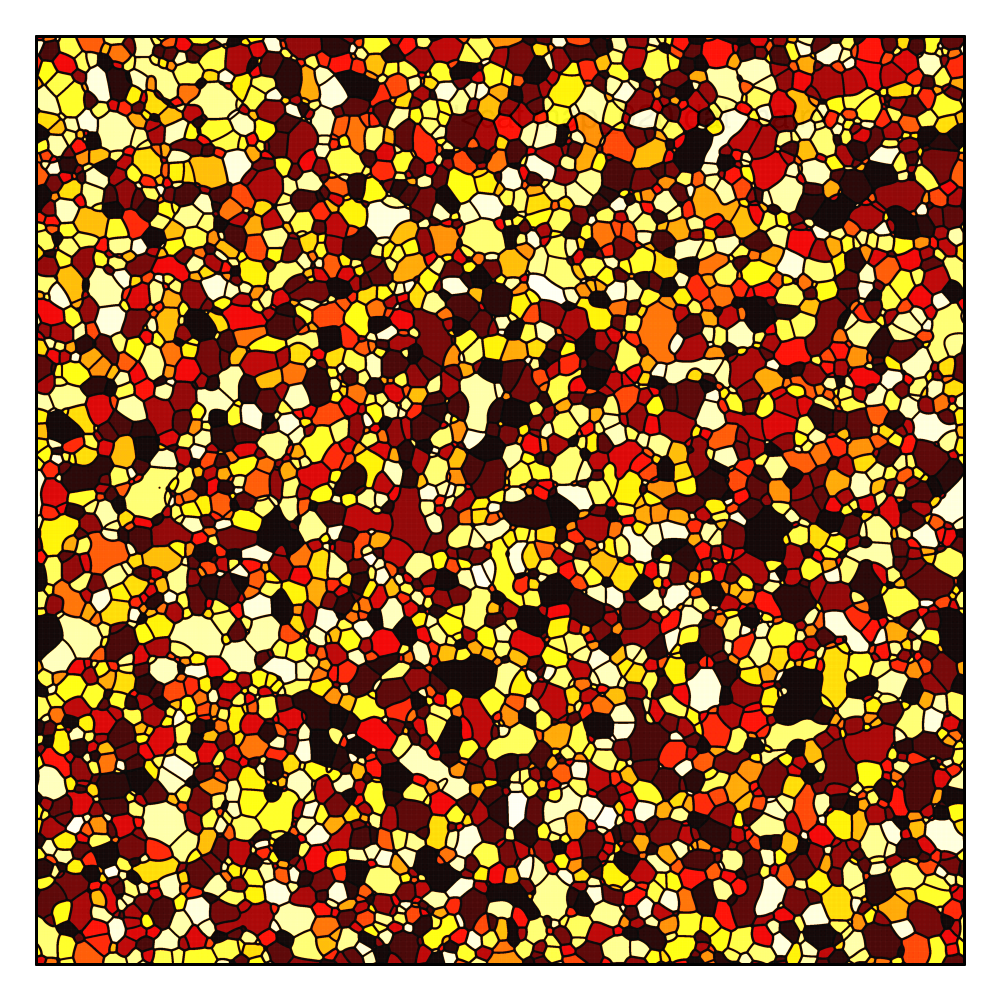}
\caption{{(Color online) Snapshot of grain/phase-map taken at dimensionless rime, $t = 10^4$, for a simulation relying on misorientation dependent grain boundary energy obeying the Read-Shockley relationship\cite{ReadShockley1950}. The simulation was performed on a $4096^2$ rectangular grid. About 4000 grains can be distinguished at this stage.}}
\end{figure}

\begin{figure}
\includegraphics[width=1.0\linewidth]{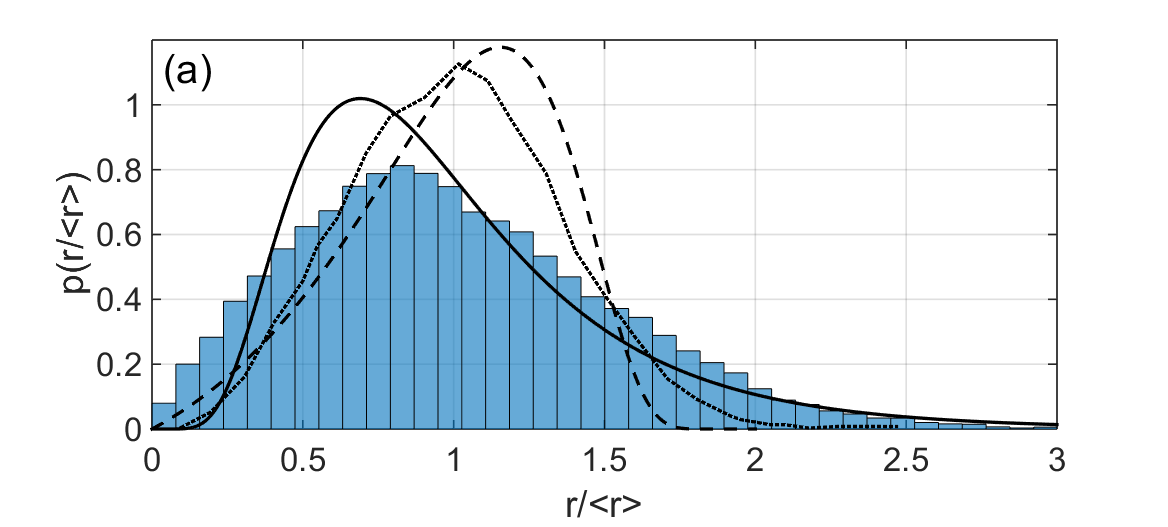}
\includegraphics[width=1.0\linewidth]{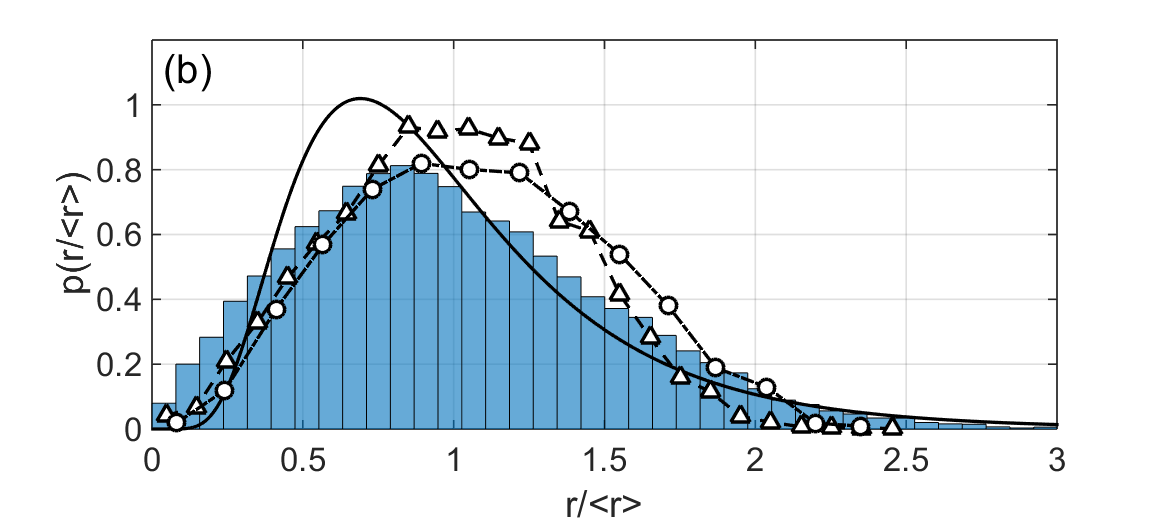}
\includegraphics[width=1.0\linewidth]{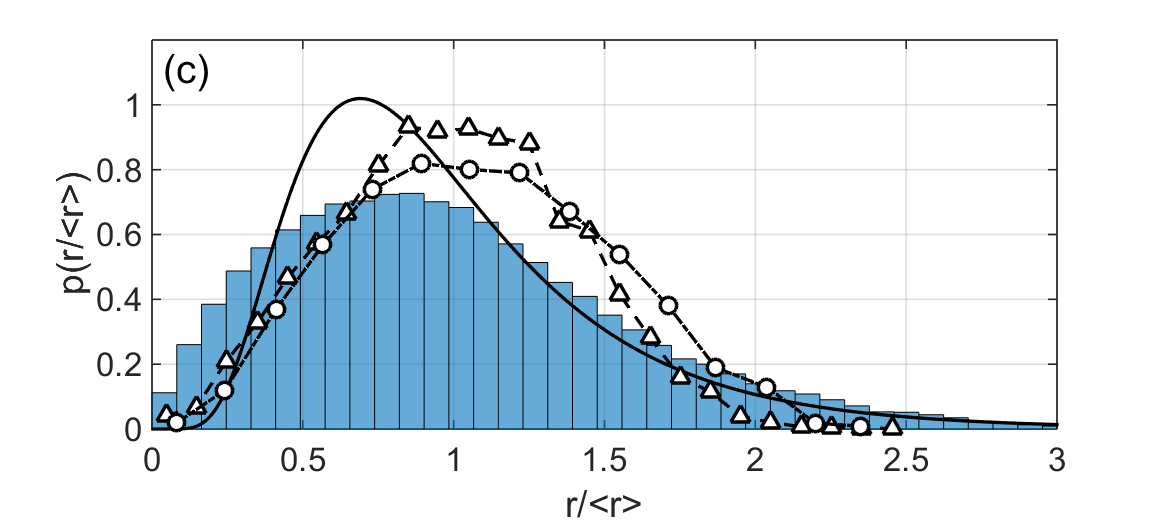}
\caption{{(Color online) Limiting grain size distributions (LGSD):  (a) for the isotropic simulation on a $8192^2$ grid. For comparison,  experimental results\cite{Barmak2013} for metallic films (solid line: lognormal distribution fitted to experimental data\cite{Barmak2013}), and predictions by the theories of Mullins\cite{Mullins1956} and Hillert\cite{Hillert1965} (dashed and dotted lines, respectively) are also shown. (b) Comparison of the LGSD from XMPF (histogram) with predictions from previous MPFs by Kim {\it et al.}\cite{Kim2006} (triangles) and Schaffnit {\it et al.}\cite{Schaffnit2007} (circles). The solid line indicates a lognormal fit to the experimental results\cite{Barmak2013}. (c) The same as panel (b), except that LGSD from an asymmetric XMPF model (the simulation shown in Fig. 6) is displayed (histogram), in which the grain boundary energy follows the Read-Shockley relationship\cite{ReadShockley1950}. Note that some improvement relative to the previous MPF models has been achieved, but the population of the small grains is larger than desirable.}} 
\end{figure}

\subsection{Grain coarsening}

In this subsection, we apply the XMPF model for grain coarsening in a two-dimensional (2D) polycrystalline system that contains a large number of differently oriented crystal grains that have equal free energy, therefore, the time evolution of the system is driven by the grain boundary energy. For the sake of simplicity, we distinguish only 30 orientations represented by $N=30$ fields. The respective non-conservative equations of motions read as:
\begin{equation}
-\frac{\partial u_i}{\partial t} = \sum_{j\neq i}\kappa_{ij} \left(\frac{\delta F}{\delta u_i}-\frac{\delta F}{\delta u_j} \right) \enskip ,
\end{equation}
which have been solved numerically on a rectangular grid, using a finite difference discretization and Euler forward time-stepping. As starting condition, the simulation box was covered by a large number of small random field patches arranged on a square lattice ($512\times512$ and $256\times256$, respectively, for the larger and smaller size simulations), mimicking athermal nucleation on a fine grid. [We note that, during time evolution, the initial condition is fast forgotten: for example, after a transient period, very similar results were obtained with a uniform $1/N$ starting, while adding a small pixelwise Gaussian noise (a spatially random initiation).] Two cases were investigated: (a) with an isotropic grain boundary energy (on a $8192^2$ grid), and (b) the misorientation dependence of the grain boundary energy follows the Read-Shockley relationship \cite{ReadShockley1950} (on a $4096^2$ grid). The corresponding mobilities were $\kappa_{ij}=1$, and $\kappa_{ij}=|u_i/(1-u_i)|\,|u_j/(1-u_j)|$, respectively.

A typical grain map displaying the result of the simulation for case (b) at a dimensionless time $t=10^4$, when $\sim4000$ grains exist,  is shown in Fig. 6.  Similar grain maps have been obtained for the other case, except that there most of the trijunctions are close to symmetric, displaying angles $\sim120^{\circ}$.      

As observed in other MPF models\cite{Kim2006,Schaffnit2007}, in the experiments\cite{Barmak2013}, and predicted by theory\cite{Feltham1950,Mullins1956,Hillert1965}, after a transient period a limiting grain size distribution (LGSD) is established, which in the case of experiments on metallic film can be accurately fitted\cite{Barmak2013} by the lognormal distribution proposed by Feltham\cite{Feltham1950}. The models of Mullins and Hillert predict significantly different LGSDs (Fig. 7). The LGSD predicted by the XMPF model approximates the experimental results somewhat better than the previous MPF models\cite{Kim2006,Schaffnit2007} [see Figs. 7(b) and 7(c)], and practically coincides with the results from the {\it mutli order parameter} approaches\cite{ChenYang1994,FanChen1997a,Moelans2014}, yet the agreement is not particularly good with the experiments at small grain sizes. Apparently, in the experiments the small grains disappear faster than in the XMPF simulations. In the investigated cases, the time dependence of the average grain size [$\langle r \rangle = A (t-t_0)^q$, where $A$ is a constant and $t_0$ the freezing time] is described by an exponent $q = 0.5 (1 \pm 0.05)$, indicating an essentially diffusion controlled grain growth. 

We mention in this respect that a simple dynamical density functional theory, the Phase-Field Crystal approach\cite{Elder2002,Emmerich2012}, which incorporates a broad range of physical phenomena (elasticity, dislocation dynamics, grain rotation, etc.), reproduces the experimental LGSD fairly well\cite{Backofen2014}. Unfortunately, in the PFC studies, as in the case of experiments, the effect of different physical phenomena on the LGSD cannot be easily separated. It is expected, however, that the comparison of different models may contribute to the identification of the governing phenomena. Along these lines, the present study determined the LGSD the physically consistent XMPF model predicts. Apparently, further efforts are needed to improve the agreement between MPF models and experiments. Work is, underway\cite{Korbuly2015} to evaluate LGSD from phase-field models relying on orientation field(s)\cite{Warren2003,Plapp2012,Granasy2014} in describing different crystallographic orientations.

\color{black}

\section{Comparison with other models}

Having presented the essential properties of the XMPF model, it is desirable to compare it with other models from the following viewpoints:\\

A. For a few of the most important multiphase-field models, we investigate whether the trivial $N = 3$ extension of the equilibrium binary solution is a stationary solution of the $N = 3$ dynamic problem [part of criterion (vi), which in addition requires that the trivial extension be a solution of the $N = 3$ Euler-Lagrange equation too].\\ 

B. We explore, furthermore, for the best behaving models identified in sub-section A. whether the free energy decreases indeed monotonically with time [criterion (v)].\\   

\begin{figure}
 (a)\includegraphics[width=0.9\linewidth]{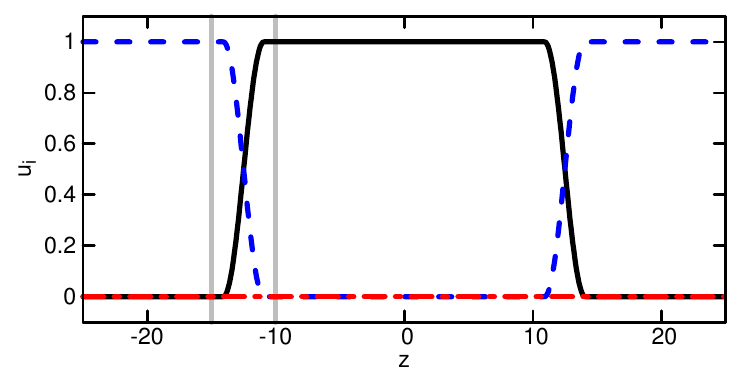}\\
 (b)\includegraphics[width=0.44\linewidth]{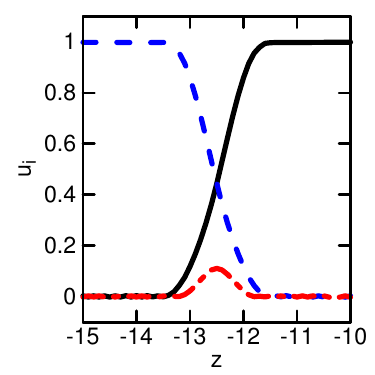}
 (c)\includegraphics[width=0.44\linewidth]{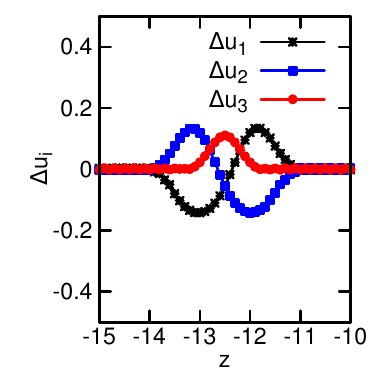}\\   	
 (d)\includegraphics[width=0.44\linewidth]{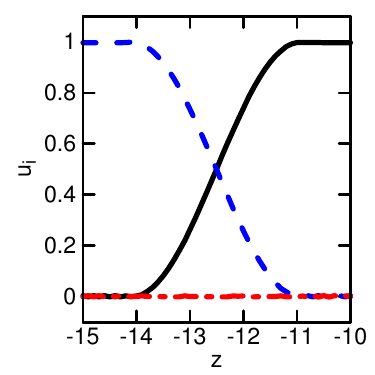}
 (e)\includegraphics[width=0.44\linewidth]{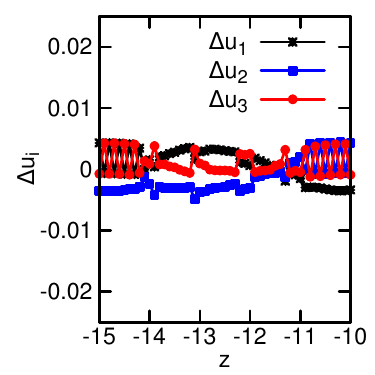}\\
 (f)\includegraphics[width=0.44\linewidth]{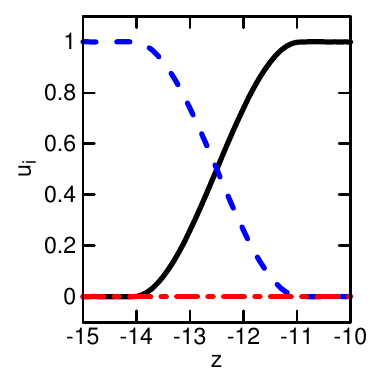}
 (g)\includegraphics[width=0.44\linewidth]{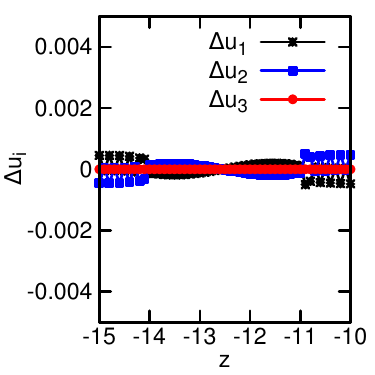}\\
  \caption{\label{fig:planar_sin} (Color online) $u(z)=[1+\sin(z)]/2$ models. (a) Initial condition (top panel), (b)-(g) final states after $10^6$ time steps. From second to  bottom row, respectively: the model of Nestler and Wheeler $p=1$ \cite{Nest2,Nest4,Nest5}, the model of Steinbach and Pezzola \cite{SteinbachPezzola1999}, and the model of Steinbach and Pezzola with non-variational dynamics \cite{Steinbach2009}. (Left column: final states, right column: difference of final state and initial condition.)}
\end{figure}

\begin{figure}
 (a)\includegraphics[width=0.9\linewidth]{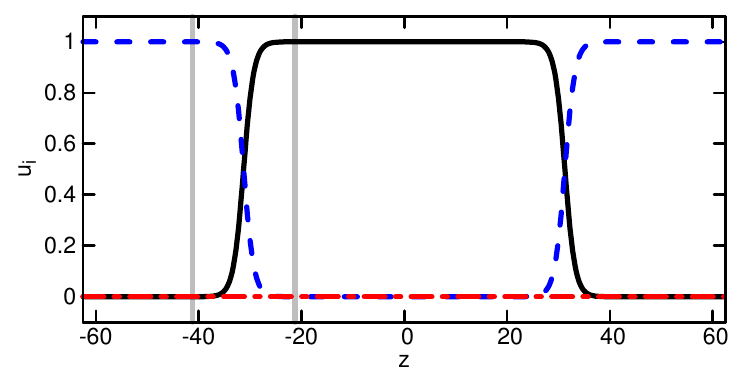}
 (b)\includegraphics[width=0.44\linewidth]{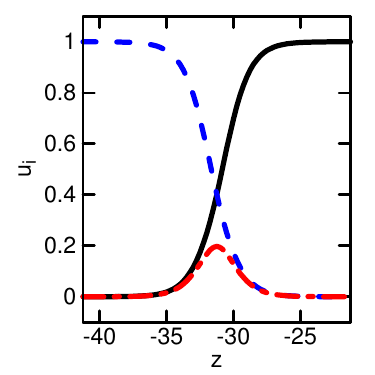}
 (c)\includegraphics[width=0.44\linewidth]{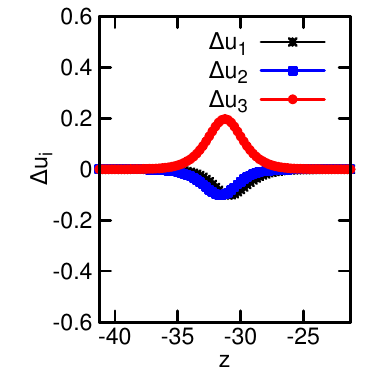}\\
 (d)\includegraphics[width=0.44\linewidth]{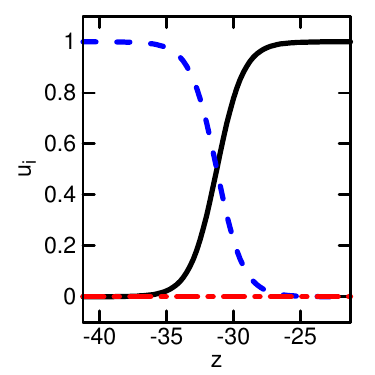}
 (e)\includegraphics[width=0.44\linewidth]{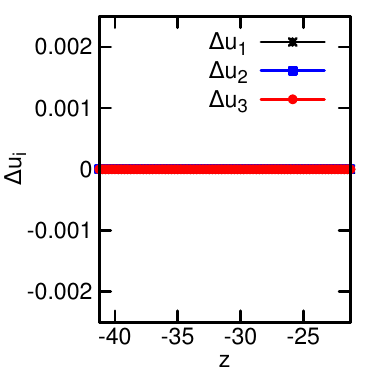}\\
 (f)\includegraphics[width=0.44\linewidth]{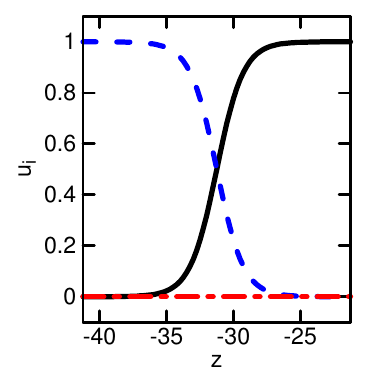}
 (g)\includegraphics[width=0.44\linewidth]{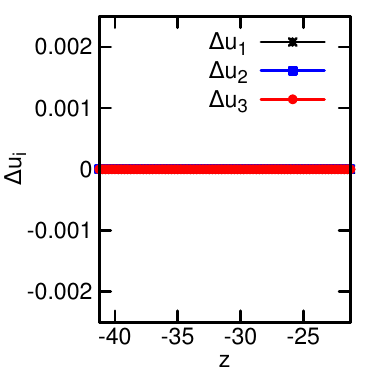}\\
  \caption{\label{fig:planar_tanh} (Color online) $u(z)=[1+\tanh(z/2)]/2$ models. (a) Initial condition (top panel), (b)-(g) final states after $2.5 \times 10^5$ time steps. From second to bottom row, respectively: the model of Steinbach {\it et al.} \cite{Steinbach1996} (coincides with the model of Nestler and Wheeler for $p=2$ \cite{Nest1,Nest2,Nest3,Nest4,Nest5}), the model of Steinbach {\it et al.} with non-variational dynamics \cite{Steinbach1996}, and the XMPF model proposed in this work. (Left column: final states, right column: difference of final state and initial condition.)}
\end{figure}

\subsection{Planar interfaces: \\ comparing the MPF models}

Here, we investigate for several MPF models, whether a trivial $N = 3$ extension of the equilibrium interface of the two-phase problem (obtained by adding $u_3 = 0$ to the two-phase equilibrium solution) behaves like a local free energy minimum of the multiphase-field model: starting from the extended solution, we explore whether the $N = 3$ equations of motion keep the solution equal to the starting condition, or drive it away. For this test, we adopt non-conservative dynamics
\begin{equation*}
-\frac{\partial u_i}{\partial t} = \sum_{j \neq i} \left( \frac{\delta F}{\delta u_i}-\frac{\delta F}{\delta u_j} \right) \enskip ,
\end{equation*}

\noindent The initial condition is a liquid-solid-liquid slab, with periodic boundary condition at the two ends, while employing the respective analytic solutions in the interface regions, accompanied with $u_3(z) = 0$ throughout the computation box. (See Figs. 8(a) and 9(a) for the initial conditions used for the models that have binary equilibrium solutions of the forms $u(z)=[1+\sin(z)]/2$ and $u(z)=[1+\tanh(z/2)]/2$, respectively).\\ 

The one-dimensional dynamic equations were solved numerically, using finite difference method, while employing dimensionless time and spatial steps, $h$ and $\Delta t$, as specified below.\\ 

\subsubsection{Models with sinusoidal equilibrium profile} 

The long-time solutions of the dynamic equations ($10^6$ time steps, beyond which no further changes were perceptible) are shown in Fig. 8 for the Nestler-Wheeler $p=1$ model \cite{Nest2,Nest4,Nest5}[Figs. 8(b) and 8(c)], for the Steinbach--Pezzola model \cite{SteinbachPezzola1999} [Figs. 8(d) and 8(e)], and for the Steinbach--Pezzola model with non-variational dynamics \cite{Steinbach2009} [Figs. 8(f) and 8(g)] ($h = 0.1$ and $\Delta t = 0.0025$ were used.) The long-time interfacial field profiles are shown on the left [Figs. 8(b), 8(d), and 8(f)], together with their difference relative to the initial conditions on the right [Figs. 8(c), 8(e), and 8(g)]. While in the first and second cases, third-phase generation can be seen at the interface, the application of non-variational dynamics in the Pezzola-Steinbach model suppressed this phenomenon ($u_3 \approx 0$ was retained).      

\subsubsection{Models with hyperbolic tangential equilibrium profile} 
  
The long-time solutions ($2.5 \times 10^5$ time steps, beyond which no perceptible changes were seen) of the EOMs are shown in Fig. 9 for the following models: Figs. 9(b) and 9(c) -- the model of Steinbach {\it et al.} \cite{Steinbach1996}; Figs. 9(d) and 9(e) -- the model of Steinbach {\it et al.} \cite{Steinbach1996} with non-variational dynamics; Figs. 9(f) and 9(g) -- the model proposed in the present paper. ($h = 0.25$ and $\Delta t = 0.01$.) The long-time interfacial field profiles are shown on the left [Figs. 9(b), 9(d), and 9(f)], together with their difference relative to the initial conditions on the right [Figs. 9(c), 9(e), and 9(g)]. While the model of Steinbach {\it et al.} \cite{Steinbach1996} leads to third-phase generation, the other two approaches are free of this problem. Remarkably, the predictions from the latter two models fall very close to each other. Yet, in the model of Steinbach {\it et al.} \cite{Steinbach1996}, the trivial three-phase extension of the binary equilibrium solution is not a solution of the three-phase Euler-Lagrange equation (see Section III.C). In other words: although the same solution is a stationary solution of the non-variational EOM, stabilized by the non-variational dynamics, it is not a free energy minimum of the three-phase problem.        

While in this test, the results of the model of Steinbach {\it et al.} (with non-variational dynamics) are practically indistinguishable from those of the XMPF model proposed in this work, under other conditions significant differences can be seen.

\begin{figure}
 (a)\includegraphics[width=0.43\linewidth]{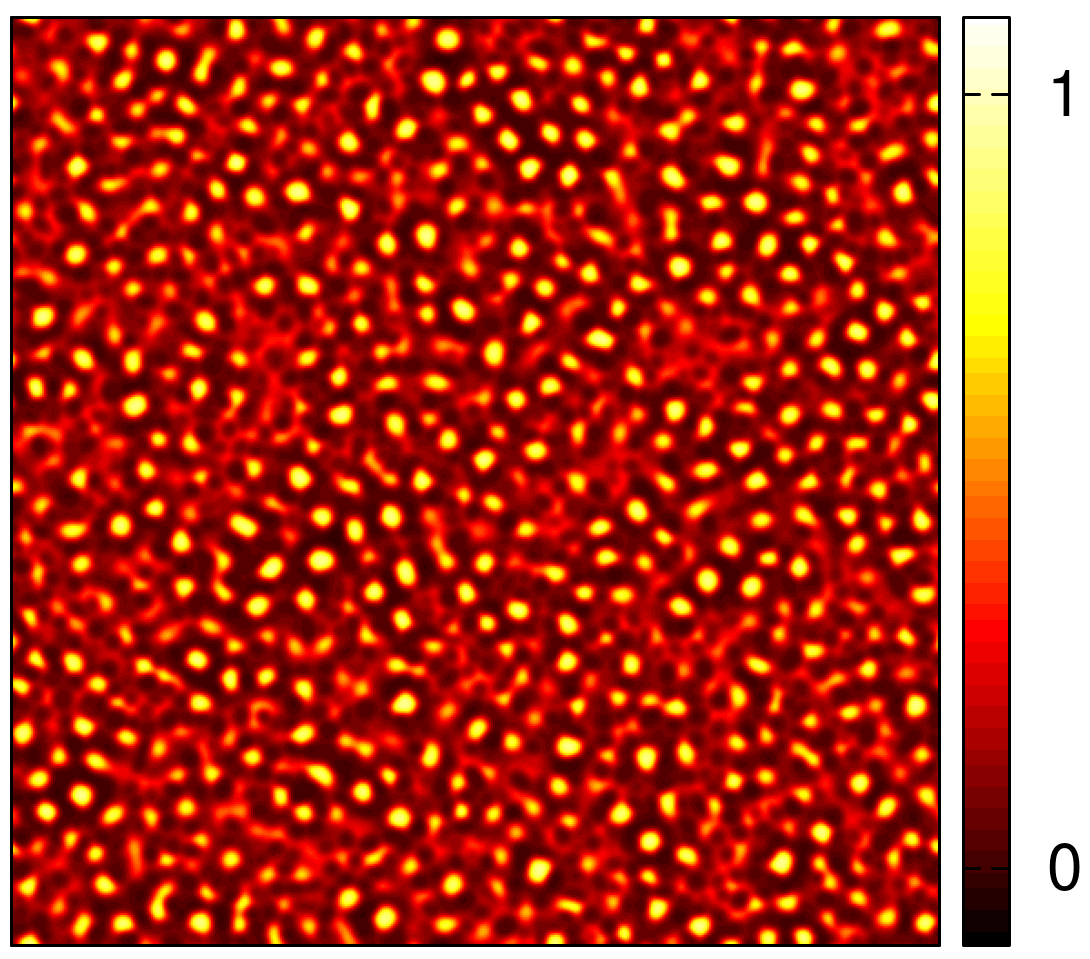}  (b)\includegraphics[width=0.43\linewidth]{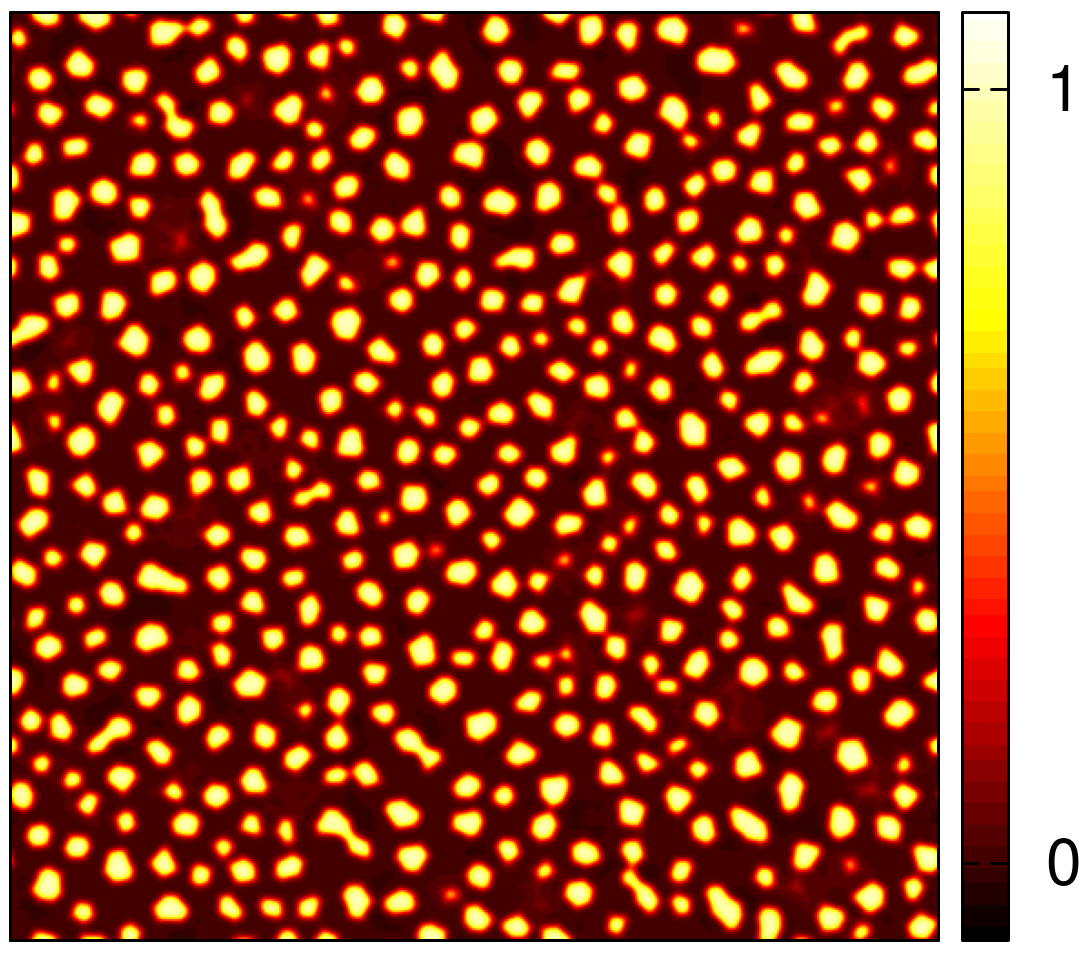}\\
 (c)\includegraphics[width=0.43\linewidth]{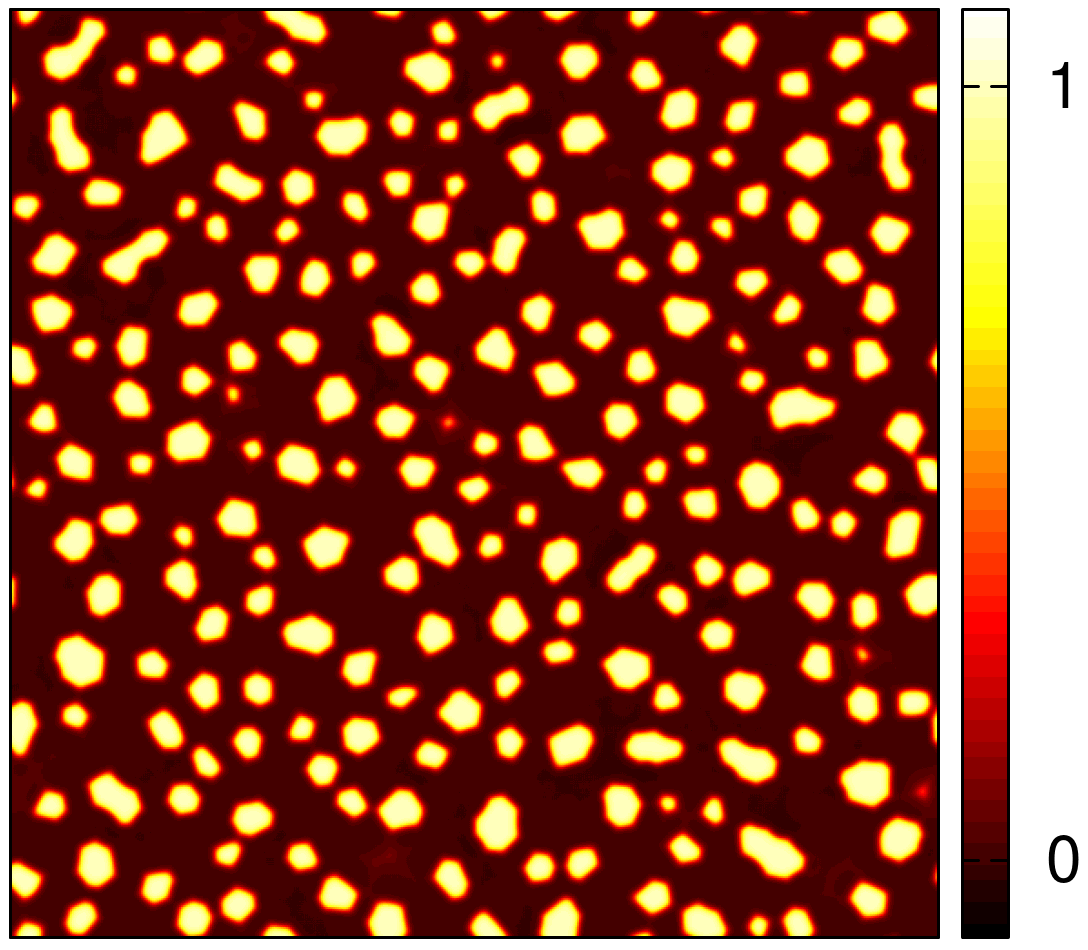} (d)\includegraphics[width=0.43\linewidth]{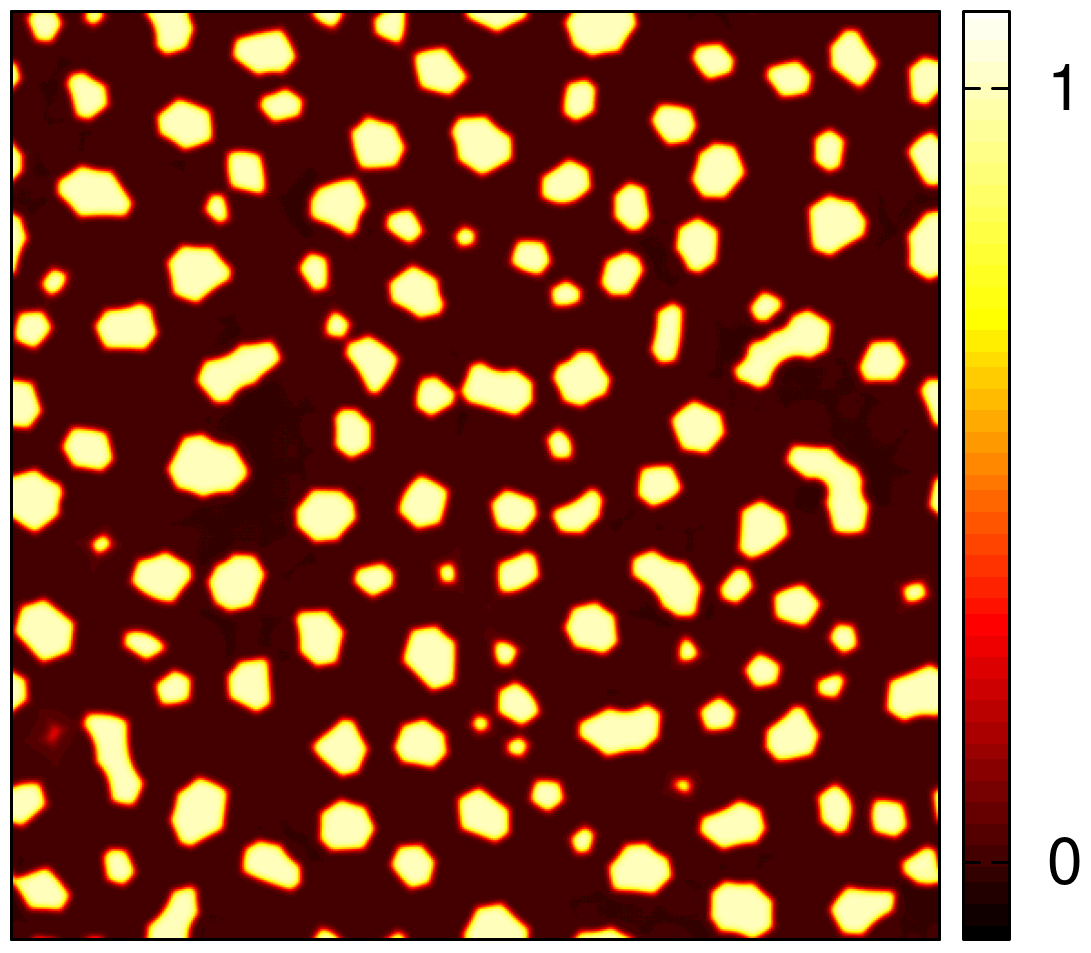}\\
 (e)\includegraphics[width=0.9\linewidth]{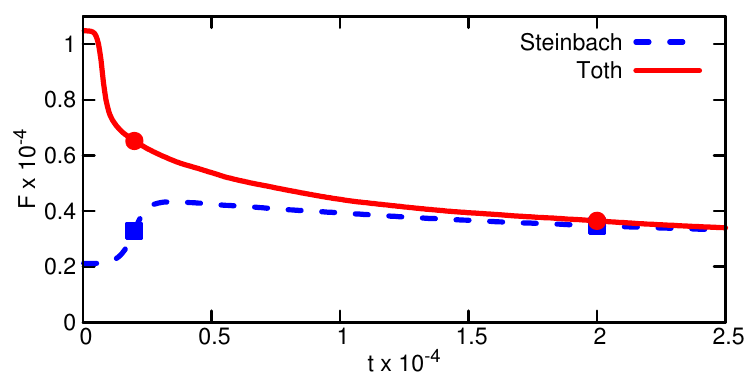}
  \caption{\label{fig:energy} (Color online) Comparison of the time evolution of phase transition (a), (c) in the non-variational model of Steinbach {\it et al.} \cite{Steinbach1996} (left), and (b), (d) in the present model (right). Snapshots of one of the fields are displayed. The time dependence of the free energy is shown in panel (e): dashed line -- model of Steinbach {\it et al.}; solid line -- the XMPF model.  The symbols correspond to snapshots shown in panels (a)-(d).}
\end{figure}

\subsection{Time dependence of the free energy}
 
In this test, we investigate the time evolution of the system in the non-variational model of Steinbach {\it et al.} \cite{Steinbach1996}, and in the XMPF model presented in this work. A symmetric $N = 4$ Cahn-Hilliard model and Lagrangian mobility matrix have been chosen for the demonstration. 

The results are summarized in Fig. 10, which shows the map of one of the fields (the others are qualitatively similar) at dimensionless times $t = 2000$ and $20000$, computed on a grid $512 \times 512$. While in the XMPF model proposed here, the free energy decreases monotonically with time as expected, in the model of Steinbach {\it et al.} (when relying on a non-variational formalism), the free energy increases initially, reaching then a maximum at about $t = 3000$, followed by a slow decrease beyond. This behavior is presumably a consequence of the applied 'binary' approximation, in which various terms of the variational equations of motion are omitted: once the free energy functional (Lyapunov functional) is defined, the variational dynamics ensures a monotonic reduction of the free energy with time. Any deviation from this approach raises the possibility of a non-monotonic time evolution of the free energy.                

\section{Summary}

In this work, we formulated a physically consistent multiphase-field theory for describing interface driven multi-domain processes. First, we identified a set of criteria, a physically consistent multiphase-field approach has to satisfy. These are: (i) the sum of the fields is 1 everywhere;  (ii) the physical results should be invariant to exchanging pairs of field indices, $i \leftrightarrow j$; (iii) a trivial multiphase extension of the equilibrium binary solution should represent an equilibrium solution of the multiphase problem, which in turn should be a stationary solution of the dynamic equations towards which the time dependent solutions evolve; (iv) variational dynamic equations shall be used to ensure non-negative entropy production;  (v) reduction/extension of the $N$-field theory to $N-1$ or $N+1$ fields should be straightforward, and happen consistently within the formalism; (vi) there should be no spurious third phase appearance at the equilibrium binary boundaries, and once a field is not present, it should not appear at any time in the dynamic equations; finally, (vii) freedom to choose the interfacial and kinetic properties for individual phase pairs. 

Next, considering these requirements, we have reviewed a range of the existing multiphase field models, and identified their advantageous and less advantageous features. 

Combining the advantageous features of the earlier multiphase-field models, we have constructed a multiphase-field approach (termed the XMPF model) obeying all criteria defined above. In addition, we performed illustrative simulations for $N = 4$ and 5 multiphase-field models that rely on conserved dynamics, describing thus multiphase separation problems ($N$-component Cahn-Hilliard problems). Symmetric (identical interface properties), asymmetric (pairwise different interface properties), and anisotropic (orientation dependent interfacial properties) cases were addressed, and it has been shown that using a suitable mobility matrix (Bollada-Jimack-Mullis type), the XMPF model avoids dynamic spurious phase generation. {We have performed further illustrative simulations for grain coarsening in polycrystalline systems using an $N=30$ XMPF model relying on non-conserved dynamics. While the predicted limiting grain size distribution is closer to the experimental results than those from the previous MPF models, further works is needed to improve the agreement.}  

The present work opens up the way towards physically consistent computations for microstructure evolution in multiphase / multigrain / multicomponent structures, and shall serve as a basis for developing a physically consistent {\it quantitative} multiphase-field approach that might be combined with melt flow and elasticity, and extended to fast processes along the lines described in Refs. \cite{Galenko1,Archer2009,Galenko2,TothGranasyTegze2014}, leading towards developing improved tools for knowledge based materials design.      

Work is underway to incorporate a phase-dependent thermodynamic driving force (a multiphase analogy of the 'tilting function' in Ref. \cite{FolchPlapp2005}) into the XMPF model, which will be presented in a separate paper. 

{We note in this respect that the inclusion of thermodynamic driving force via a tilting function has no effect on the present results concerning the two- and multiphase equilibria. The existence of equilibrium two-phase planar interfaces in the multiphase problem is a basic requirement, which needs to be satisfied by a physically consistent model.}   

\acknowledgments{This work has been supported by the VISTA basic
research programme of the Norwegian Academy of Science and Letters
and the Statoil, under project No. 6359 "Surfactants for
water/CO2/hydrocarbon emulsions for combined CO2 storage
and utilization", by the ESA PECS contracts of No. 40000110756/11/NL/KML and 40000 110759/11/NL/KML, and by the EU FP7 project EXOMET (contract No. NMP-LA-2012-280421, co-funded by ESA).}

\section*{Appendix A1: Invariance of results to exchanging pairs of field indices, $i \leftrightarrow j$}
 
The general dynamic equations of a multiphase model read as:
\begin{equation*}
-\frac{\partial u_i}{\partial t} = \sum_{j\neq i} \kappa_{ij} \left( \frac{\delta F}{\delta u_i}-\frac{\delta F}{\delta u_j} \right) \enskip ,
\end{equation*}
where there are $N(N-1)$ mobilities ($\kappa_{ij}$). \textit{The principle of formal indistinguishability} of the variables means that the variables are not "labeled", i.e. none of them is distinguished formally on the basis of its index. This is true if the dynamic equations are invariant for the re-labeling of the variables, i.e. re-labeling of the variables on the level of the free energy functional results in the same as re-labeling the variables in the dynamic equations. This criterion is satisfied by symmetric mobility matrices, namely,
\begin{equation*}
\kappa_{ij}=\kappa_{ji} \enskip .
\end{equation*}

\textit{Proof.} The dynamic equation for $u_J$ reads as
\begin{equation}
\label{eq:AppA1}
-\frac{\partial u_J}{\partial t} = \sum_{k\neq J} \kappa_{Jk} \left( \frac{\delta F}{\delta u_J}-\frac{\delta F}{\delta u_k} \right) \enskip .
\end{equation}  
The variables can be re-labeled by using the variable transformation $v_k:=u_k$ for $k\neq I,J$, $v_I:=u_J$, and $v_J:=u_I$. Using this in Eq. (\ref{eq:AppA1}) yields then
\begin{equation*}
-\frac{\partial v_I}{\partial t} = \sum_{k\neq I,J} \left(  \frac{\delta F}{\delta v_I}-\frac{\delta F}{\delta v_k} \right) + \kappa_{JI} \left( \frac{\delta F}{\delta v_I} - \frac{\delta F}{\delta v_J} \right) \enskip ,
\end{equation*}
where the chain rule for the functional derivative has also been used (see Appendix A2). Furthermore, re-labeling variables in the free energy functional first ($F[\mathbf{u}]\to F[\mathbf{v}]$), then deriving the dynamic equations simply results in
\begin{equation*}
-\frac{\partial v_I}{\partial t} = \sum_{k\neq I,J} \left(  \frac{\delta F}{\delta v_I}-\frac{\delta F}{\delta v_k} \right) + \kappa_{IJ} \left( \frac{\delta F}{\delta v_I} - \frac{\delta F}{\delta v_J} \right) \enskip .
\end{equation*}  
Comparing the two equations yields then $\kappa_{IJ}=\kappa_{JI}$.\\

In order to illustrate the "no labeling" condition in practice, we choose a typical example of labeling the variables. Some authors eliminate of one of the variables even at the level of the free energy functional, i.e. they introduce the independent variables $v_i:=u_i$ for $i=1\dots N-1$, thus resulting in $u_N=1-\sum_{i=1}^{N-1}v_i$. Then, the following dynamic equations are used:
\begin{equation*}
-\tau_i\frac{\partial v_i}{\partial t}=\frac{\delta F}{\delta v_i} \enskip .
\end{equation*}
These can be written in terms of the old variables as:
\begin{equation}
\label{eq:elim1}
-\frac{\partial u_i}{\partial t} = \frac{1}{\tau_i} \left( \frac{\delta F}{\delta u_i} - \frac{\delta F}{\delta u_N} \right)
\end{equation}
for $i=1\dots N-1$, and
\begin{equation}
\label{eq:elim2}
-\frac{\partial u_N}{\partial t} = - \sum_{j=1}^{N-1}\frac{1}{\tau_i}\left( \frac{\delta F}{\delta u_i} - \frac{\delta F}{\delta u_N} \right) \enskip .
\end{equation}
It is straightforward to see, that Eqs. (\ref{eq:elim1}) and (\ref{eq:elim2}) prescribe the following mobility matrix:
\begin{equation*}
L_{ii}=\tau_i^{-1} \quad \text{and} \quad L_{iN}=-\tau_i^{-1}
\end{equation*}  
for $i=1\dots N-1$, while the last row reads as
\begin{equation*}
L_{Ni}=-\tau_i^{-1} \quad \text{and} \quad L_{NN}=\sum_{i=1}^{N-1}\tau_i^{-1} \enskip ,
\end{equation*}  
where the form $-\partial u_i/\partial t=\sum_{j=1}^N L_{ij} (\delta F/\delta u_j)$ is used. It is trivial that the elements of $\mathbb{L}$ sum up to 0 in each row and column, but the matrix is not symmetric! It means that the concept of eliminating a variable on the level of the free energy functional labels the variables, i.e. the eliminated variable is formally distinguished. Indeed, exchanging variables $I$ and $N$, deriving the dynamic equations, then exchanging $I$ and $N$ back result in a mobility matrix similar to the one described by Eqs. (\ref{eq:elim1}) and (\ref{eq:elim2}), however, the $N^{th}$ and the $I^{th}$ rows are exchanged. On the one hand, it means that the formal variable exchange corresponds to the elimination of phase $I$ instead of phase $N$. On the other hand, since the resulting mobility matrix is not identical to the original one, the eliminated variable is always labeled, therefore, this concept does not satisfy the condition of no labeling.\\

\section*{Appendix A2: Chain rule for functional differentiation}

Mathematically speaking, \textit{the solution of the Euler-Lagrange equations is invariant to the variable transformation $\mathbf{Q}=\mathbf{T}[\mathbf{q}]$, if the transformation $\mathbf{T}[.]$ is unambiguous, i.e., if the inverse transform $\mathbf{T}^{-1}[.]$ also exists}.

\textit{Proof.} The Euler-Lagrange equations for the new variables read as:
\begin{equation}
\label{eq:genFD}
\frac{\delta F}{\delta Q_i} = \frac{\partial I}{\partial Q_i} - \nabla \frac{\partial I}{\partial \nabla Q_i} = 0 \enskip ,
\end{equation}
where $I$ denotes the full integrand of Eq. (\ref{eq:genfunc}). The terms on the right-hand side can be expanded as follows:
\begin{eqnarray}
\label{eq:ELtr1} \frac{\partial I}{\partial Q_i} &=& \sum_j \frac{\partial I}{\partial q_j}\frac{\partial q_j}{\partial Q_i} + \frac{\partial I}{\partial \nabla q_j}\frac{\partial \nabla q_j}{\partial Q_i}  \\
\label{eq:ELtr2} \nabla \frac{\partial I}{\partial \nabla Q_i} &=& \nabla \left[ \sum_j \frac{\partial I}{\partial q_j}\frac{\partial q_j}{\partial \nabla Q_i} + \frac{\partial I}{\partial \nabla q_j}\frac{\partial \nabla q_j}{\partial \nabla Q_i} \right] \enskip .
\end{eqnarray}
Since $q_j=T^{-1}_j(\mathbf{Q})$, $\partial q_j/\partial \nabla Q_i =0$. In addition, formally $\nabla q_j = \sum_i \frac{\partial q_j}{\partial Q_j} \nabla Q_i$, therefore, $\frac{\partial \nabla q_j}{\partial \nabla Q_i}=\frac{\partial q_j}{\partial Q_i}$. Using these together with Eqs. (\ref{eq:ELtr1}) and (\ref{eq:ELtr2}) in Eq. (\ref{eq:genFD}) yields
\begin{equation}
\label{eq:ELexpand}
\begin{split}
\frac{\delta F}{\delta Q_i} &= \sum_j \left( \frac{\partial I}{\partial q_j} - \nabla \frac{\partial I}{\partial \nabla q_j} \right) \frac{\partial q_j}{\partial Q_i} +\\
&+ \sum_j  \frac{\partial I}{\partial \nabla q_j} \left( \frac{\partial \nabla q_j} {\partial Q_i} - \nabla \frac{\partial q_j}{\partial Q_i} \right) \enskip 
\end{split}
\end{equation}
Finally, $\nabla q_j = \sum_k \frac{\partial q_j}{\partial Q_k} \nabla Q_k \Rightarrow \frac{\partial \nabla q_j}{\partial Q_i} = \sum_k \frac{\partial^2 q_j}{\partial Q_k \partial Q_i} \nabla Q_k$ and $\nabla \frac{\partial q_j}{\partial Q_i} = \sum_k \frac{\partial^2 q_j}{\partial Q_i \partial Q_k} \nabla Q_k$, therefore, the second sum on the right hand side of Eq. (\ref{eq:ELexpand}) vanishes. The final result then reads as:
\begin{equation}
\label{eq:ELfinal}
\frac{\delta F}{\delta Q_i} = \sum_j \frac{\delta F}{\delta q_j} \frac{\partial q_j}{\partial Q_i} \enskip ,
\end{equation}
i.e. \textit{the chain rule of differentiation also applies for the functional derivative}. Let now $\mathbf{q}^*(\mathbf{r})$ denote the solution of $\delta F/\delta \mathbf{q}=0$. Apparently, the right hand side of Eq. (\ref{eq:ELfinal}) vanishes at $\mathbf{q}^*(\mathbf{r})$. Formally $\mathbf{q}^*(\mathbf{r})=\mathbf{T}^{-1}[\mathbf{Q}^*(\mathbf{r})]$, indicating that $\mathbf{Q}^*(\mathbf{r}) = \mathbf{T}[\mathbf{q}^*(\mathbf{r})]$, i.e. the solution in $\mathbf{Q}$ is just the transformation of the solution in $\mathbf{q}$. In other words, the solution of the Euler-Lagrange equations is invariant to the choice of the generalized variables.\\

\section*{Appendix B: Numerical method}

The dynamic equations were solved numerically on a periodic, two-dimensional domain by using an operator-splitting based, quasi-spectral, semi-implicit time stepping scheme as follows. The dynamic equations can be re-written in the form
\begin{equation}
\label{eq:solve}
\frac{\partial \mathbf{u}}{\partial t} = \mathbf{f}(\mathbf{u},\nabla\mathbf{u}) \enskip ,
\end{equation} 
where $\mathbf{f}(\mathbf{u},\nabla\mathbf{u})$ is the general, non-linear right-hand side. During time stepping $\mathbf{f}(\mathbf{u},\nabla\mathbf{u})$ is calculated at time point $t$, while $\partial u_i/\partial t$ is approximated as
\begin{equation}
\label{eq:disct}
\frac{\partial u_i}{\partial t} \approx \frac{u_i^{t+\Delta t}-u_i^t}{\Delta t} \enskip .
\end{equation}
Next, we add a suitably chosen linear term $\hat{s}[u_i]= \sum_{i=1}^\infty (-1)^i s_i \nabla^{2i} u_i$ (where $s_i \geq 0$) to both sides of Eq. (\ref{eq:solve}). We consider this term at $t+\Delta t$ at the left-hand side, but at $t$ on the right-hand side of the equation. This concept, together with Eq. (\ref{eq:disct}) results in the following, explicit time stepping scheme in the spectrum:
\begin{equation}
\label{eq:timestep}
u_i^{t+\Delta t}(\mathbf{k}) = u_i^t(\mathbf{k}) + \frac{\Delta t}{1+s_i(\mathbf{k})\Delta t} \mathcal{F}\{f_i[\mathbf{u}^t(\mathbf{r}),\nabla\mathbf{u}^t(\mathbf{r})]\} \enskip ,
\end{equation}
where $s_i(\mathbf{k})=\sum_{j=1}^\infty s^{(i)}_j (\mathbf{k}^2)^j$, and $\mathcal{F}\{.\}$ stands for the Fourier transform. The \textit{splitting constants} $\{s^{(i)}_j\}$ must be chosen so that Eq. (\ref{eq:timestep}) to be stable. 

It is important to note that our numerical scheme is \textit{unbounded}, which means that the spatial solution $u_i(\mathbf{r},t)$ can go under 0 or above 1 because of the numerical errors. The construction of the free energy functional and the modified Bollada-Jimack-Mullis mobility matrix, however, ensure that the system converges to equilibrium. This means that no artificial modification of the solution is needed after a time step, which could lead to instabilities in the spectral method. 


\end{document}